\documentclass[a4paper,fleqn]{cas-sc}

\usepackage[numbers]{natbib}

\def\tsc#1{\csdef{#1}{\textsc{\lowercase{#1}}\xspace}}
\tsc{WGM}
\tsc{QE}

\usepackage{amssymb}
\usepackage{latexsym}
\usepackage{amsmath}
\usepackage{bm}

\usepackage{subcaption}
\usepackage{float}

\usepackage{caption}
\usepackage{booktabs}
\usepackage{tabularx}

\usepackage{siunitx}

\usepackage{placeins}

\DeclareMathOperator{\grad}{\nabla}
\DeclareMathOperator{\dive}{\nabla\cdot}

\newcommand{\vel}{\bm{u}}
\newcommand{\disc}{\bm{q}}

\newcommand{\ave}[1]{\left\{\!\left\{#1\right\}\!\right\}}
\newcommand{\jump}[1]{\left[\!\left[#1\right]\!\right]}
\newcommand{\tjump}[1]{\left<\!\left<#1\right>\!\right>}
\newcommand{\pad}[2]{\frac{\partial{#1}}{\partial{#2}}}

\newcommand{\rpth}[1]{\left(#1\right)}
\newcommand{\spth}[1]{\left[#1\right]}
\newcommand{\cpth}[1]{\left\{#1\right\}}

\definecolor{dartmouthgreen}{rgb}{0.05, 0.5, 0.06}

\usepackage{todonotes}

\begin{document}

\let\WriteBookmarks\relax
\def\floatpagepagefraction{1}
\def\textpagefraction{.001}

\shorttitle{High-order adaptive discontinuous finite elements for SW equations with sub-grid bathymetry}
	
\shortauthors{L. Arpaia et al.}  
	
\title[mode = title]{High-order adaptive discontinuous finite elements for the shallow water equations with sub-grid irregular bathymetry}

\author[1]{Luca Arpaia}[orcid=0000-0002-5518-1377]
\cormark[1]
\ead{luca.arpaia@cnr.it}

\author[2]{Giuseppe Orlando}[orcid=0000-0002-7119-4231]
\cormark[2]
\ead{giuseppe.orlando@polytechnique.edu}

\author[1]{Christian Ferrarin}[orcid=0000-0003-1172-1463]
\ead{christian.ferrarin@cnr.it}

\author[3]{Luca Bonaventura}[orcid=0000-0002-1994-0217]
\ead{luca.bonaventura@polimi.it}

\affiliation[1]{organization={Consiglio Nazionale delle Ricerche, Istituto di Scienze Marine},
addressline={Arsenale – Tesa 105, Castello 2737/F}, 
city={Venezia},
postcode={30122}, 
country={Italy}}

\affiliation[2]{organization={CMAP, CNRS, \'{E}cole polytechnique, Institut Polytechnique de Paris},
addressline={Route de Saclay}, 
city={Palaiseau},
postcode={91120}, 
country={France}}	
	
\affiliation[3]{organization={Dipartimento di Matematica, Politecnico di Milano},
addressline={Piazza Leonardo da Vinci 32}, 
city={Milano},
postcode={20133}, 
country={Italy}}
	
\cortext[1]{Corresponding author}
	
\begin{abstract}
We present a discontinuous finite element method for the shallow water equations which exploits high-resolution realistic bathymetry data without any regularity assumption, also in the case of high-order discretizations. We prove a number of mathematical properties specific to the proposed method that is well-balanced, mass-conserving and positivity-preserving under a mild CFL condition also in the presence of wet-dry fronts.
The method includes a consistent conservative discretization for passive tracers. 
We use a high-order Discontinuous Galerkin (DG) method as implemented in the \texttt{deal.II} library. This environment provides efficient and native parallelization techniques and automatically handles non-conforming meshes to implement adaptive strategies which are tested in a coastal environment. Idealized test cases show the robustness in presence of irregular bathymetries also with under-resolved features at the grid scale. 
A benchmark with realistic bathymetry and a complex domain shows the potential of the proposed discretization for adaptive simulations of coastal flows.
\end{abstract}
	
	
	
	
\maketitle


\section{Introduction}
\label{sec:intro}

The shallow water equations are a simplified model of hydrostatic flows at small depth-to-width aspect ratio, which is widely used for modelling circulations in well-mixed basins, lagoons, and coastal seas. A rigorous derivation of this model from the incompressible Navier-Stokes equations can be found, e.g., in \cite{bresch:2007, decoene:2009}. A great number of different approximation techniques for their numerical solution has been proposed in the past, often tailored to address the specific difficulties posed by different flow regimes. For low Froude number regimes, which will be the focus of this work, efficient semi-implicit finite difference methods have been proposed, e.g., in the seminal papers \cite{casulli:1990, casulli:1992}. These methods are extremely popular in the coastal sea modelling community and they have been extended to high-order finite elements, e.g. in \cite{busto:2022, dumbser:2013, tavelli:2014, tumolo:2013}. In \cite{casulli:2009}, the semi-implicit finite difference method proposed in \cite{casulli:1990} has been modified to exploit high-resolution bathymetric datasets, which are now available for coastal areas at spatial resolutions that exceed the computational mesh resolution by orders of magnitude, and are significantly finer than those typically used in conventional coastal simulations based on low-order numerical schemes. The purpose of this work is to propose a high-order adaptive finite element solver for the shallow water equations for coastal sea modelling, which is also capable of exploiting high-resolution bathymetries.

While high-order finite element methods are appealing for applications in coastal environments, they face a number of challenges. As mentioned, bathymetry datasets are available at very high resolutions, whereas high-order methods typically employ coarser meshes. Such a disparity implies that the mesh may not be aligned with the direction of large bathymetric gradients or jumps and the resulting discretization must be able to handle such variations within an element or along an edge. On such coarse meshes, it may also be important to capture relevant bathymetric features that are under-resolved at the grid scale. For finite element methods, if the water depth is chosen as the prognostic variable in the continuity equation, in order to guarantee consistency with continuity and well-balancing, also the bathymetry must be approximated using the same polynomial basis as the water depth. In the case of irregular bathymetries and polynomials of order higher than one, smoothing or TVD limiting of the bathymetry is then needed to avoid or cure the well-known Gibbs phenomenon. This is particularly unfortunate when highly accurate bathymetric data are available.

We propose a robust and accurate approach that uses realistic bathymetry data without any regularity assumption. First, we choose as prognostic variable the free-surface elevation $\zeta$, similarly to the formulation considered in \cite{casulli:2009} and used in many coastal ocean models \cite{klingbeil:2018}. In subcritical regimes $\zeta$ is a smooth variable and thus suitable for a finite element representation. The bathymetry depth $z_{b}$ at the quadrature nodes is evaluated directly from the reference data, without any post-processing. When a higher order polynomial degree is employed, the number of quadrature nodes is increased and accurate sub-grid bathymetric features can be recovered. As in \cite{casulli:1992}, we consider the hydrostatic pressure gradient in a non-conservative form. This is fully justified in the low Froude regimes that are the target applications of the present work. The hydrostatic well-balancing and the ability to preserve steady states with uniform $\zeta$ and $\vel = \bm{0}$, known in the literature as the {\cal C}-property \cite{vazquez:1999}, follow automatically from this formulation, independently of the regularity of the bathymetry and the chosen quadrature formulae. We use the discretization of the non-conservative pressure term proposed in \cite{tumolo:2015}. Its relationship to the conservative formulation is discussed, showing that, for piecewise polynomial bathymetries with no jumps at the element boundaries, the approach proposed in \cite{tumolo:2015} is equivalent to a conservative discretization.

With respect to the conservation of mass, if the water depth $h = \zeta + z_{b}$ is strictly positive and the bathymetry $z_{b}$ does not change in time, evolving $h$ or the free-surface elevation $\zeta$ is equivalent. This is no longer true in presence of wetting and drying regions, because the relationship between $\zeta$ and $h$ becomes nonlinear and is given by $h = \max\rpth{\zeta + z_{b},0}$. The choice of evolving the water depth assures mass conservation straightforwardly and, for this reason, combined with a positivity-preserving method, it is the preferred choice for flooding applications, see e.g., the well-balanced \cite{bunya:2009, xing:2010} or pre-balanced discontinuous finite elements \cite{duran:2014}.
However, as mentioned, some of the assumptions underlying discretizations based on $h$ as a prognostic variable are often in contrast with the irregular bathymetries that are common in coastal modelling. Moreover, at wet/dry fronts it is not trivial to verify the well-balancing property with respect to hydrostatic equilibrium. In fact, if the mesh is not aligned to the wet-dry front, the water depth $h$ is not regular in those elements for which $h=0$ at some location. Even in the lake-at-rest state, this forces the activation of an additional bathymetry limiter that nullifies spurious waves from wet-dry elements, or other \textit{ad hoc} computations of the free-surface gradient in those elements \cite{bonev:2018, vater:2019}. A simpler, yet effective option for Discontinuous Galerkin (DG) is to degrade the formal order of accuracy in wet-dry elements, switching to a first-order, cell-centered Finite-Volume approximation, which is better suited for irregular solutions \cite{filippini:2024}. Recent studies have attempted to go beyond a piecewise constant bathymetry at wet-dry fronts by using the subcells technique with an \emph{a posteriori} \cite{haidar:2022} or \emph{a priori} limiting \cite{meister:2016} of the the free-surface and of the bathymetry.

If the free-surface elevation is used instead as prognostic variable, which is a common choice for coastal ocean models, \emph{ad hoc} techniques are often employed to restore mass conservation. These techniques, based on adding to the dry region a very thin layer of water and/or switching off singular terms, are not mathematically rigorous and may develop spurious oscillations along wet-dry fronts see, e.g., \cite{klingbeil:2018} for a review. A mass-conserving formulation that still keeps the free-surface elevation as prognostic variable has been proposed in \cite{casulli:2009} and is based on an iterative algorithm to solve the nonlinear relationship $h = \max\rpth{\zeta + z_{b},0}$. As mentioned above, using the free-surface elevation leverages an accurate  representation of irregular bathymetry also at wet-dry fronts which are of key importance for flooding. Recently, this technique has been adapted to a second-order Finite Volume method in \cite{bitsch:2026}. Our treatment of wetting and drying consists in an adaptation of the same technique proposed in \cite{casulli:2009} to high-order finite elements. To cope with the irregular discharge/velocity at the wet-dry fronts, we employ a degree adaptation strategy, reverting the solution to piecewise constant polynomials, based on \emph{a priori} heuristic criteria that detect wet-dry fronts. We establish two properties of the resulting numerical scheme: well-balancing in presence of wet-dry fronts and water depth positivity under a CFL condition that depends on the accuracy of the quadrature formula, i.e., the resolution of the sub-grid bathymetry. To the best of our knowledge, this approach yields the first high-order method that simultaneously combines well-balancing, mass conservation, and positivity preservation without any assumption on the bathymetry, i.e., without the need for regularization, smoothing, or TVD limiting.

Irregular bathymetry has additional implications on other model components, such as concentrations of biogeochemical species. Following the analysis in \cite{gross:2002, lin:1996}, we take special care to guarantee that the conservative discretizations of the continuity and species concentration equations are consistent. Failure of this consistency with continuity (CWC) property was shown in \cite{gross:2002} to cause macroscopic violations of monotonicity, even for completely monotonic methods. Indeed, we show that, for an irregular bathymetry not approximated with polynomials, a naive conservative update of the depth-integrated concentrations will fail to guarantee such a consistency. We then propose a consistent discretization that guarantees this property also for arbitrary bathymetries.

The spatial discretization is based on a high-order Discontinuous Galerkin (DG) approach. Applications of the DG method to the shallow water equations have been proposed by many authors since the early 2000s, on Cartesian \cite{tassi:2007, xing:2010} and on unstructured meshes, see \cite{giraldo:2002, kubatko:2006} among many others. We have used the DG method as implemented in the \texttt{deal.II} library \cite{arndt:2023, bangerth:2007}. This modelling framework has proven to be extremely useful for the implementation of numerical models for environmental flows, see, e.g., \cite{orlando:2022, orlando:2023a, orlando:2024a}. The \texttt{deal.II} library provides efficient and native parallelization techniques and automatically handles non-conforming meshes to implement dynamic $hp-$adaptation \cite{bangerth:2009} with excellent scalability results \cite{bangerth:2023}, as also shown in the previously cited papers. A numerical method for the shallow water equations based on \texttt{deal.II} has been recently proposed in \cite{guermond:2025}. 
Our longer term goal is to implement efficient Implicit--Explicit Runge--Kutta (IMEX-RK) time discretization schemes with an implicit treatment of the pressure gradient terms, as already done in \cite{orlando:2022, orlando:2023a} for atmospheric models, thus guaranteeing time step restrictions based on velocity rather than celerity values. However, in this work we will restrict our attention to IMEX-RK time discretization methods that treat the pressure gradient terms explicitly, while applying the implicit scheme to the frictional terms, that can become very stiff when realistic friction models are employed on shallow bathymetries. The IMEX scheme is based on the combination of the \texttt{L}-stable TR-BDF2 for the implicit part and an explicit three-stage second-order Runge--Kutta scheme designed to match the coupling conditions \cite{giraldo:2020,orlando:2022}.

Finally, we test the dynamic Adaptive Mesh Refinement (AMR) that can be implemented in \texttt{deal.II} for coastal applications. Well known applications of AMR to the shallow water equations have been presented, among many others, in \cite{leveque:2011, popinet:2011} and coupled to DG methods in \cite{gerhard:2015, kesserwani:2023}. As discussed, e.g., in \cite{arpaia:2018, donat:2014, liang:2015, popinet:2011}, when performing dynamic mesh adaptation, guaranteeing simultaneously well-balancing and mass conservation is not straightforward. Mass conservation can be achieved with conservative remaps of the bathymetry field that conserve the bathymetry average across the different meshes. We discuss the case of irregular sub-grid bathymetry where, to conserve the bathymetry integral, we cannot rely on exact integration of polynomial functions. In the discretization proposed here, the bathymetry values at the quadrature nodes for the new resolution are sampled at run-time from an analytical function or extracted from a bathymetry database.
In the case of dynamical AMR, the {\cal C}-property is guaranteed automatically, but mass conservation is not preserved. We quantify the mass conservation error and we evaluate it for short-term simulations, while demonstrating the potential efficiency gains achieved through dynamic AMR approaches.

The outline of the paper is the following. In Section \ref{sec:modeleq}, the model equations are discussed. Their numerical approximation is introduced in Section \ref{sec:numerical}. The validation of the properties and of the robustness of the proposed method in presence of irregular bathymetry (including under-resolved features at the grid scale) is presented in Section \ref{sec:tests}. In Section \ref{sec:realistic} we consider a realistic bathymetry and a complex geometry, reporting results of semi-realistic simulations of the tidal circulations in the Venice Lagoon. We use this test to demonstrate the effectiveness of the proposed AMR approach in enhancing local solution features at reduced computational cost. Some conclusions and perspectives for the future developments mentioned before are reported in Section \ref{sec:conclu}.

\section{Model equations}
\label{sec:modeleq}

We consider the shallow water equations for $\bm{x} \in \Omega$, $t \in (0,T],$ which can be formulated as
\begin{align}\label{eq:shw}
    &\pad{h}{t} + \dive\disc = 0, \nonumber \\
    &\pad{\disc}{t} + \dive\rpth{\disc \otimes \vel} + gh\grad\zeta = -\gamma\rpth{q, h}\disc, \\
    &\pad{hc_{k}}{t} + \dive\rpth{\disc c_{k}} = h \phi_{k}\rpth{c_{1},\dots,c_{M}}, \quad k=1,\dots,M, \nonumber
\end{align}
endowed with suitable initial and boundary conditions. Here $\Omega $ is a two-dimensional spatial domain, $T$ is the final time, $g = \SI{9.81}{\meter\per\second\squared}$ is the gravitational acceleration, $\zeta$ denotes the free-surface elevation and $z_{b}$ denotes the bathymetry depth, both computed with respect to a fixed reference height, so that $h = \zeta + z_{b}$ is the total water depth. The quantities $c_{k}, k=1,\dots,M$ represent concentration of passive tracers or, in presence of reaction rates $\phi_{k}$, the concentrations of chemical or biological species. Notice that no assumption is made on the regularity of the bathymetry $z_{b}$, that can have jumps even inside the elements of the mesh.

In presence of wetting or drying in certain regions of the domain, the relationship between the water depth and the free-surface is nonlinear. More specifically, it reads
\begin{equation}\label{eq:zeta2depth-2}
    h(\zeta)= \max\rpth{\zeta + z_{b}, 0}
\end{equation}
We denote by $\disc = (hu, hv)^{\top}$ the mass flux or discharge per unit area, while $\vel = (u, v)^{\top} = \disc/h$ denotes the vector of vertically averaged velocities, which in our formulation are diagnostic variables. The coefficient $\gamma$ denotes a non-negative friction coefficient, which will depend on the absolute value of the discharge $q = \left|\disc\right|$ and on the water depth $h$. More specifically, we will consider the friction model
\begin{equation}\label{eq:manning}
    \gamma\rpth{q,h} = \frac{g n^{2} q}{h^{7/3}},
\end{equation}
where $n$ denotes the so-called Manning friction coefficient. 

\section{Numerical discretization}
\label{sec:numerical}

In this section, we outline the numerical methods employed for the spatial and temporal discretization of system \eqref{eq:shw} and we analyze in detail the properties of the resulting scheme.

\subsection{Time discretization}
\label{ssec:timedisc}

For the time discretization, an IMplicit--EXplicit (IMEX) Runge--Kutta (RK) method \cite{kennedy:2003} is used. These methods are useful for time dependent problems that can be formulated as $\bm{y}' = \bm{f}_{\mathrm{S}}(\bm{y},t) + \bm{f}_{\mathrm{NS}}(\mathbf{y},t)$, where the $\mathrm{S}$ and $\mathrm{NS}$ subscripts denote the stiff and non-stiff components of the system, respectively, to which the implicit and explicit companion methods are applied. IMEX-RK methods are represented compactly by the two Butcher tableaux \cite{butcher:2008}
\begin{center}
    \begin{tabular}{c|c}
	   $\mathbf{c}$ & $\mathbf{A}$ \\
	   \hline
      \rule{0pt}{2.5ex}
	   & $\mathbf{b}^{\top}$
    \end{tabular}
    \qquad
    \begin{tabular}{c|c}
	   $\tilde{\mathbf{c}}$ & $\tilde{\mathbf{A}}$  \\
	   \hline
      \rule{0pt}{2.5ex}
	   & $\tilde{\mathbf{b}}^{\top}$
    \end{tabular}
\end{center}
with $\mathbf{A} = \cpth{a_{lm}}, \mathbf{b} = \cpth{b_{l}}, \mathbf{c} = \cpth{c_{l}}, \tilde{\mathbf{A}} = \cpth{\tilde{a}_{lm}}, \tilde{\mathbf{b}} = \cpth{\tilde{b}_{l}}$, and $\tilde{\mathbf{c}} = \cpth{\tilde{c}_{l}}$. Coefficients $a_{lm}, b_{l}$, and $c_{l}$ denote the coefficients of the explicit companion method, while $\tilde{a}_{lm}, \tilde{b}_{l}$, and $\tilde{c}_l$ denote the coefficients of the implicit companion method. The coefficients are determined so that the method is consistent of a given order. In particular, beyond the order conditions associated with each sub-method, they must satisfy additional coupling conditions, as discussed in detail in \cite{pareschi:2005}, to which we refer for a complete presentation (see also the discussion in Appendix \ref{app:IMEX_coeffs}). We consider Diagonal Implicitly Runge-Kutta (DIRK) methods for the implicit companion scheme, so that $\tilde{a}_{lm} = 0$ for $l < m$. If $\bm{v}^{n} \approx \bm{y}(t^{n})$, the generic $l-$stage IMEX-RK method can be defined therefore as
\begin{equation*}
    \bm{v}^{(l)} = \bm{v}^{n} + \Delta t \sum\limits_{m=1}^{l-1} \spth{a_{lm}\bm{f}_{\mathrm{NS}}\rpth{\bm{v}^{(m)}, t + c_{m}\Delta t} + \tilde{a}_{lm}\bm{f}_{\mathrm{S}}\rpth{\bm{v}^{(m)}, t + \tilde{c}_{m}\Delta t}} + \Delta t \,\tilde{a}_{ll}\,\bm{f}_{\mathrm{S}}\rpth{\bm{v}^{(l)}, t + \tilde{c}_{l}\Delta t},
\end{equation*}
where $l=1, \dots, s$. After the computation of the intermediate stages, $\bm{v}^{n+1}$ is computed as
$$\bm{v}^{n+1} = \bm{v}^{n} + \Delta t \sum_{l=1}^{s}\spth{b_{l}\,\bm{f}_{\mathrm{NS}}\rpth{\bm{v}^{(l)}, t + c_{l}\Delta t} + \tilde{b}_{l}\,\mathbf{f}_{\mathrm{S}}\rpth{\bm{v}^{(l)}, t + \tilde{c}_{l}\Delta t}}.$$
The generic semi-discrete IMEX-RK stage for system \eqref{eq:shw} with implicit treatment of the friction term then reads as follows
\begin{align}\label{eq:imex_shw}
    \max\rpth{\zeta^{(l)} + z_{b}, 0} &= \max\rpth{\zeta^{n} + z_{b}, 0} - \Delta t\sum_{m=1}^{l-1} a_{lm}\dive\disc^{(m)}, \nonumber \\
    h^{(l)} &= \max\rpth{\zeta^{(l)} + z_{b}, 0}, \nonumber \\
    \disc^{(l)} &= \disc^{n} - \Delta t\sum_{m=1}^{l-1} a_{lm}\dive\rpth{\disc^{(m)} \otimes \vel^{(m)}} - \Delta t\sum_{m=1}^{l-1}a_{lm}g h^{(m)}\grad\zeta^{(m)} \\
    &- \Delta t\sum_{m=1}^{l-1}\tilde{a}_{lm}\gamma\rpth{q^{(m)},h^{(m)}}\disc^{(m)} - \Delta t\tilde{a}_{ll}\gamma\rpth{q^{(l-1)}, h^{(l-1)}}\disc^{(l)}, \nonumber \\
    h^{(l)}c_{k}^{(l)} &= \nonumber h^{n}c_{k}^{n} - \Delta t\sum_{m=1}^{l-1} a_{lm} \dive\rpth{\disc^{(m)}c_{k}^{(m)}} + \Delta t\sum_{m=1}^{l-1} a_{lm} h^{(m)} \phi_{k}\rpth{c_{1}^{(m)},\dots,c_{M}^{(m)}}, \qquad k = 1 \dots M. \nonumber
\end{align}
Notice that the source terms representing biogeochemical reactions have been discretized in time with the explicit part of the IMEX method. In most commonly used biogeochemical models, these terms are non-stiff and often simply discretized by the forward Euler method in practical applications. However, inclusion of stiff chemical reactions discretized by the implicit part of the IMEX scheme would be possible along the same lines, resulting in a set of decoupled $M \times M$ nonlinear systems for each discrete degree of freedom. The full discretization of these terms will not be considered further in this paper, but it can be included without changes in the general structure of the proposed method and will be taken into account in more realistic simulations to be carried out in future work.

The nonlinear dependence of the friction term on the discharge and water depth values is linearized in time, so as to avoid the solution of a nonlinear system at each time step. Since the stability restrictions based on celerity do not allow the use of very large time steps for this approach, and considering also the uncertainties associated with the empirical friction formulae usually employed to define $\gamma$, this simplification does not imply a significant loss of accuracy. All the details of the Butcher tableaux used for the time discretization are reported in Appendix \ref{app:IMEX_coeffs}.

We stress the fact that we solve for the depth-averaged concentration $c$ and not for the depth-integrated mass $hc$, as it would be natural for a conservative formulation. As explained later in this section, this allows us to obtain a discretization that is consistent with that of the continuity equation. This choice also simplifies the implementation of the initial and boundary conditions, as well as the output of the concentration variables.

\subsection{Spatial discretization}
\label{ssec:spacedisc}

For the spatial discretization, we employ a nodal DG method \cite{giraldo:2020}, which is implemented in the framework of the \texttt{deal.II} library \cite{arndt:2023, bangerth:2007}. We consider a decomposition of the domain $\Omega$ into a family of quadrilaterals $\mathcal{T}_{\mathcal{H}}$ and denote each element by $K$ with mesh size $\mathcal{H}_{K}$. The skeleton $\mathcal{E}$ denotes the set of all element faces and $\mathcal{E} = \mathcal{E}^{I} \cup \mathcal{E}^{B}$, where $\mathcal{E}^{I}$ is the subset of interior faces and $\mathcal{E}^{B}$ is the subset of boundary faces. Then, suitable jump and average operators have to be defined for discontinuous finite element discretizations. A face $\varGamma \in \mathcal{E}^{I}$ shares two elements that we denote by $K^{+}$ with outward unit normal $\bm{n}^{+}$ and $K^{-}$ with outward unit normal $\bm{n}^{-}$, whereas for a face $\varGamma \in \mathcal{E}^{B}$ we denote by $\bm{n}$ the outward unit normal. For a scalar function $\varphi$ the jump is defined as
$$\jump{\varphi} = \varphi^{+}\bm{n}^{+} + \varphi^{-}\bm{n}^{-} \quad \text{if }\varGamma \in \mathcal{E}^{I}, \qquad \jump{\varphi} = \varphi\bm{n} \quad \text{if }\varGamma \in \mathcal{E}^{B}.$$
The average is defined as
$$\ave{\varphi} = \frac{1}{2}\rpth{\varphi^{+} + \varphi^{-}} \quad \text{if }\varGamma \in \mathcal{E}^{I}, \qquad \ave{\varphi} = \varphi \quad \text{if }\varGamma \in \mathcal{E}^{B}.$$
Similar definitions apply for a vector function $\boldsymbol{\varphi}$
\begin{align*}
    &\jump{\boldsymbol{\varphi}} = \boldsymbol{\varphi}^{+}\cdot\bm{n}^{+} + \boldsymbol{\varphi}^{-}\cdot\bm{n}^{-} \quad \text{if }\varGamma \in \mathcal{E}^{I}, \qquad 
    \jump{\boldsymbol{\varphi}} = \boldsymbol{\varphi}\cdot\bm{n} \quad \text{if }\varGamma \in \mathcal{E}^{B}, \\
    &\ave{\boldsymbol{\varphi}} = \frac{1}{2}\rpth{\boldsymbol{\varphi}^{+} + \boldsymbol{\varphi}^{-}} \quad \text{if }\varGamma \in \mathcal{E}^{I}, \qquad \ave{\boldsymbol{\varphi}} = \boldsymbol{\varphi} \quad \text{if }\varGamma \in \mathcal{E}^{B}.
\end{align*}
For vector functions, it is also useful to define a tensor jump as
$$\tjump{\boldsymbol{\varphi}} = \boldsymbol{\varphi}^{+}\otimes\bm{n}^{+} + \boldsymbol{\varphi}^{-}\otimes\bm{n}^{-} \quad \text{if }\varGamma \in \mathcal{E}^{I}, \qquad \tjump{\boldsymbol{\varphi}} = \boldsymbol{\varphi}\otimes\bm{n} \quad \text{if }\varGamma \in \mathcal{E}^{B}.$$
We also introduce the following finite element spaces
$$Q_{r} = \cpth{v \in L^{2}(\Omega) : v\rvert_{K} \in \mathbb{Q}_{r} \quad \forall K \in \mathcal{T}_{\mathcal{H}}}, \qquad \mathbf{V}_{r} = \spth{Q_{r}}^{2},$$
where $\mathbb{Q}_{r}$ is the space of polynomials of degree $r$ in each coordinate direction. We then denote by $\boldsymbol{\varphi}_{i}(\bm{x})$ the basis functions for the space $\mathbf{V}_{r}$ and by $\psi_{i}(\bm{x})$ the basis functions for the space $Q_{r}$, the finite element spaces chosen for the discretization of the discharge and of the free-surface elevation (as well as the tracer concentration), respectively, so that
$$\disc \approx \sum_{j = 1}^{2\left|\mathcal{T}_{\mathcal{H}}\right|(r+1)^{2}}q_{j}(t)\boldsymbol{\varphi}_j(\bm{x}), \qquad \zeta \approx \sum_{j = 1}^{\left|\mathcal{T}_{\mathcal{H}}\right|(r+1)^{2}}\zeta_{j}(t)\psi_{j}(\bm{x}).$$
The shape functions correspond to the products of Lagrange interpolation polynomials for the support points of $\rpth{r + 1}$-order Gauss--Legendre--Lobatto quadrature rule in each coordinate direction. 
It is important to remark that the total water depth $h$ is not expanded in terms of the basis functions, but rather computed only at the quadrature nodes using the second equations in \eqref{eq:imex_shw}. This allows us to use non-polynomial representations of the bathymetry and to preserve well-balancing.

Given these definitions, the algebraic system for the continuity equation \eqref{eq:imex_shw} at each stage reads as follows
\begin{eqnarray}\label{eq:algebraic_elevation}
    \mathbf{H}(\boldsymbol{\zeta}^{(l)}) = \mathbf{H}(\boldsymbol{\zeta}^{n}) + \Delta t\,\sum_{m=1}^{l-1} a_{lm}\,\mathbf{F}^{(m)}
\end{eqnarray}
Here, $\boldsymbol{\zeta}^{(l)}$ and $\boldsymbol{\zeta}^{n}$ denote the vectors of the degrees of freedom associated to the free-surface height, respectively at the $l-$stage and at time $t^{n}$. More specifically, the elements of the vectors $\mathbf{H}$ and $\mathbf{F}^{(m)}$ are defined as
\begin{subequations}
\begin{align}
    H_{i}(\boldsymbol{\zeta})
    &= \sum_{K \in \mathcal{T}_\mathcal{H}} \int_{K} \psi_{i}\,\max\rpth{\sum_{j = 1}^{\left|\mathcal{T}_{\mathcal{H}}\right|(r+1)^{2}} \zeta_{j}\,\psi_{j} + z_{b},\,0}\,\mathrm{d}\Omega, \label{eq:depth_rhs} \\
    F_{i}^{(m)} &= \sum_{K \in \mathcal{T}_{\mathcal{H}}} \int_{K} \disc^{(m)} \cdot \grad\psi_{i}\,\mathrm{d}\Omega \nonumber \\
    &- \sum_{\varGamma \in \mathcal{E}} \int_{\varGamma} \rpth{\ave{\disc^{*,(m)}} + \frac{\lambda^{(m)}}{2}\jump{\zeta^{*,(m)}}} \cdot \jump{\psi_{i}}\,\mathrm{d}\Sigma. \label{eq:mass_rhs}
\end{align}
\end{subequations}
In the computation of the integral at the interface between two elements  a Rusanov flux \cite{rusanov:1962} is used, so that
\begin{equation}\label{eq:lxf_parameter}
    \lambda^{(m)} = \max\rpth{\left|\vel^{+,(m)} \cdot \bm{n}^{+}\right| + \sqrt{gh^{+,(m)}}, \left|\vel^{-,(m)} \cdot \bm{n}^{-}\right| + \sqrt{gh^{-,(m)}}}.
\end{equation}
In the above formula and in all other computations in the right-hand side, the velocity is computed with a modified water depth in order to avoid the division by a very small number in shallow areas \cite{kurganov:2007}:
\begin{equation}\label{eq:velocity}
    \vel = \frac{\disc}{h_{\mathrm{des}}} = \frac{\sqrt{2}h}{\sqrt{\max\left(h^{4},\epsilon^{4}\right) + h^{4}}}\disc
\end{equation}
Here, $\epsilon$ is a test case-dependent small coefficient that will be specified in the corresponding section for each benchmark. 
In the numerical flux, we employ the so-called hydrostatic reconstruction \cite{audusse:2004, gross:2002}, which consists in selecting a single-valued bathymetry at the interface when the bathymetry is discontinuous, i.e.
\begin{equation}\label{eq:hydro_rec}
    z^{*}_{b} = \min\rpth{z_{b}^{-}, z_{b}^{+}},
\end{equation}
and modifying the water depth, the discharge and the free-surface, accordingly
\begin{eqnarray}\label{eq:hydro_rec_vars}
    h^{*,+} &=& \max\rpth{\zeta^{+} + z^{*}_{b}, 0}, \qquad h^{*,-} = \max\rpth{\zeta^{-} + z^{*}_{b}, 0}, \nonumber \\
    \disc^{*,+} &=& h^{*,+} \vel^{+},\qquad\qquad\qquad\,\, \disc^{*,-} = h^{*,-} \vel^{-},\\
    \zeta^{*,+} &=& \max\rpth{\zeta^{+}, -z^{*}_{b}}, \quad \textrm{and} \quad \zeta^{*,-} = \max\rpth{\zeta^{-}, -z^{*}_{b}}.
\end{eqnarray}
The application of the above reconstruction avoids spurious mass, momentum and pressure fluxes at the wet-dry interfaces, also in presence of a discontinuous bathymetry at the element edge. In the latter case, hydrostatic reconstruction will also be used to determine the positivity of the water depth.

Equation \eqref{eq:algebraic_elevation} is a weakly nonlinear system with unknown $\boldsymbol{\zeta}^{(l)}$ that can be solved using, e.g., fixed point iterations \cite{casulli:2009} or the Newton method \cite{bitsch:2026}. We use the subscript $k$ to denote the Newton iteration index, in order to distinguish it from the Runge–Kutta stage index, which is denoted by the superscript $(l)$. Hence, we obtain
\begin{equation}\label{eq:newton}
    \boldsymbol{\zeta}^{(l)}_{k+1} =
    \boldsymbol{\zeta}^{(l)}_{k} - \boldsymbol{\mathcal{M}}^{-1}\rpth{\boldsymbol{\zeta}^{(l)}_{k}}\rpth{\mathbf{H}(\boldsymbol{\zeta}^{(l)}_{k}) - \mathbf{H}(\boldsymbol{\zeta}^{n}) - \Delta t\sum_{m=1}^{l-1}a_{lm}\mathbf{F}^{(m)}},
\end{equation}
where
$$\boldsymbol{\mathcal{M}}\rpth{\boldsymbol{\zeta}^{(l)}_{k}} = \pad{\mathbf{H}}{\boldsymbol{\zeta}^{(l)}_{k}}$$
denotes the Jacobian matrix. In order to account for the null contribution of dry regions to the integral in \eqref{eq:depth_rhs}, we define a land-sea mask
\begin{equation}\label{eq:mask}
    \ell^{(l)}_{k}(\bm{x}) = 
    \begin{cases}
        1, & \text{if }\bm{x}:\,\,\zeta^{(l)}_{k}(\bm{x}) + z_{b}(\bm{x}) > 0 \\
        0, & \text{otherwise.}
    \end{cases}
\end{equation}
Elements with (resolved) wet/dry interfaces are characterized by zero values of $\ell^{(l)}_{k}$ at some quadrature node. This allows us to compute the entries of the Jacobian matrix as
\begin{equation}\label{eq:mass_matrix}
    \mathcal{M}_{ij}\rpth{\boldsymbol{\zeta}^{(l)}_{k}}= \sum_{K \in \mathcal{T}_{\mathcal{H}}} \int_{K} \psi_{j}\psi_{i}\ell^{(l)}_{k}(\bm{x})\,\mathrm{d}\Omega.
\end{equation}
When the $l^{1}-$norm of the subset of the increment vector $\boldsymbol{\zeta}^{(l)}_{k+1}-\boldsymbol{\zeta}^{(l)}_{k}$, containing the degrees of freedom on the element, falls below $\num{1e-16}$, the Newton iterations are stopped and we set $\boldsymbol{\zeta}^{n+1} = \boldsymbol{\zeta}^{(l)}_{k+1}$ on that element.

The algebraic system for the discharge equation at each stage reads as follows
\begin{equation}\label{eq:algebraic_discharge}
    \mathbf{A}^{(l)}\mathbf{Q}^{(l)} = \widetilde{\mathbf{Q}}^{n} + \Delta t \sum_{m=1}^{l-1}a_{lm}\mathbf{G}^{(m)} + \Delta t\sum_{m=1}^{l-1}\tilde{a}_{lm}\mathbf{L}^{(m)},
\end{equation}
where $\mathbf{Q}^{(l)}$ denotes the vector of the degrees of freedom associated to the discharge field, $\mathbf{A}^{(l)}$ denotes a modified mass matrix, $\mathbf{G}^{(l)}$ and $\mathbf{L}^{(l)}$ represent the DG approximations of the non-stiff and stiff components of the right-hand side of the discharge equation, according to \eqref{eq:imex_shw}, evaluated at the previous stages. Here, we have therefore set
\begin{subequations}
\begin{align}
    A_{ij}^{(l)} &= \sum_{K \in \mathcal{T}_{\mathcal{H}}} \int_{K} \spth{1 + \tilde{a}_{ll}\Delta t\gamma\rpth{\left|\disc^{(l-1)}\right|,h^{(l-1)}}}\boldsymbol{\varphi}_{j} \cdot \boldsymbol{\varphi}_{i}\,\mathrm{d}\Omega, \label{eq:mass_matrix_discharge} \\
    \widetilde{Q}_{i}^{n} &= \sum_{K \in \mathcal{T}_{\mathcal{H}}} \int_{K} \disc^{n} \cdot \boldsymbol{\varphi}_{i}\,\mathrm{d}\Omega, \\
    G_{i}^{(m)} &= \sum_{K \in \mathcal{T}_{\mathcal{H}}} \int_{K} \disc^{(m)} \otimes \vel^{(m)} : \grad\boldsymbol{\varphi}_{i}\,\mathrm{d}\Omega \nonumber \\
    &- \sum_{\varGamma \in \mathcal{E}} \int_{\varGamma} \rpth{\ave{\disc^{*,(m)} \otimes \vel^{(m)}} + \frac{\lambda^{(m)}}{2}\tjump{\disc^{*,(m)}}} : \tjump{\boldsymbol{\varphi}_{i}}\,\mathrm{d}\Sigma \nonumber \\
    &- \sum_{K \in \mathcal{T}_{\mathcal{H}}} \int_{K}g h^{(m)}\grad\zeta^{(m)} \cdot \boldsymbol{\varphi}_{i} \mathrm{d}\Omega + \sum_{\varGamma \in \mathcal{E}} \int_{\varGamma} g\ave{h^{*,(m)}}\ave{\boldsymbol{\varphi}_{i}} \cdot \jump{\zeta^{*,(m)}}\,\mathrm{d}\Sigma \label{eq:discharge_rhs} \\
    L_{i}^{(m)} &= - \sum_{K \in \mathcal{T}_{\mathcal{H}}} \int_{K} \gamma\rpth{\left|\disc^{(m)}\right|, h^{(m)}_{des}}\disc^{(m)} \cdot \boldsymbol{\varphi}_{i}\,\mathrm{d}\Omega. \label{eq:discharge_rhs_implicit}
\end{align}
\end{subequations}

Finally, the algebraic formulation for the equations of the concentrations at each stage can be written in a similar manner
\begin{equation}\label{eq:algebraic_concentration}
    \boldsymbol{\mathcal{H}}(\boldsymbol{\zeta}^{(l)})\mathbf{C}^{(l)} = \widetilde{\mathbf{C}}^{n} + \Delta t \sum_{m=1}^{l-1}a_{lm}\mathbf{R}^{(m)},
\end{equation}
where now the reaction source terms have been omitted. Here, $\mathbf{C}^{(l)}$ denotes the vector of degrees of freedom associated with the concentration $c$ of a generic species. $\boldsymbol{\mathcal{H}}(\boldsymbol{\zeta}^{(l)})$ denotes a modified mass matrix in which the water depth has been condensed, and $\mathbf{R}^{(l)}$ represents the DG approximation of the right-hand side of the concentration equation, evaluated at the previous stages according to \eqref{eq:imex_shw}. More precisely,
\begin{subequations}
\begin{align}
    \mathcal{H}_{ij}(\boldsymbol{\zeta}^{(l)}) &= \sum_{K \in \mathcal{T}_{\mathcal{H}}} \int_{K} \,\max\left( \zeta^{(l)} + z_b, \, 0 \right) \, \psi_{j}\psi_{i}\,\mathrm{d}\Omega, \label{eq:mass_matrix_concentration} \\
    \widetilde{C}_{i}^{n} &= \sum_{K \in \mathcal{T}_{\mathcal{H}}} \int_{K} h^{n}c^{n}\psi_{i}\,\mathrm{d}\Omega, \label{eq:concentration_rhs} \\
    R_{i}^{(m)} &= \sum_{K \in \mathcal{T}_{\mathcal{H}}} \int_{K} \disc^{(m)}c^{(m)} \cdot \grad\psi_{i}\,\mathrm{d}\Omega \nonumber \\
    &- \sum_{\varGamma \in \mathcal{E}} \int_{\varGamma} \rpth{\ave{\disc^{*,(m)}c^{(m)}} + \frac{\lambda^{(m)}}{2}\jump{h^{*,(m)}c^{(m)}}} \cdot \jump{\psi_{i}}\,\mathrm{d}\Sigma.
    \label{eq:r_concentration}
\end{align}
\end{subequations}
For coherence, in all the numerical fluxes we have used the reconstructed variables \eqref{eq:hydro_rec_vars}.

Equations \eqref{eq:shw} do not include either physical viscosity nor diffusion beyond the friction term and all numerical experiments in Sections \ref{sec:tests} and \ref{sec:realistic} have been carried out without adding numerical viscosity or diffusion. However, for the tests with realistic bathymetry in Section \ref{sec:realistic}, a small numerical diffusion term has been added to the concentration equation, as a simple monotonization mechanism. This is necessary to control spurious oscillations that arise in regions with large tracer gradients. This numerical dissipation has been included adding to \eqref{eq:r_concentration} the term
\begin{equation}\label{eq:numdiff}
    -\sum_{K \in \mathcal{T}_{\mathcal{H}}} \int_{K} \nu_{c}h^{(m)}\grad c^{(m)} \cdot \grad\psi_{i}\,\mathrm{d}\Omega,
\end{equation}
where $\nu_{c}$ is a flow dependent eddy diffusivity scaled by the element area $|K|$ and by the polynomial degree $r$, i.e. $\nu_{c} = C|K|r^{-2}\mathcal{S}$ where $\mathcal{S}$ is a characteristic strain rate of the resolved flow and the constant $C$ is to be chosen empirically for each case.

\subsection{Computation of the integrals}
\label{ssec:spacedisc_integral}

In high-order finite element methods for the shallow water equations, the choice of the numerical quadrature formula is typically related to the need to integrate accurately the nonlinear advection term, which consists of non-polynomial terms. Moreover, in well-balanced and pre-balanced finite element methods, the quadrature rule is often chosen to integrate exactly the pressure gradient term. Since the water depth $h$ is not a polynomial, exact integration is less relevant in our case. On the other hand, the accuracy of the quadrature formula is closely related to the ability of recovering accurate sub-grid bathymetric features. In particular, one may wish to employ highly accurate quadrature rules on coarse meshes. In this work we use a minimal number of quadrature nodes that are chosen according to the polynomial order, so that, in the particular case of a polynomial bathymetry, the degree of exactness is enough for exact integration of the pressure gradient term. For this reason, in the integrals over elements $K$ in the right-hand sides \eqref{eq:mass_rhs}, \eqref{eq:discharge_rhs} and \eqref{eq:r_concentration}, a standard tensor-products of Gauss--Legendre formula is used with $(\lfloor3/2\,r\rfloor + 1)^{2}$ quadrature nodes for each element. For instance, this choice results in $16 $ nodes per element for $r=2$ and $25$ nodes per element for $r=3$. More accurate quadrature formulae could easily be employed, but this is beyond the scope of the present work.

Wetting and drying poses instead a constraint on the choice of the quadrature formula for the inter-element boundary term and for the mass matrices in the continuity equation, defined respectively in \eqref{eq:mass_rhs} and \eqref{eq:depth_rhs},\eqref{eq:mass_matrix}. For reasons related to positivity preservation that will be discussed in Section \ref{ssec:spacedisc_prop}, for the integrals over the inter-element boundaries $\varGamma$ we use a Gauss--Legendre--Lobatto formula with $r+2$ quadrature nodes, whose degree of exactness is $3r$. As in the case of volume integrals, more accurate quadrature formulae could also be applied to the line integrals.
We use the same quadrature formula in all numerical fluxes, including those in the momentum and concentration equations \eqref{eq:discharge_rhs}-\eqref{eq:r_concentration}. The same Gauss--Legendre--Lobatto quadrature formula is applied in the mass matrix computations by considering the tensor product on the reference quadrilateral of the one dimensional formula. For coherence, the same quadrature formula is also applied to the mass matrices of the concentration equation \eqref{eq:mass_matrix_concentration} and  \eqref{eq:concentration_rhs}.

Note that the Gauss--Legendre--Lobatto quadrature used for the mass matrix is computationally more expensive than the Gauss--Legendre to obtain exact integration. Indeed, we are forced to use $(r+2)^{2}$ points instead of $(r+1)^{2}$ points used for the standard DG scheme with polynomial bathymetry. This means that $2r+3$ more points are needed to compute the mass matrix entries. For this reason, the discharge mass matrix \eqref{eq:mass_matrix_discharge}, which does not play any role in the positivity proof, can be computed with the usual Gauss--Legendre quadrature formula with $(r+1)^{2}$ points for each element.
The use of the same number of quadrature nodes and degrees of freedom for each element yields, after a change of basis, diagonal blocks in the algebraic structure of the mass matrix \cite{kronbichler:2019}. More specifically, as explained in \cite{kronbichler:2019, kronbichler:2016}, the discharge mass matrix is inverted as $\mathbf{M}^{-1} = \mathbf{S}^{-T}\mathbf{J}^{-1}\mathbf{S}^{-1} $. Here $\mathbf{J} $ is a diagonal matrix, whose entries are equal to the determinant of the Jacobian times the quadrature weight, while $\mathbf{S}$ is a square matrix with basis functions in the row and quadrature nodes in columns, i.e. $S_{ij} = \psi_{i}(\bm{x}_{j}) $. The matrix $\mathbf{S} $ is constructed as the Kronecker product (tensor product) of small one-dimensional matrices that can be inverted efficiently on the fly at each stage. Notice that, if friction is present and for a general bathymetry, such entries can be no longer represented in general as the product of polynomials. Table~\ref{tab:quadrature} summarizes the quadrature formulas used in this work.

\begin{table}[h!]
   \begin{tabular}{|c|c|c|c|}
   \hline
   terms in finite elements & volume & boundary & mass matrix \\
   \hline
   continuity & GL, $\rpth{ \lfloor 3/2\,r\rfloor+1}^{2}$ & GLL, $r+2$ & GLL, $\rpth{r+2}^{2}$ \\
   \hline
   momentum & GL, $\rpth{\lfloor 3/2\,r\rfloor +1}^{2}$ & GLL, $r+2$ & GL, $\,\,\rpth{r+1}^{2}$ \\
   \hline
   tracer & GL, $\rpth{\lfloor 3/2\,r \rfloor +1}^{2}$ & GLL, $r+2$ & GLL, $\rpth{r+2}^{2}$ \\
   \hline
   \end{tabular}
   \caption{Summary of the quadrature formulae used in the integrals of the different terms in the finite element method. GL stands for Gauss--Legendre and GLL stands for Gauss--Legendre--Lobatto.}
   \label{tab:quadrature}
\end{table}

A matrix-free approach is employed \cite{arndt:2023}, meaning that no global sparse matrix is built and only the action of the linear operators on a vector is actually implemented. This choice provides a superior efficiency with respect to matrix-based implementations, in particular for high-order polynomial representations \cite{kronbichler:2018}.

\subsection{Degree and $h-$ adaptivity}

Discharge or velocity variables are not guaranteed to be differentiable over wet-dry elements.
To cope with this regularity issue, the previously described discretization can be extended to a locally variable polynomial degree, in order to revert to a first-order DG scheme with $r=0$ on the dry and wet-dry elements. We use the water depth as a simple proxy to identify such elements. Setting
\begin{equation}
    \mu_{K} = \min_{q \in K} h(\bm{x}_{q}),
\end{equation}
where $q$ is the index of the set of the quadrature nodes of the Gauss--Legendre--Lobatto formula on the element, we reduce/increase the polynomial order on an element based on the following criteria
\begin{eqnarray}\label{eq:criteria_p}
    \textrm{if   }& \mu_{K} > h_{\mathrm{lim}} &\quad\textrm{refine degree, } r_{K}=r, \label{eq:p-criteria} \\
    \textrm{if   }&
    \mu_{K} < h_{\mathrm{lim}}
    & \quad \textrm{coarsen degree, } r_{K}=0, \nonumber
\end{eqnarray}
We continue to denote by $r$ the global polynomial degree while we denote by $r_{K}$ the polynomial degree of the element, which can be either $r_{K}=r$ or $r_{K}=0$ in wet-dry and dry elements. The depth threshold for coarsening $h_{\mathrm{lim}}$ is case-dependent. After degree coarsening, the mass matrix reduces to the ``wet'' element area
$$\mathcal{M}(\boldsymbol{\zeta}^{(l)}_{k})= \sum_{K \in \mathcal{T}_{\mathcal{H}}} \int_{K} \ell^{(l)}_{k}(\bm{x})\,\mathrm{d}\Omega.$$
As the above integral is computed at Gauss--Legendre--Lobatto points, the mass-matrix is always invertible until the land-sea mask is null at all the Gauss--Legendre--Lobatto points $\ell^{(l)}_{k}(\bm{x}_{q}) = 0$. This is equivalent to a dry element. When, on an element, the Newton method converges to a dry state, for which $\mathcal{M}(\boldsymbol{\zeta}^{(l)}_{k})=0$, we simply set $\boldsymbol{\zeta}^{(l)} = \boldsymbol{\zeta}^{(l)}_{k}$ on that element skipping the division by zero. Notice that, to start-up the nonlinear procedure we choose as initial guess $\boldsymbol{\zeta}^{(l)}_{0}$, the computed value at the beginning of the time step, $\boldsymbol{\zeta}^{n}$. For dry elements, the wet element area is zero 
and the Newton method cannot be initiated with the consequence that flooding of that element would be always prevented. To desingularize the mass matrix in dry elements we choose the initial guess such that at least one Gauss--Legendre--Lobatto quadrature node for each element is wet, so that $\mathcal{M}(\boldsymbol{\zeta}^{(l)}_{0})\neq 0$. This is achieved by simply lifting the free-surface by a small value beyond the maximum bathymetry value. The initial guess for the degrees of freedom on the element $j \in K$ are
\begin{equation}\label{eq:initial_guess}
    \zeta^{(l)}_{0,j} =
    \begin{cases}
        \zeta^{n}_{j}, & \text{if } \,\exists\,q \in K: \zeta(\bm{x}_{q}) > z_{b}(\bm{x}_{q})\\[1ex]
        -\max\limits_{q} z_{b}(\bm{x}_{q}) + \epsilon_{0}, & \text{otherwise}.
    \end{cases}
\end{equation}
where $q$ is the index of the set of the quadrature nodes of the Gauss--Legendre--Lobatto formula on the element and $\epsilon_{0}$ any small positive value.

It can be observed that, if there is no resolved wet-dry front, all elements are wet and $\ell^{(l)}_{k}(\bm{x}_q)=1$ at all quadrature nodes. Then, $\boldsymbol{\mathcal{M}}(\boldsymbol{\zeta}^{(l)}_{k})$ is constant and coincides with the standard finite element mass matrix $\mathbf{M}$. Starting from the intial guess $\boldsymbol{\zeta}^{(l)}_{0} = \boldsymbol{\zeta}^{n}$, the first iteration reads
\begin{equation*}\label{eq:ite-1}
    \boldsymbol{\zeta}^{(l)}_{1} =
    \boldsymbol{\zeta}^{n}
    + \mathbf{M}^{-1} \Delta t \sum_{m=1}^{l-1} a_{lm} \mathbf{F}^{(m)}
\end{equation*}
and the second shows convergence $\boldsymbol{\zeta}^{(l)}_{2} = \boldsymbol{\zeta}^{(l)}_{1}$. As a consequence, the same algorithm and code path can be used for the update of wet and dry elements without the need to employ conditional statements in the implementation.

It is important to stress that we do not lower the accuracy of the quadrature formulae at those elements $K$ where $r_{K}=0$: the same quadrature formulae are used during the whole simulation. This choice, which seems inefficient for piecewise constant polynomials, has two consequences. Firstly, it allows to maintain a sub-grid bathymetry also close to wet-dry front, thus reducing to the Finite Volume method of \cite{casulli:2009}. Furthermore, it also allows to conserve the water mass across the remaps from high-order to low-order and \emph{vice versa}, as we will show later.

The previously described discretization can be extended seamlessly to non-conforming, locally refined meshes, thanks to the flexible implementation of DG available in \texttt{deal.II} \cite{bangerth:2009}. The $h-$adaptive capabilities of \texttt{deal.II} have been already successfully exploited in \cite{orlando:2023c, orlando:2023a, orlando:2024a} for the discretization of the Euler equations. For the purposes of this work, degree and mesh adaptivity must be independent, each governed by its own refinement/coarsening criteria. For the $h-$adaptive simulations, we denote a case-dependent mesh refinement indicator on each element by $\eta_{K} > 0$. We refine and coarsen an element based on the following criteria:
\begin{eqnarray}
    \textrm{if   }& \eta_{K} > \theta_{r}\,\eta_{\text{max}}
    \textrm{ and } n_{K} < n_{\text{max}}
    \textrm{ and } \mathcal{H}_{K} > \mathcal{H}_{\text{min}} &\quad\textrm{refine } K, \nonumber \\
    \textrm{if   }&
    \eta_{K} < \theta_{c}\,\eta_{\text{max}}
    \textrm{ and } n_{K} > 1
    & \quad \textrm{coarsen } K, \nonumber 
\end{eqnarray}
where
$$\eta_{\text{max}} = \max_{\mathcal{K\in T_{H}}}\eta_{K},$$
whereas $\theta_{r}$ and $\theta_{c}$ are the mesh refinement and coarsening thresholds, respectively. We denote by $n_{K}$ the mesh level of the element, ranging from the coarsest level of the initial mesh, $n_{K} = 1$, to the finest level $n_{K} = n_{\text{max}}$. For initial meshes with a variable mesh size, we also set a minimum target mesh size $\mathcal{H}_{\text{min}}$, below which an element cannot be refined, even if $n_{K} < n_{\text{max}}$. The definition of the mesh refinement indicator and the values of the different thresholds are case-dependent and will be specified in the corresponding section for each benchmark. 

\subsection{Properties of the spatial discretization}
\label{ssec:spacedisc_prop}

We discuss here in detail the properties of the DG approximation. Firstly, the treatment of the non-conservative pressure term is based on a double integration by parts, as inspired by \cite{bassi:1997}. More specifically, as in \cite{tumolo:2015}, we have
\begin{align}\label{eq:pressure_tumolo}
    \sum_{K \in \mathcal{T}_{\mathcal{H}}} \int_{K} ghD_{h}\zeta \cdot \boldsymbol{\varphi}_{i}\,\mathrm{d}\Omega &= \sum_{K \in \mathcal{T}_{\mathcal{H}}} \int_{K} gh\grad\zeta \cdot \boldsymbol{\varphi}_{i}\,\mathrm{d}\Omega \nonumber \\
    &+ \sum_{\varGamma \in \mathcal{E}} \int_{\varGamma} g\ave{h}\ave{\zeta}\jump{\boldsymbol{\varphi}_{i}}\,\mathrm{d}\Sigma - \sum_{\varGamma \in \mathcal{E}} \int_{\varGamma} g\ave{h}\jump{\zeta\boldsymbol{\varphi}_{i}}\,\mathrm{d}\Sigma \nonumber \\
    &= \sum_{K \in \mathcal{T}_{\mathcal{H}}} \int_{K} gh\grad\zeta \cdot \boldsymbol{\varphi}_{i}\,\mathrm{d}\Omega - \sum_{\varGamma \in \mathcal{E}} \int_{\varGamma} g\ave{h}\ave{\boldsymbol{\varphi}_{i}} \cdot \jump{\zeta}\,\mathrm{d}\Sigma,
\end{align}
where in the last relation we have employed the so-called \emph{magic formula} \cite{arnold:1982}
\begin{equation}\label{eq:magic_formula}
    \jump{\zeta\boldsymbol{\varphi}_{i}} = \ave{\zeta}\jump{\boldsymbol{\varphi}_{i}} + \jump{\zeta} \cdot \ave{\boldsymbol{\varphi}_{i}}.
\end{equation}
For a continuous bathymetry, formulation \eqref{eq:pressure_tumolo} can be reinterpreted as a DG approximation of the pressure term written in flux form under the so-called strong form, which simply stands for a double integration by parts \cite{arpaia:2022, bonev:2018}. More specifically, starting from the conservative formulation of the pressure gradient and considering a centered flux for the pressure gradient, after a double integration by parts we get
\begin{align}
    \sum_{K \in \mathcal{T}_{\mathcal{H}}} \int_{K} \rpth{D_{h}\rpth{P\mathbf{I}} + ghD_{h}z_{b}} \cdot \boldsymbol{\varphi}_{i}\,\mathrm{d}\Omega &= \sum_{K \in \mathcal{T}_{\mathcal{H}}} \int_{K} g h\grad\zeta \cdot \boldsymbol{\varphi}_{i}\,\mathrm{d}\Omega \nonumber \\
    &+ \sum_{\varGamma \in \mathcal{E}} \int_{\varGamma} \ave{P}\jump{\boldsymbol{\varphi}_{i}}\mathrm{d}\Sigma - \sum_{\varGamma \in \mathcal{E}} \int_{\varGamma} \jump{P\boldsymbol{\varphi}_{i}}\,\mathrm{d}\Sigma \nonumber \\
    &= \sum_{K \in \mathcal{T}_{\mathcal{H}}} \int_{K} gh\grad\zeta \cdot \boldsymbol{\varphi}_{i}\,\mathrm{d}\Omega - \sum_{\varGamma \in \mathcal{E}} \int_{\varGamma} \jump{P} \cdot \ave{\boldsymbol{\varphi}_{i}}\,\mathrm{d}\Sigma, \nonumber
\end{align}
with $P = \frac{1}{2}gh^{2}$ denoting the shallow water pressure. Thanks to \eqref{eq:magic_formula}, in the case of continuous bathymetry, we obtain
$$\jump{P} = g\ave{h}\jump{h} = g\ave{h}\jump{\zeta},$$ 
so as to recover relation \eqref{eq:pressure_tumolo}.

For shallow water flows, an important role is played by the so-called lake-at-rest steady state. We are often interested in solutions which can be regarded as perturbations of the lake-at-rest solution given as
$$\zeta = \zeta_{0}=\textrm{constant}, \quad \disc = \bm{0}.$$
The exact preservation of this initial steady state at the discrete level is known as the {\cal C}-property, or well-balancedness. For our purposes we give the following definition of a well-balanced scheme.
\\
\\
\textbf{Definition \ref{ssec:spacedisc_prop}.1} Let the following solution denotes the discrete DG representation of the lake-at-rest state:
\begin{subequations}
\begin{align}
    \zeta_{j} &=
    \begin{cases}
        \zeta_{0}, & \text{if } \,\exists\,q \in K: \zeta(\bm{x}_{q}) > -z_{b}(\bm{x}_{q})\\[1ex]
        -\max\limits_{q} z_{b}(\bm{x}_{q}), & \text{otherwise}.
    \end{cases}\qquad \forall j \in K,\, \forall K \in \mathcal{T}_\mathcal{H} \label{eq:lake-at-rest-zeta} \\
    \disc &= \bm{0}, \label{eq:lake-at-rest-q}
\end{align}
\end{subequations}
where $j\in K$ is the index of the set of degrees of freedom for the free-surface on the element $K$ and $q\in K$ is the index of the set of the quadrature nodes of the Gauss--Legendre--Lobatto formula with $\left(r+2\right)^2$ nodes. The above initialization sets, in dry elements, the free-surface equal to the maximum bathymetry value on the Gauss--Legendre--Lobatto quadrature nodes. Then a scheme is well-balanced, if it exactly preserves the discrete lake-at-rest solution \eqref{eq:lake-at-rest-zeta}-\eqref{eq:lake-at-rest-q}.
\\
\\
\textbf{Proposition \ref{ssec:spacedisc_prop}.1} The DG scheme \eqref{eq:algebraic_elevation}-\eqref{eq:algebraic_discharge}, with non-conservative pressure gradient discretization as in  \eqref{eq:discharge_rhs} and with hydrostatic reconstruction at the element edges \eqref{eq:hydro_rec_vars}, is well-balanced for any bathymetry function $z_{b}(\bm{x})$.
\\
\\
\textbf{Proof.} We show that, for the initial lake-at-rest state defined in \eqref{eq:lake-at-rest-zeta} and \eqref{eq:lake-at-rest-q}, the right-hand side of the continuity and of the momentum equation is zero. With a non-conservative discretization of the pressure gradient, we have simply to verify that the right-hand side is equal to zero  at all quadrature nodes of the Gauss--Legendre--Lobatto formula. Thus, all the expressions that follow must be intended at quadrature nodes. At the RK $l-$stage we assume that for $m < l$
$$\ave{\mathbf{q}^{*,(m)}} = \bm{0}$$
at the edge quadrature nodes. Also, $\jump{\zeta^{*,(m)}} = 0$. This is trivially verified for the ``wet'' quadrature nodes where the solution is wet on both sides of the edge
$$\zeta^{*,+} = \zeta^{*,-} = \zeta_{0}.$$
For the "dry" quadrature nodes, i.e. where the solution is dry on both sides of the edge, owing to the initialization \eqref{eq:lake-at-rest-zeta} and thanks to the definition of the reconstructed bathymetry \eqref{eq:hydro_rec}, the free-surface verifies
\begin{align*}
    \zeta^{+} &= -\max_{q\in K^+} z_{b}(\bm{x}_{q}) \le -z_{b}^{+} \le -z_{b}^{*}, \\
    \zeta^{-} &= -\max_{q\in K^-} z_{b}(\bm{x}_{q}) \le -z_{b}^{-} \le -z_{b}^{*},
\end{align*}
so that, after applying the variables reconstruction \eqref{eq:hydro_rec_vars}, one obtains
\begin{align*}
    \zeta^{*,+} &= \max\rpth{\zeta^{+}, -z^{*}_{b}} = -z_{b}^{*}, \\
    \zeta^{*,-} &= \max\rpth{\zeta^{-}, -z^{*}_{b}} = -z_{b}^{*}.
\end{align*}
The case of a ``wet-dry" quadrature node, i.e., where the solution is wet on one side of the edge only, corresponds to a wet-dry edge with a discontinuous bathymetry. On the wet side, the free-surface is initialized as
$$\zeta^{+} = \zeta_{0} \le -z_{b}^{*}$$
and, after the reconstruction, reads
$$\zeta^{*,+} = \max\rpth{\zeta^{+}, -z^{*}_{b}} =-z_{b}^{*}.$$
It follows that $\jump{\zeta^{*,(m)}} = \bm{0}$ and \eqref{eq:mass_rhs} is zero, i.e.
$$F_{i}^{(m)} = 0.$$
Similarly, in the right-hand side of the momentum equation \eqref{eq:discharge_rhs} and \eqref{eq:discharge_rhs_implicit}, we have that
$$\ave{\disc^{*,(m)} \otimes \vel^{(m)}} = \bm{0},\quad \tjump{\disc^{*,(m)}} = \bm{0},\quad \grad\zeta^{(m)} = \bm{0},$$
and, as already shown, $\jump{\zeta^{*,(m)}} = \bm{0}$, so that
$$G_{i}^{(m)}=0 \quad \textrm{ and }\quad L_{i}^{(m)} = 0.$$
It follows $\mathbf{Q}^{(l)} = \bm{0}$ and $\mathbf{H}(\boldsymbol{\zeta}^{(l)}) = \mathbf{H}(\boldsymbol{\zeta}_{0})$ or $\boldsymbol{\zeta}^{(l)} = \boldsymbol{\zeta}_{0}$. The scheme is therefore well-balanced at each $l-$stage $\quad\Box$.
\\
\\
As common in Discontinuous Galerkin discretizations, a constraint on the time step is required to guarantee positivity of the water depth values in the wet regions. When employing the nonlinear relationship \eqref{eq:zeta2depth-2}, positivity becomes ultimately related to mass conservation since any positivity violation, after the application of \eqref{eq:zeta2depth-2}, translates into a mass loss. For our purposes, it is enough to prove that, under a CFL-like condition, the water mass contained in an element $K$ is always non negative at each RK stage $l$. The water mass contained in an element at the $l-$stage is obtained by summing the DG scheme \eqref{eq:algebraic_elevation} over all degrees of freedom $i \in K$:
\begin{equation}\label{eq:positivity-6}
    \overline{H}_{K}(\boldsymbol{\zeta}^{(l)}) = \sum_{i\in K} \rpth{H_{i}(\boldsymbol{\zeta}^{n}) + \Delta t \sum^{l-1}_{m=1} a_{lm} F^{m}_{i}}
    \ge 0.
\end{equation}
This simply states that a given element cannot lose more mass, within a given $l-$stage, than the mass contained in the element at the beginning of the time step.

In Section \ref{ssec:spacedisc_integral}, following  \cite{xing:2010, zhang:2011}, we have used, for the mass matrices, Gauss--Legendre--Lobatto quadrature formulae. The quadrature nodes of the such Gauss--Legendre--Lobatto formulae that lie on the element boundary coincide with the quadrature nodes used for the inter-element boundary integrals. This is the key to end up with a positivity result valid for a general non polynomial representation of the bathymetry. In particular we have introduced a $\widehat{Q}-$points Gauss--Legendre--Lobatto quadrature, with $\widehat{Q}=r+2$, defined on the unit interval $[0,1]$ with weights $\widehat{w}_{\beta}$. Similar quadrature formulae for the quadrilaterals are generated as tensor products of this 1-dimensional Gauss--Legendre--Lobatto quadrature formula. 

To investigate the positivity of the RK-DG scheme with irregular bathymetry, we need to study its counterpart with explicit Euler, so we set $s=2$ and $a_{21}=1$ in the scheme \eqref{eq:algebraic_elevation} which, for the $i-$th degree of freedom, reads therefore
\begin{equation}\label{eq:algebraic_elevation_forwardeuler}
    H_{i}(\boldsymbol{\zeta}^{n+1}) = H_{i}(\boldsymbol{\zeta}^{n}) + \Delta t F^{n}_{i}.
\end{equation}
\\
\\
\textbf{Proposition \ref{ssec:spacedisc_prop}.2} (Forward Euler time integrator) For the DG scheme \eqref{eq:algebraic_elevation_forwardeuler} to be positivity preserving, i.e., $\overline{H}_{K}(\boldsymbol{\zeta}^{n+1}) \ge 0$ for any bathymetry function $z_{b}(\bm{x})$, a sufficient
condition is the CFL condition
\begin{eqnarray}\label{eq:deltat_fe}
    \Delta t &\le& \Delta t_{FE} = \widehat{w}_{1} \frac{J_{K}}{\lambda^{n}\rpth{x_{\alpha}}l_{\varGamma}},\qquad \alpha=1,...,\widehat{Q},\quad \forall \, \varGamma \in \partial {K},
\end{eqnarray}
where $\widehat{w}_{1}$ is the quadrature weight of the end-point of the Gauss--Legendre--Lobatto rule on $[0,1]$, $J_{K}$ is the absolute value of the determinant of the Jacobian from the reference element to the  element $K$, $l_{\varGamma}$ is the length of the element face $\varGamma \in \partial{K}$ and $\lambda^{n}\rpth{x_{\alpha}}$ is the Lax--Friedrichs parameter evaluated at the Gauss--Legendre--Lobatto points of the face quadrature formula.
\\
\\
\textbf{Proof.} The evolution of the water mass within an element $K$ is obtained by summing the DG scheme \eqref{eq:algebraic_elevation_forwardeuler} over all degrees of freedom $i \in K. $ Exploiting the property of the Lagrange shape functions that $\sum\limits_{i \in K} \psi_{i} = 1$, one obtains
\begin{equation}\label{eq:algebraic_elevation_forwardeuler_average}
    \overline{H}_{K}(\boldsymbol{\zeta}^{n+1}) = \overline{H}_{K}(\boldsymbol{\zeta}^{n}) + \Delta t \overline{F}^{n}_{K},
\end{equation}
where $\overline{H}_{K}$ and $\overline{F}^{n}_{K}$ are computed as
\begin{subequations}
\begin{align}
    \overline{H}_{K}(\boldsymbol{\zeta}^{n})
    &= \int_{K} \max(\zeta^{n} + z_{b}, 0)\,\mathrm{d}\Omega, \label{eq:positivity-7} \\
    \overline{F}_{K}^{n}
    &= - \int_{\partial K} \rpth{\ave{\disc^{*,n}} + \frac{\lambda^{n}}{2}\jump{\zeta^{*,n}}} \cdot \bm{n}^{-}\,\mathrm{d}\Sigma.
\end{align}
\end{subequations}
Here, the minus superscript denotes the inner side of the edge, and $\bm{n}^{-}$ is the outward unit normal to the element $K$. We need to prove that the right-hand side can be written as a convex combination of positive water depth values. We compute the integrals with the quadrature formula and we can write the right-hand side of \eqref{eq:algebraic_elevation_forwardeuler_average} as
\begin{eqnarray*}\label{eq:positivity-8}
    \sum_{\beta=1}^{\widehat{Q}}\sum_{\alpha=1}^{\widehat{Q}} && \max\rpth{\zeta^{n} + z_{b}\rpth{x_{\alpha},y_{\beta}}, 0}J_{K} \widehat{w}_{\alpha}\widehat{w}_{\beta} \\
    - \Delta t \sum_{\varGamma \in \partial{K}} \sum_{\alpha=1}^{\widehat{Q}}  && \rpth{\ave{\disc^{*,n}\rpth{x_{\alpha}}} + \frac{\lambda^{n}\rpth{x_{\alpha}}}{2}\jump{\zeta^{*,n}\rpth{x_{\alpha}}}} \cdot \bm{n}^{-}l_{\varGamma}\widehat{w}_{\alpha}.
\end{eqnarray*}
We reformulate the above expression introducing some definitions. Firstly, we set $h^{n} = \max\rpth{\zeta^{n} + z_{b}, 0}$, see \eqref{eq:zeta2depth-2}. Then, we introduce the reconstructed variables \eqref{eq:hydro_rec_vars}. Since the reconstructed bathymetry \eqref{eq:hydro_rec} is always single valued at element edges we replace $\jump{\zeta^{*,n}}$ with $\jump{h^{*,n}}$ in the numerical flux, so as to obtain
\begin{eqnarray*}\label{eq:positivity-9}
    \sum_{\beta = 1}^{\widehat{Q}}\sum_{\alpha=1}^{\widehat{Q}} && h^{n}\rpth{x_{\alpha},y_{\beta}} J_K \widehat{w}_{\alpha}\widehat{w}_{\beta} \\
    - \Delta t \sum_{\varGamma \in \partial{K}} \sum_{\alpha=1}^{\widehat{Q}} && \rpth{\ave{h^{*,n}\rpth{x_{\alpha}}\vel^{n}\rpth{x_{\alpha}}} + \frac{\lambda^{n}\rpth{x_{\alpha}}}{2}\jump{h^{*,n}\rpth{x_{\alpha}}}} \cdot \bm{n}^{-}l_{\varGamma}\widehat{w}_{\alpha}
\end{eqnarray*}
We rewrite $h^{*,n} = \Theta^{n}h^{n}$, with $\Theta^{n} = \frac{h^{*,n}}{h^{n}}$. From the definition \eqref{eq:hydro_rec}, $0 \le \Theta^{n} \le 1$. Developing the product $\jump{h} \cdot \bm{n}^{-} = h^{-} - h^{+}$, we get
\begin{eqnarray*}\label{eq:positivity-10}
    \sum_{\beta=1}^{\widehat{Q}}\sum_{\alpha=1}^{\widehat{Q}} &&h^{n}\rpth{x_{\alpha},y_{\beta}} J_{K} \widehat{w}_{\alpha}\widehat{w}_{\beta} \\
    - \Delta t \sum_{\varGamma \in \partial{K}} \sum_{\alpha=1}^{\widehat{Q}}
    &&\left(\frac{1}{2}\rpth{h^{n,+}\Theta^{n,+}\rpth{x_{\alpha}}\vel^{n,+}\rpth{x_{\alpha}} \cdot \bm{n}^{+} + h^{n,-}\Theta^{n,-}\rpth{x_{\alpha}}\vel^{n,-}\rpth{x_{\alpha}} \cdot \bm{n}^{-}} \right. \\
    &+& \left. \frac{\lambda^{n}\rpth{x_{\alpha}}}{2}\rpth{h^{n,-}\Theta^{n,-}\rpth{x_{\alpha}} - h^{n,+}\Theta^{n,+}\rpth{x_{\alpha}}}\right) l_{\varGamma}\widehat{w}_{\alpha}.
\end{eqnarray*}
We can regroup common terms so as to obtain
\begin{eqnarray*}\label{eq:positivity-11}
    \sum_{\beta=1}^{\widehat{Q}}\sum_{\alpha=1}^{\widehat{Q}} &&h^{n}\rpth{x_{\alpha},y_{\beta}}J_{K} \widehat{w}_{\alpha}\widehat{w}_{\beta} \\
    - \Delta t \sum_{\varGamma \in \partial{K}} \sum_{\alpha=1}^{\widehat{Q}}
    && \left(\frac{h^{n,+}\Theta^{n,+}}{2}\rpth{x_{\alpha}}\rpth{
    -\vel^{n,+}\rpth{x_{\alpha}} \cdot \bm{n}^{+}-\lambda^{n}\rpth{x_{\alpha}}} \right. \\
    &+& \left. \frac{h^{n,-}\Theta^{n,-}}{2}\rpth{x_{\alpha}}\rpth{\vel^{n,-}\rpth{x_{\alpha}} \cdot \bm{n}^{-} + \lambda^{n}\rpth{x_{\alpha}}}
    \right) l_{\varGamma}\widehat{w}_{\alpha}.
\end{eqnarray*}
We can move the boundary points of the volume integral with the line integral. Hence, the right-hand side reduces to
\begin{eqnarray*}\label{eq:positivity-12}
    \sum_{\beta=2}^{\widehat{Q}-1}\sum_{\alpha=2}^{\widehat{Q}-1} &&h^{n}\rpth{x_{\alpha},y_{\beta}}J_{K}\widehat{w}_{\alpha}\widehat{w}_{\beta} \\
    + \Delta t \sum_{\varGamma \in \partial{K}} \sum^{\widehat{Q}}_{\alpha = 1}
    && \left(\frac{h^{n,+}\Theta^{n,+}}{2}\rpth{(x_{\alpha}}\rpth{\vel^{n,+}\rpth{x_{\alpha}} \cdot \bm{n}^{+}+\lambda^{n}\rpth{x_{\alpha}}} \right. \\
    &+& \left. \frac{h^{n,-}}{2}\rpth{x_{\alpha}}\rpth{ \frac{2\widehat{w}_{1}J_{K}}{l_{\varGamma}\Delta t} -\Theta^{n,-}\vel^{n,-}\rpth{x_{\alpha}} \cdot \bm{n}^{-} -\Theta^{n,-}\lambda^{n}\rpth{x_{\alpha}}}
    \right) l_{\varGamma}\widehat{w}_{\alpha}.
\end{eqnarray*}
The right-hand side is a convex combination of positive quantities, namely the water depths $h^{n}(x_{\alpha}) \ge 0$, if the sign of the coefficients multiplying the water depths is positive
\begin{eqnarray*}\label{eq:positivity-13}
    J_{K}\widehat{w}_{\alpha}\widehat{w}_{\beta} &>& 0, \qquad \alpha = 2,\cdots,\widehat{Q}-1,\,\beta=2,\cdots\widehat{Q}-1, \\
    \vel^{n,+}\rpth{x_{\alpha}} \cdot \bm{n}^{+} + \lambda^{n}\rpth{x_{\alpha}} &\ge& 0,\qquad \alpha=1,\cdots,\widehat{Q}, \\
    \frac{\widehat{w}_{1} J_{K}}{l_{\varGamma}} + \frac{\Delta t}{2}\rpth{-\Theta^{n,-}\vel^{n,-}\rpth{x_{\alpha}} \cdot \bm{n}^{-} - \Theta^{n,-}\lambda^{n}
    \rpth{x_{\alpha}}} &\ge& 0,\qquad \alpha=1,\cdots,\widehat{Q},\quad \forall \, \varGamma \in \mathcal{E}_{K}.
\end{eqnarray*}
The first expression is of course positive. The second expression is non-negative because of the definition of the Lax--Friedrichs parameter \eqref{eq:lxf_parameter}. The third expression is non-negative under the following CFL condition:
\begin{equation}\label{eq:positivity-15}
    \Delta t \le \frac{2\widehat{w}_{1} J_{K}}{\Theta^{n,-}\rpth{x_{\alpha}}\rpth{\lambda^{n}\rpth{x_{\alpha}} + \vel^{n,-}\rpth{x_{\alpha}} \cdot \bm{n}^{-}}l_{\varGamma}} \le \widehat{w}_{1} \frac{J_{K}}{\lambda^{n}\rpth{x_{\alpha}}l_{\varGamma}},\qquad \alpha=1,\cdots,\widehat{Q}, \quad \forall \varGamma \in \mathcal{E}_{K},
\end{equation}
which concludes the proof $\Box$.
\\
\\
Noticing that the radius of absolute monotonicity for the RK32 time integrator is $\frac{1}{a_{21}}$ \cite{giraldo:2020,orlando:2022}, we prove the positivity of the DG scheme with general bathymetry and with explicit RK32 time integrator.
\\
\\
\textbf{Proposition \ref{ssec:spacedisc_prop}.3} (Explicit RK32 time integrator) For the DG scheme \eqref{eq:algebraic_elevation} to be positivity preserving, i.e., $\overline{H}_{K}(\boldsymbol{\zeta}^{n+1}) \ge 0 $ for any bathymetry function $z_{b}(\bm{x})$, a sufficient
condition is the CFL condition
\begin{equation}
    \Delta t \le \frac{1}{a_{21}}\Delta t_{FE}
\end{equation}
where $a_{21}$ is the RK coefficient of the second stage and $\Delta t_{FE}$ is the time step restriction for the Forward Euler time integrator established in \eqref{eq:deltat_fe}.
\\
\\
Notice that $\frac{1}{a_{21}}\approx 1.7$, resulting in a significant increase of the  allowed time step for the three stage RK32 integrator, with respect to the Forward Euler integrator.
In Table \ref{tab:cfl} we report the derived CFL condition for the DG scheme with general sub-grid bathymetry and RK32 time integrator, where we set $\textrm{CFL} = \frac{\widehat{w}_{1}}{a_{21}}$. Accounting for a sub-grid bathymetry condition requires similar CFL numbers to those found for SSP-DG with a high-order (but limited) polynomial representation of the bathymetry \cite{xing:2010}. The CFL upper bound is close to the estimated linear stability restriction for the high-order scheme, $\textrm{CFL} = \frac{1}{2r+1}$ given in \cite{cockburn:2001}. The use of more accurate quadrature formulae is instead more penalizing in terms of time step restrictions; this issue is briefly discussed in the Conclusions.

\begin{table}[h!]
    \centering
    \begin{tabular}{|c|c|c|}
     \hline
     Property, scheme & Positivity, RK32-DG & Linear Stability, SSP-DG \\
     \hline
     CFL & $\frac{\widehat{w}_1}{a_{21}}$ & $\frac{1}{2r+1}$ \\
     \hline
     $r=1$ & $0.28$  & $1/3$ \\
     \hline
     $r=2$ & $0.14$  & $1/5$ \\
     \hline
     $r=3$ & $0.085$ & $1/7$ \\
     \hline
    \end{tabular}
    \caption{Comparison of the CFL number derived in Proposition \ref{ssec:spacedisc_prop}.3 for the positivity of the water depth and the CFL for linear stability estimated in \cite{cockburn:2001} for SSP-DG.}
    \label{tab:cfl}
\end{table}

Another important theoretical property that is easy to verify for the proposed discretization is the consistency of the discrete forms of the concentration equation and continuity equation. As remarked in \cite{gross:2002}, this property is essential to guarantee a monotonic solution for tracer transport coupled to free-surface flow, whenever flux form discretizations are employed. Inserting a constant concentration into the concentration equation \eqref{eq:algebraic_concentration}, and comparing \eqref{eq:mass_rhs} and \eqref{eq:r_concentration} with $\jump{h^{*,(m)}} = \jump{\zeta^{*,(m)}}$, one obtains $R^{(m)}_{i} = F^{(m)}_{i}$. Furthermore, comparing \eqref{eq:depth_rhs} with \eqref{eq:concentration_rhs} and \eqref{eq:mass_matrix_concentration}, one has $\widetilde{C}^{n}_{i} = H_{i}(\boldsymbol{\zeta}^{n})$ and
\begin{equation}\label{eq:cwc1}
    \sum_{j=1}^{\left|\mathcal{T}_{\mathcal{H}}\right|(r+1)^{2}} \mathcal{H}_{ij}(\boldsymbol{\zeta}^{(l)}) = H_{i}(\boldsymbol{\zeta}^{(l)}), \quad H_{i}(\boldsymbol{\zeta}^{(l)}) = H_{i}(\boldsymbol{\zeta}^{n}) + \Delta t \sum_{m=1}^{l-1}a_{lm} F^{(m)}_{i},
\end{equation}
which is identical to the discrete continuity equation \eqref{eq:algebraic_elevation}. If instead the vertically integrated concentration $m = ch$ had been chosen as prognostic variable to be expanded in terms of the local polynomial basis, this would have entailed the need to compute, in the right-hand side $R^{(m)}_{i}$, the quotients $m/h$ to recover the concentration variables. Due to the non-polynomial nature of $h$ in our formulation, this would lead to the impossibility to verify $c = m/h = \textrm{const}$, and to guarantee theoretically the consistency with the fully discrete formula \eqref{eq:algebraic_elevation}, resulting in spurious monotonicity violations.

Finally, we discuss in detail the global mass conservation properties of the adaptive version of our method. We consider two independent $h-$ and degree refinement/coarsening step that yield a new mesh $\mathcal{T}^{n+1}_{\mathcal{H}}$ and basis functions with an updated polynomial degree. Both the mesh and the polynomial degree differ from the mesh $\mathcal{T}^{n}_{\mathcal{H}}$ and from the degree used at the previous time step. We denote by $\zeta^{n+1}$ the free-surface updated on the new mesh and with the new polynomial degree, by $\zeta^{n}$ the free-surface at the previous time step computed on the old mesh and with the old polynomial degree. We also denote by $\widetilde{\zeta}^{n}$ the free-surface at the previous time step remapped on the new mesh and onto the  basis functions with the new polynomial degree. We assume for simplicity an explicit Euler time discretization, since the choice of the time discretization does not affect the discussion. The evolution of the total mass over one time step can be represented as
\begin{equation*}
    \sum_{K \in \mathcal{T}^{n+1}_{\mathcal{H}}} \overline{H}_K(\boldsymbol{\zeta}^{n+1}) = \sum_{K \in \mathcal{T}^{n+1}_{\mathcal{H}}} \overline{H}_K(\widetilde{\boldsymbol{\zeta}}^{n}) - \Delta t\sum_{\varGamma \in \mathcal{E}^{B}} \int_{\varGamma} \rpth{\ave{\disc^{*,n}} + \frac{\lambda^{n}}{2}\jump{\widetilde{\zeta}^{*,n}}} \cdot \bm{n}\,\mathrm{d}\Sigma
\end{equation*}
This is obtained after summing the continuity equation \eqref{eq:algebraic_elevation} over all the degrees of freedom on each element and over all the mesh elements. Replacing in \eqref{eq:positivity-7} the maximum operator with the land-sea mask, $\max\rpth{\widetilde{\zeta}^{n}+z_{b},0} = \rpth{\widetilde{\zeta}^{n} + z_{b}}\ell^{n}$, we obtain
\begin{equation*}
    \sum_{K \in \mathcal{T}^{n+1}_{\mathcal{H}}} \overline{H}_{K}(\boldsymbol{\zeta}^{n+1}) = \sum_{K \in \mathcal{T}^{n+1}_{\mathcal{H}}} \int_{K} \rpth{\widetilde{\zeta}^{n} + z_{b}}\ell^{n}\,\mathrm{d}\Omega - \Delta t\sum_{\varGamma \in \mathcal{E}^{B}} \int_{\varGamma} \rpth{\ave{\disc^{*,n}} + \frac{\lambda^{(n)}}{2}\jump{\widetilde{\zeta}^{*,n}}} \cdot \bm{n}\,\mathrm{d}\Sigma.
\end{equation*}
If a conservative remapping is performed for the free-surface $\zeta$, the integral of the free-surface over the wet area is conserved across the refinement/coarsening step
\begin{equation}\label{eq:cons-remap}
    \sum_{K \in \mathcal{T}^{n+1}_{\mathcal{H}}} \int_{K}\widetilde{\zeta}^{n}\ell^{n}\,\mathrm{d}\Omega = \sum_{K \in \mathcal{T}^{n}_{\mathcal{H}}} \int_{K}\zeta^{n}\ell^{n}\,\mathrm{d}\Omega.
\end{equation}
We denote by $Q^{n}$ the difference between the bathymetry integrals computed in the new time step and in the old one, i.e.
\begin{equation}\label{eq:mass-loss}
    Q^{n} = \sum_{K \in \mathcal{T}^{n+1}_{\mathcal{H}}} \int_{K} z_{b}\ell^{n}\,\mathrm{d}\Omega - \sum_{K \in \mathcal{T}^{n}_{\mathcal{H}}} \int_{K} z_{b}\ell^{n}\,\mathrm{d}\Omega.
\end{equation}
Notice that, each of the above element integrals is computed with the Gauss--Legendre--Lobatto formula with $r+2$ points that has been widely discussed in the previous paragraphs. Then it follows that
\begin{equation*}
    \sum_{K \in \mathcal{T}^{n+1}_{\mathcal{H}}} \overline{H}_K(\boldsymbol{\zeta}^{n+1}) = \sum_{K \in \mathcal{T}^{n}_{\mathcal{H}}} \overline{H}_K(\boldsymbol{\zeta}^{n}) - \Delta t \sum_{\varGamma \in \mathcal{E}^{B}} \int_{\varGamma} \rpth{\ave{\disc^{*,n}} + \frac{\lambda^{(n)}}{2}\jump{\zeta^{*,n}}} \cdot \bm{n}\,\mathrm{d}\Sigma + Q^{n}.
\end{equation*}
Therefore, the change in mass that occurs in the time step from $t^{n}$ to $t^{n+1}$, including degree and mesh refinement/coarsening, consists of two terms: the mass flux across the domain boundary and a spurious source/sink term $Q^{n}$. Within a pure degree adaptation strategy with $\mathcal{T}^{n+1}_{\mathcal{H}} = \mathcal{T}^{n}_{\mathcal{H}}$, the mass conservation error is null, since the mass matrix, and therefore the bathymetry integral, is computed at each time step always with the same quadrature formula independently of the polynomial degree. The error term $Q^{n},$ instead, is related to the bathymetry values used in the mass matrices computed at the different mesh levels.
During the $h-$refinement and coarsening process, the bathymetry values at the quadrature nodes for the new resolution are evaluated at run-time from an analytical function or extracted from a bathymetry database. Such sampling of the bathymetry on different meshes yields $Q^{n} \neq 0$, thus inducing a potential violation of the mass conservation,  an issue encountered in many mesh adaptation techniques \cite{arpaia:2018, donat:2014, liang:2015, popinet:2011}.
In Section \ref{sec:realistic}, we estimate $\left|Q^{n}\right|$ for an adaptive coastal application with realistic geometry and bathymetry, showing that reasonably small values can be achieved. Exact conservation could be obtained by adapting the conservative bathymetry remaps originally proposed in \cite{popinet:2011} in a Finite Volume framework to high-order finite elements, but this goes beyond the scope of this paper.

\section{Idealized numerical experiments}
\label{sec:tests}

The numerical method outlined in Section \ref{sec:numerical} is now validated in a number of relevant idealized benchmarks. If an exact solution is available, accuracy is measured by the $L^{2}$ error norm on the free-surface elevation and on the discharge, which we compute as
\begin{equation}\label{eq:errors_l2}
    \left|\left|\zeta - \zeta_{\mathrm{ex}}\right|\right|^{2}_{2} = \sum_{K \in \mathcal{T}_{\mathcal{H}}} \int_{K} \rpth{\zeta - \zeta_{\mathrm{ex}}}^{2}\,\mathrm{d}\Omega, \qquad \left|\left|\disc - \disc_{\mathrm{ex}}\right|\right|^{2}_{2} = \sum_{K \in \mathcal{T}_{\mathcal{H}}} \int_{K} \left|\disc - \disc_{\mathrm{ex}}\right|^{2}\,\mathrm{d}\Omega.
\end{equation}

\subsection{Preservation of hydrostatic equilibrium}
\label{ssec:lake_at_rest_pert}

\begin{figure}[h!]
    \centering
    \includegraphics[trim={2cm 14cm 0cm 17cm},clip,scale=0.11]{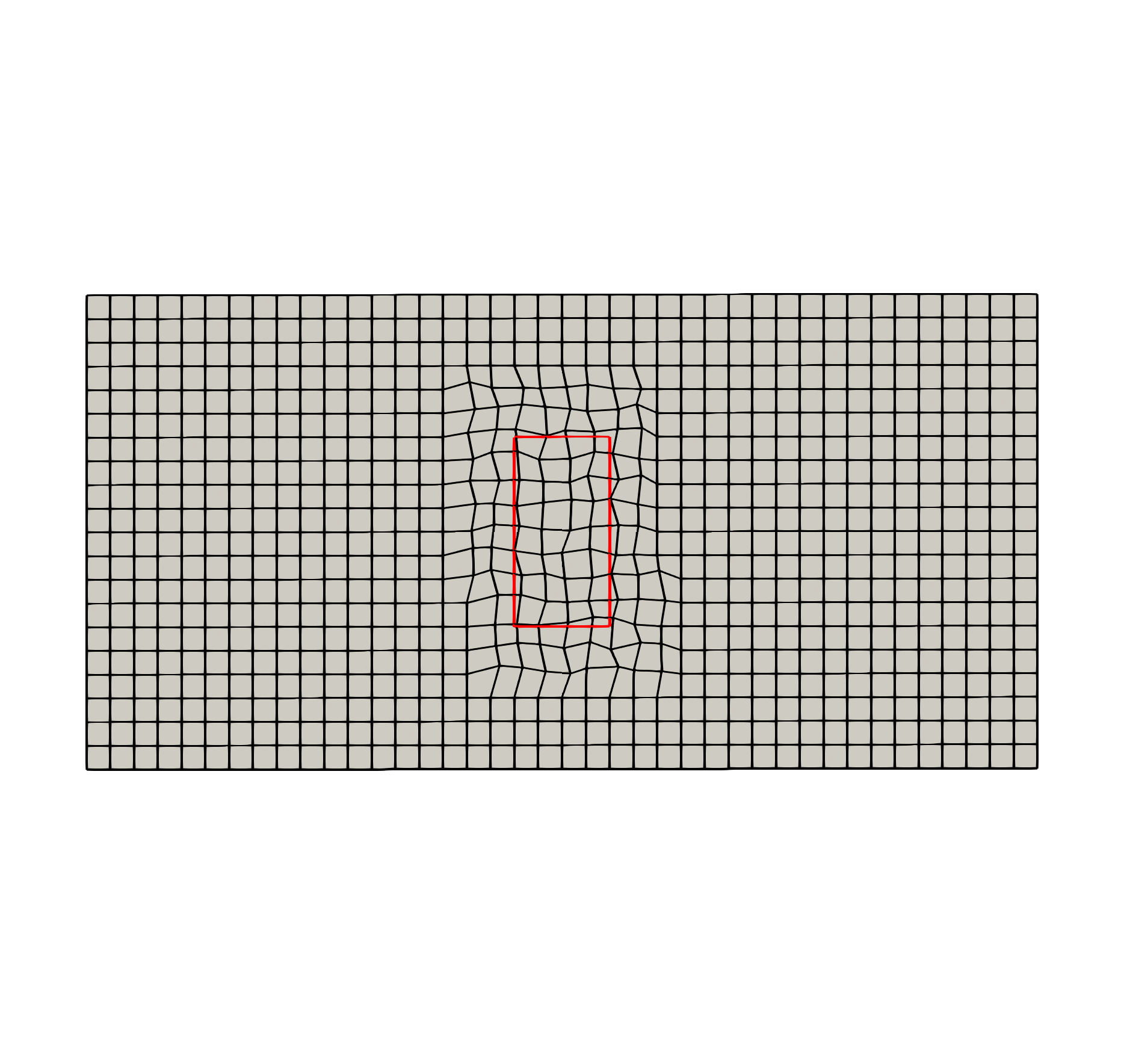}
    \includegraphics[trim={0cm 0cm 0cm 1cm},clip,scale=0.06]{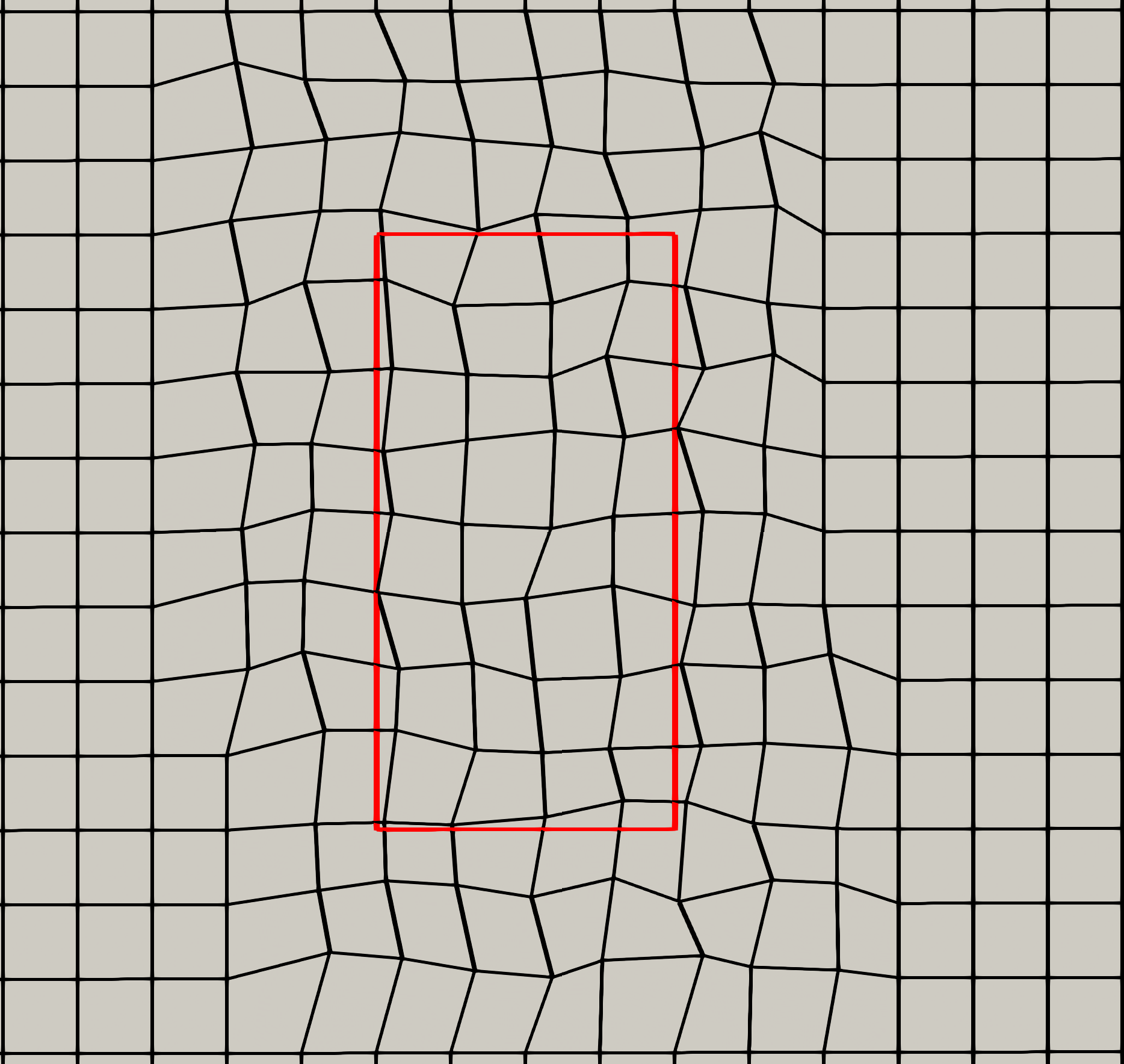}
    \caption{Lake-at-rest test. Left: distorted mesh with red line in correspondence of the bathymetry jump. Right: zoom of the area where the jump is located.}
    \label{fig:lake_at_rest_mesh}
\end{figure}

\begin{figure}[h!]
    \centering
    \includegraphics[trim={0 1cm 0 2cm},clip,scale=0.13]{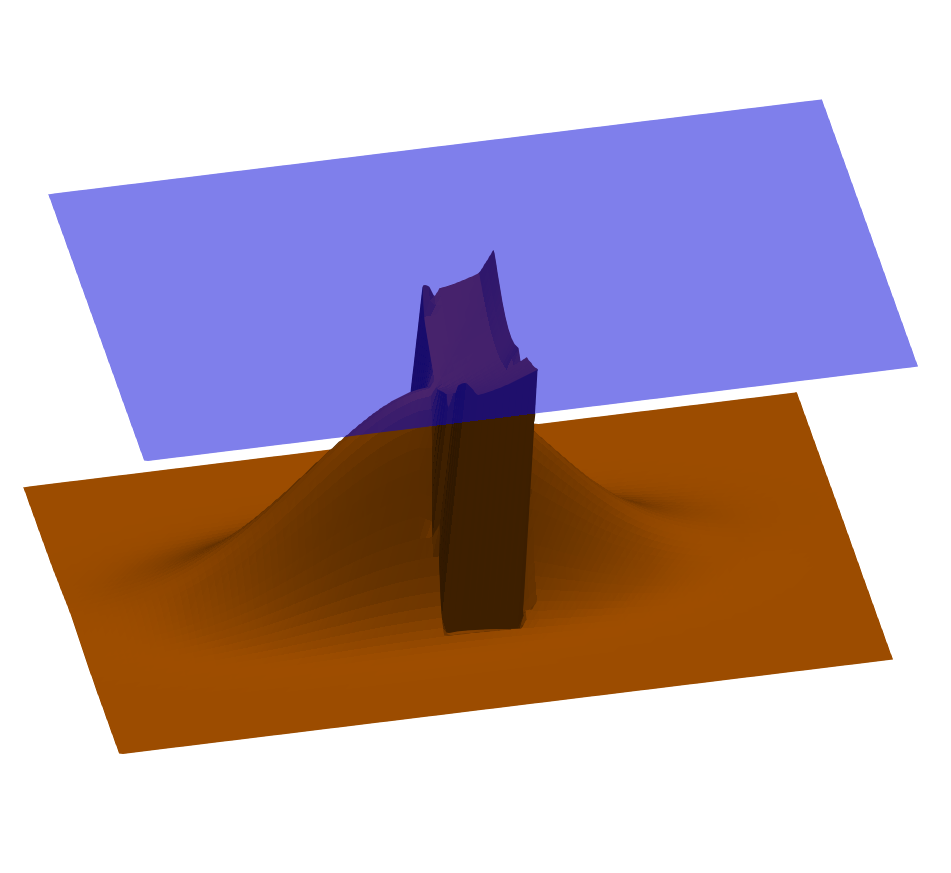}\includegraphics[trim={0cm 0cm 0cm 0cm},clip,scale=0.65]{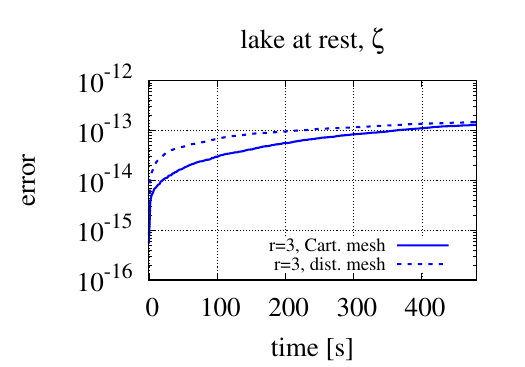}\includegraphics[trim={0cm 0cm 0cm 0cm},clip,scale=0.65]{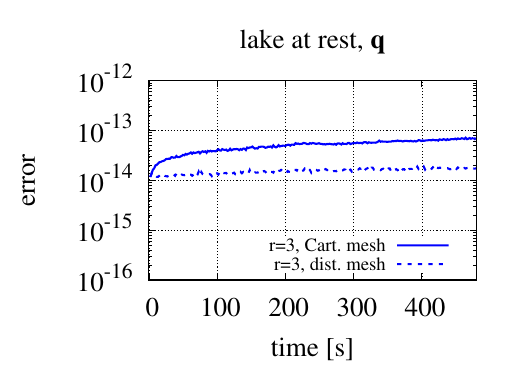}
    \caption{Lake-at-rest test. Time history of the errors for $r=3$ obtained for the submerged discontinuous bathymetry on a Cartesian mesh (blue) and on mesh with randomly distorted edges (dashed blue). Middle: $\zeta$ $L^{2}$ error history. Right: $\disc$ $L^{2}$ error history.}
    \label{fig:lake_at_rest_history}
\end{figure}

In order to check numerically the preservation of the hydrostatic equilibrium, or {\cal C}-property, that has been discussed in Section \ref{sec:numerical}, we consider the classical lake-at-rest benchmark. A rectangular basin is considered with horizontal dimensions $\SI{2}{\meter} \times \SI{1}{\meter}$ and constant free-surface elevation value of $\SI{1}{\meter}$. We consider two bathymetry profiles. The first one is completely submerged and it is given by the discontinuous function
\begin{equation*}
    z_{b}(\bm{x}) = -z_{b0}\exp\rpth{\psi(\bm{x})},
\end{equation*}
with
\begin{equation*}
    z_{b0} = \SI{0.65}{\meter}, \quad
    \psi = 
    \begin{cases}
        \sqrt{\rpth{x - 0.9}^{2} + \rpth{x - 0.5}^{2}}, & \text{if } \bm{x} \in \rpth{0.9,1.1} \times \rpth{0.3, 0.7} \\
        -5\rpth{x - 0.9}^{2} - 50\rpth{x - 0.5}^{2}, & \text{otherwise,}
    \end{cases}
\end{equation*}
where all the length values given above are in meters. The mesh has $40 \times 20$ elements arranged in a Cartesian pattern and it has been constructed so that the element edges are aligned along the discontinuity. Starting from this mesh, we consider a second mesh whose edges are randomly distorted around the discontinuity (Figure \ref{fig:lake_at_rest_mesh}). As a result, the jump in the bathymetry is not aligned with the edges, but it crosses the elements and the edges themselves. The distorted mesh is used to demonstrate the robustness of the method when the mesh is not aligned with the bathymetric gradients and the bathymetry exhibits large variations within an element. In Table \ref{tab:lake_at_rest_pert_rest} we report the errors at a very long time $T = \SI[parse-numbers=false]{480}{\second}$, which corresponds to 220 times the time it takes for the wave to travel across the basin and, more specifically, to between 145,000 and 670,000 time steps, depending on the polynomial order. Figure \ref{fig:lake_at_rest_history} reports the time history of the errors for $r=3$. It can be observed that the hydrostatic balance is preserved up to machine precision on both meshes, without any need of special treatment of the bathymetry. A slight growth of the error occurs over a very long timescale, which is of minor relevance for our short- and medium-range applications.

Next, we consider a bathymetry profile given by the smooth function with a wet-dry interface
\begin{equation*}
    z_{b}(\bm{x}) = -z_{b0}\exp\rpth{\psi(\bm{x})},
\end{equation*}
with
\begin{equation*}
    z_{b0} = \SI{1.3}{\meter}, \quad
    \psi = -5\rpth{x - 0.9}^{2} - 50\rpth{x - 0.5}^{2}
\end{equation*}
where all the length values given above are in meters. For this test we only consider the Cartesian mesh. Again, hydrostatic balance is preserved up to machine precision also in presence of wet-dry fronts. Figure \ref{fig:lake_at_rest_history} reports the time history of the errors for the different polynomial orders.
\begin{table}[h!]
    \centering
    \begin{tabular}{|c|c|c|}
        \hline
        \multicolumn{3}{|c|}{Submerged discontinuous bathymetry, Cartesian mesh} \\
        \hline
        Polynomial degree $r$ & $\zeta$ error & $\disc$ error \\
        \hline
        $r=1$ & $\num{6.28e-16}$ & $\num{0.00e+00}$ \\
        \hline
        $r=2$ & $\num{6.28e-16}$ & $\num{0.00e+00}$ \\
        \hline
        $r=3$ & $\num{1.298e-13}$ & $\num{6.947e-14}$ \\
        \hline
        \multicolumn{3}{|c|}{Submerged discontinuous bathymetry, distorted mesh} \\
        \hline
        Polynomial degree $r$ & $\zeta$ error & $\disc$ error \\
        \hline
        $r=1$ & $\num{6.28e-16}$ & $\num{0.00e+00}$ \\
        \hline
        $r=2$ & $\num{0.00e+00}$ & $\num{0.00e+00}$ \\
        \hline
        $r=3$ & $\num{1.469e-13}$ & $\num{1.733e-14}$  \\
        \hline
        \multicolumn{3}{|c|}{Partially submerged bathymetry, Cartesian mesh} \\
        \hline
        Polynomial degree $r$ & $\zeta$ error & $\disc$ error \\
        \hline
        $r=1$ & $\num{2.170e-15}$ & $\num{3.886e-14}$ \\
        \hline
        $r=2$ & $\num{2.523e-15}$ & $\num{1.189e-13}$ \\
        \hline
        $r=3$ & $\num{1.145e-13}$ & $\num{7.984e-14}$  \\
        \hline
    \end{tabular}
    \caption{$L^{2}$ errors in lake-at-rest simulations at $T = \SI[parse-numbers=false]{480}{\second}$.}
    \label{tab:lake_at_rest_pert_rest}
\end{table}
\begin{figure}[h!]
    \centering
    \includegraphics[trim={0 1cm 0 2cm},clip,scale=0.13]{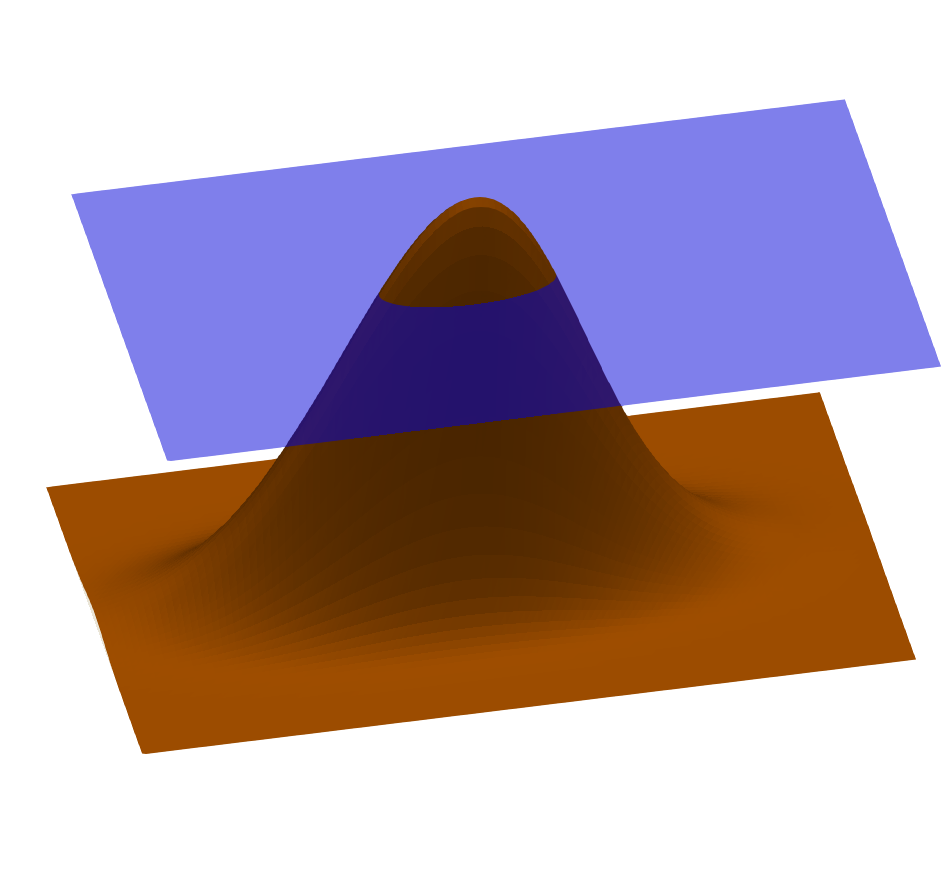}\includegraphics[scale=0.65]{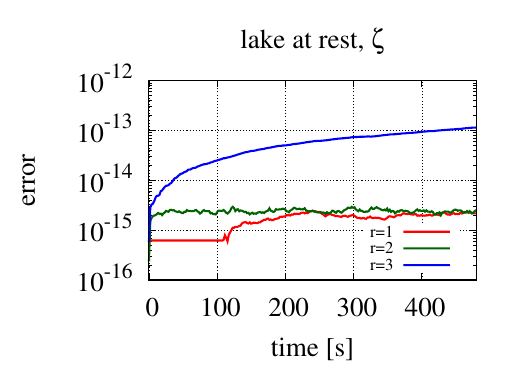}\includegraphics[scale=0.65]{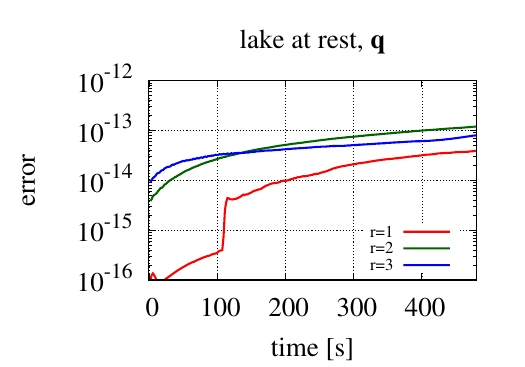}
    \caption{Lake-at-rest. Time history of the errors for $r=1,2$, and $3$ obtained for a partially submerged bathymetry on a Cartesian mesh. Middle: $\zeta$ $L^{2}$ error history. Right: $\disc$ $L^{2}$ error history.}
    \label{fig:lake_at_rest_history2}
\end{figure}

\subsection{Channel flow}
\label{ssec:channel_friction}

\begin{figure}[h!]
    \centering
    \includegraphics[scale=0.29]{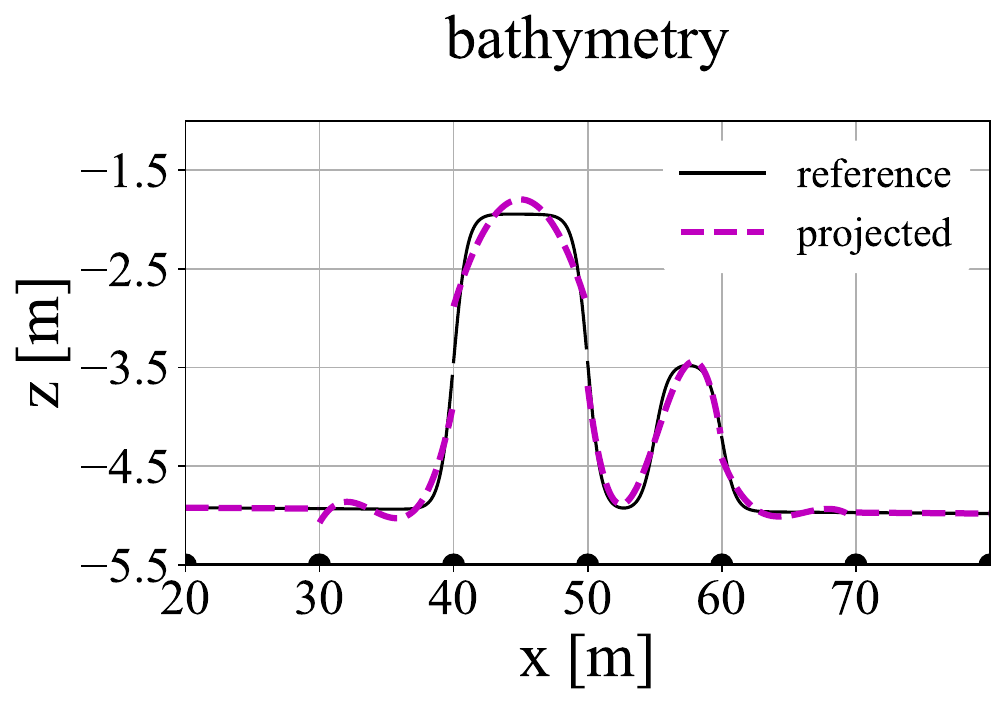}\includegraphics[scale=0.45]{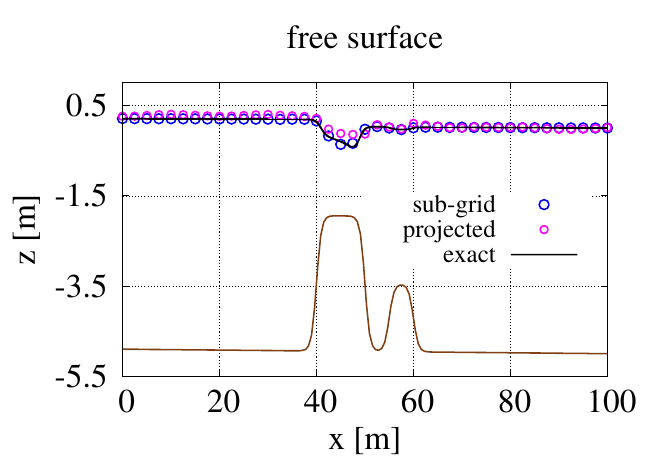}\includegraphics[scale=0.45]{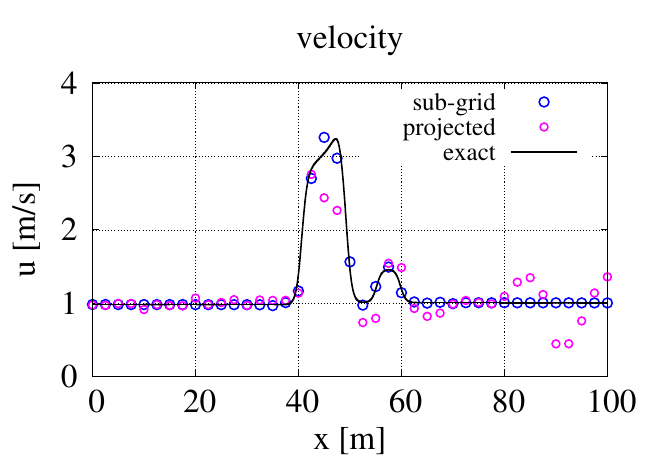}
    \includegraphics[scale=0.29]{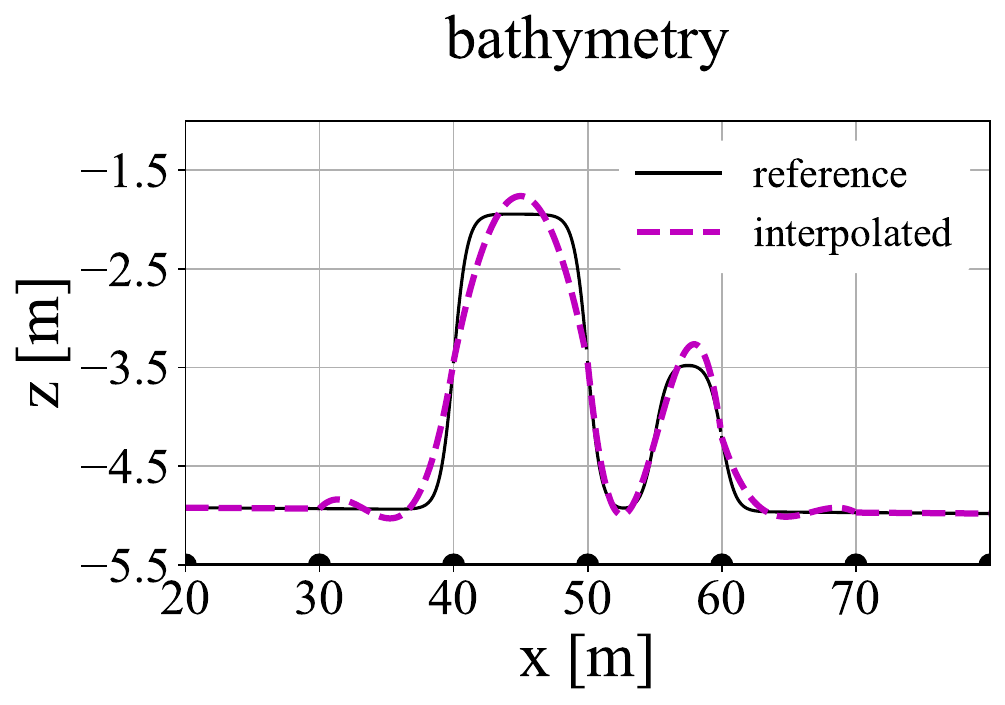}\includegraphics[scale=0.45]{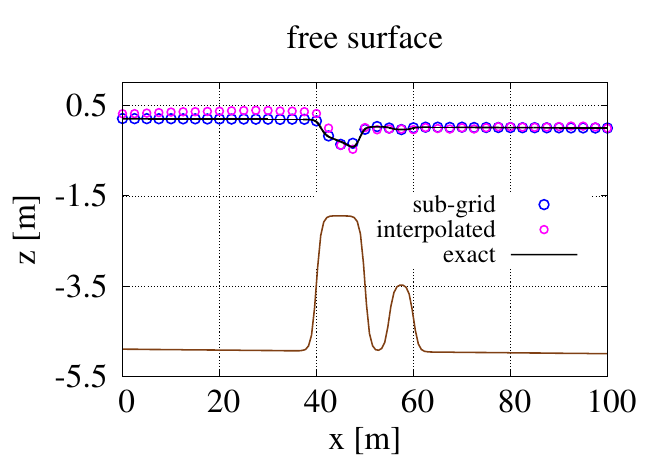}\includegraphics[scale=0.45]{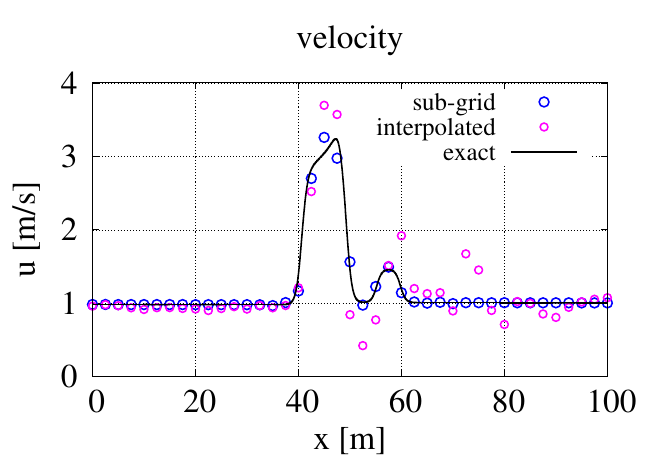}
    \includegraphics[scale=0.29]{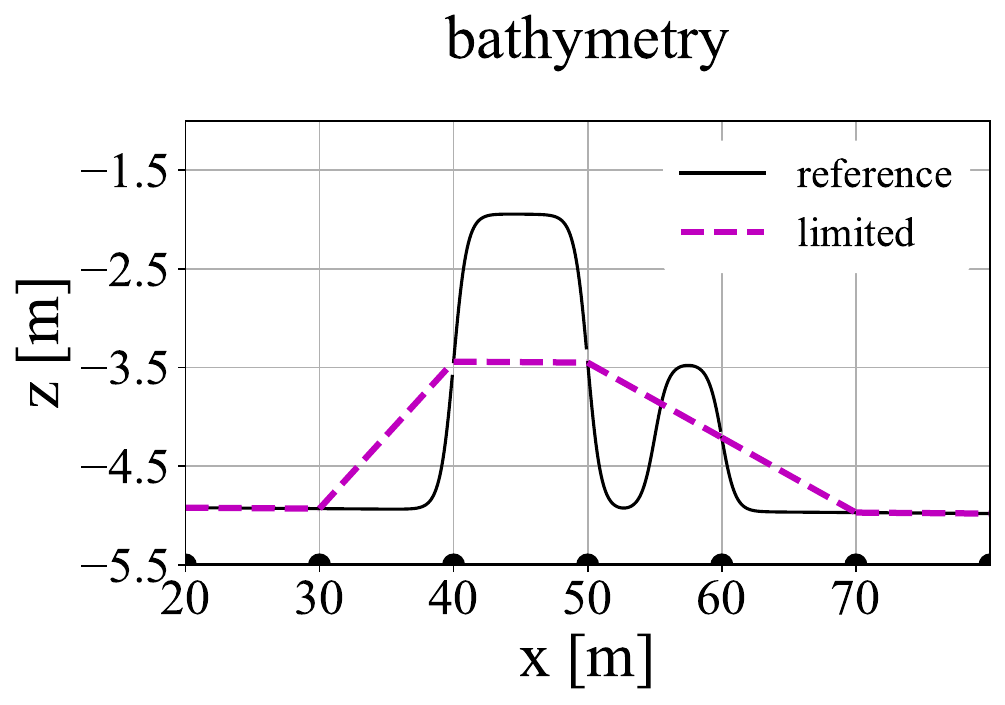}\includegraphics[scale=0.45]{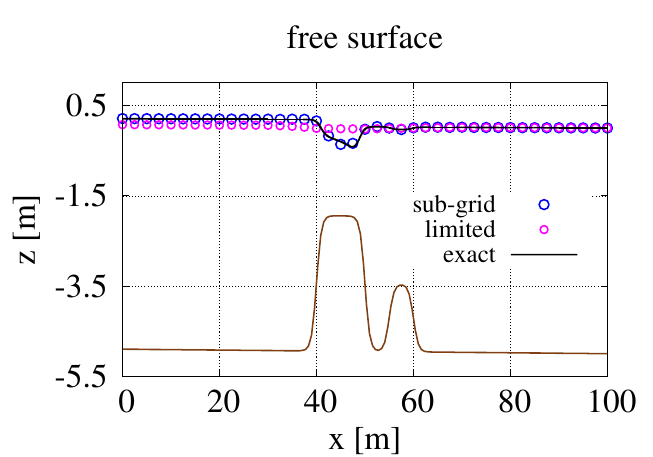}\includegraphics[scale=0.45]{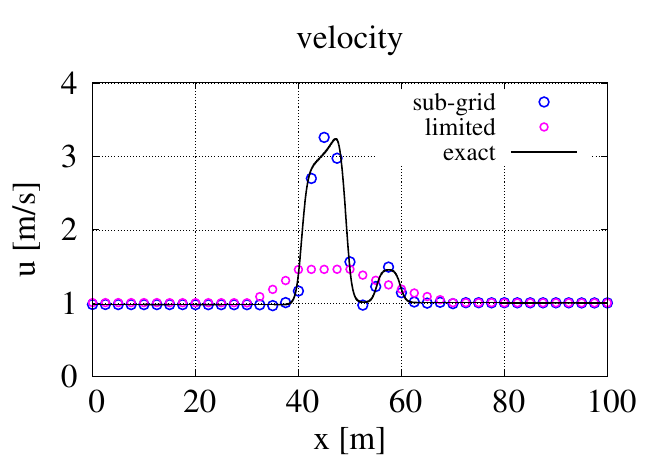}
    \caption{Channel flow, case with irregular bathymetry. Blue dots denote the results obtained using the coarsest mesh ($N = 10$) and the highest order ($r = 3$). Comparison with different bathymetry polynomial approximations (magenta dots): projected (top), interpolated (middle) and limited bathymetry (bottom). Left: bathymetry. Middle: free-surface elevation. Right: velocity.}
    \label{fig:channel_irregular}
\end{figure}

We now examine the capability of the scheme to represent bathymetric obstacles that are not fully resolved at the grid scale. For this purpose, we consider a steady state solution of the one-dimensional shallow water equations with constant discharge $hu = q_{0}$, varying topography and friction. Given a bathymetry profile $z_{b}(x)$, we compute the water depth that balances the flow from the integration of the momentum equation in \eqref{eq:shw}, which reduces to the following nonlinear ODE
\begin{equation}\label{eq:const_discharge}
    \rpth{\frac{q_{0}^{2}}{gh^{3}} - 1}\frac{\partial h}{\partial x} = -\frac{\partial z_{b}}{\partial x} + \frac{\gamma(q_{0},h)q_{0}}{gh}
\end{equation}
A reference numerical solution of the ODE is computed using the classical fourth order Runge-Kutta scheme (RK44) \cite{lambert:1991} (see Table \ref{tab:rk4} in Appendix \ref{app:IMEX_coeffs}) on a   mesh much finer than those used in the computations. For one dimensional subcritical flows we can impose only the value of one characteristic or physical variable at each boundary. Based on physical arguments, we impose, as customary, the discharge value at the inflow boundary and the water height $\zeta = \SI{0}{\meter}$ at the outflow boundary. Wall boundary conditions are applied on the $y-$direction. We employ a channel of length $L = \SI{100}{\meter}$ and the following bathymetry profile
\begin{align*}
    z_{b}(x) = 4.9 + 0.001 x &- \frac{z_{b0}}{2} \rpth{\tanh\rpth{\frac{x - 0.4L}{c_{0}}} - \tanh\rpth{\frac{x - 0.5L}{c_{0}}}} \\
    &- \frac{z_{b0}}{4}\rpth{\tanh\rpth{\frac{x - 0.55L}{c_{0}}} - \tanh\rpth{\frac{x - 0.6L}{c_{0}}}}. 
\end{align*}
that displays two obstacles with steep slopes. The values of the Manning coefficient $n$ and of $z_{b0}$ are reported in Table \ref{tab:channel_table}. We consider a very coarse two-dimensional mesh with $10$ elements along the $x-$direction. For such resolution, the obstacles are unresolved at the grid scale. In Figure \ref{fig:channel_irregular}, we compare the numerical solution, obtained with $r=3$, against the exact one. We repeat the same test using a polynomial approximation of the bathymetry. The bathymetry coefficients at the nodes can be the interpolating ones or can be computed from a projection approach, the last being more suited for irregular data. We also test a piecewise linear bathymetry over the element with the nodal coefficient interpolated at the grid nodes. For the method that we propose, the computed solution is close to the exact one, in spite of the undersampling of the bathymetry at the grid scale. If the bathymetry is approximated with high-order polynomials, it exhibits oscillations (first and second rows in Figure \ref{fig:channel_irregular}). As a result of the simulations, after $\SI{1800}{\second}$ the solution does not converge to a steady state and exhibits increasing errors in the velocity field downstream of the hills, which becomes physically inconsistent (Figure \ref{fig:channel_irregular} and Table \ref{tab:channel_irregular_errors}). If the bathymetry is approximated with piecewise linear polynomials, it does not exhibit oscillations but the hills are barely resolved (third row in Figure \ref{fig:channel_irregular}). In this case, we get convergence to a steady state but with highly inaccurate results. This result highlights the advantage of using high-order polynomials to approximate the free-surface and a high-order quadrature formula that recovers accurate information on the sub-grid bathymetric features. Similar results have been observed in \cite{orlando:2024b} for atmospheric flows, where it has been shown that the use of a high-order polynomial representation acts as a sort of sub-tessellation and significantly enhances the resolution of small-scale features.

\begin{figure}[h!]
    \centering
    \includegraphics[scale=0.7]{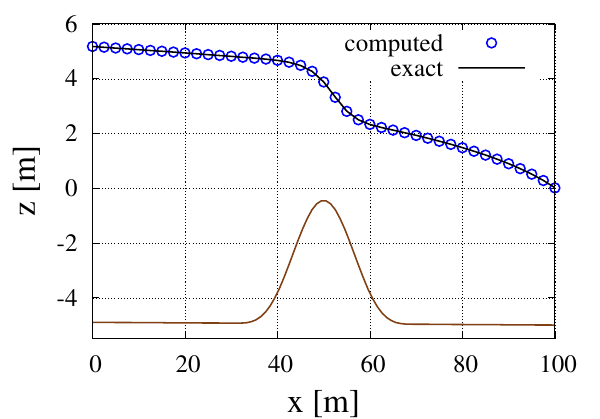}
    \includegraphics[scale=0.6]{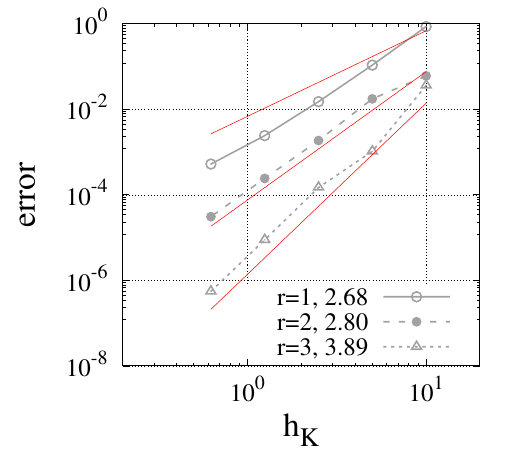}
    \caption{Channel flow, case with Manning coefficient $n=1$ (strong friction) and regular bathymetry. Left: free-surface elevation. Right: Convergence for different $r-$order accurate DG schemes with the IMEX-RK time discretization, $L^{2}$ error on the free-surface as a function of the mesh size (red lines trace the theoretical convergence rate).}
    \label{fig:channel_superviscous}
\end{figure}

If $\gamma \neq 0$, so that bottom friction is present, this benchmark allows also to check the correctness of the implementation of the friction terms and the ability of the IMEX method to use time steps beyond the stability limit of the explicit RK methods. We consider the same channel flow benchmark employed in \cite{rosatti:2011}. However, we consider the smoother depth profile
\begin{equation*}
    z_{b}(x) = 4.9 + 0.001x -
    \begin{cases}
        z_{b0}\cos^{4}\rpth{\frac{\pi\rpth{x - L/2}}{2c_{0}}}, & \text{if } -c_{0} \leq x - L/2 \leq c_{0}   \\
        0 & \text{otherwise,}
    \end{cases}
\end{equation*}
so as to achieve the correct convergence behaviour for higher polynomial degrees. Moreover we consider a much larger value of the Manning friction coefficient, thus allowing to test the performance of the implicit treatment of the friction term in the IMEX time discretization method. The specific values of $n$ and $z_{b0}$ used for this test are reported in Table \ref{tab:channel_table}. We consider five meshes with $N$ elements along the $x-$axis, where $N$ ranges from $10$ to $160$. In Figure \ref{fig:channel_superviscous}, we compare the numerical solution against the exact one obtained on the mesh with $40$ elements and $r=1$. The convergence study, also in the same figures, shows the correct empirical order of convergence for $r=1,2,3$. Notice that, since this is a steady state benchmark, the spatial convergence order is always retrieved, in spite of the second order accuracy of the time discretization method. In the case with strong friction, on the coarsest meshes the simulations run stably with a time step that is approximately 2.5 times larger than the maximum time step for which the RK32 method could run stably in the same test.
\begin{table}[h!]
    \centering
    \begin{tabular}{|c|c|c|c|c|}
    \hline
    Configuration & $q_{0}\,\spth{\SI{}{\meter^{2}\second^{-1}}}$ & $z_{b0}\,\rpth{\SI{}{\meter}}$ & $c_{0}\,\rpth{\SI{}{\meter^{-1}}}$ & $n\,\rpth{\SI{}{\meter^{-1/3}\second}}$ \\
    \hline
    Irregular bathymetry & $5$ & $3$ & $1$ & $0.065$ \\
    \hline
    Strong friction & $5$ & $4.5$ & $20$ & $1$ \\
    \hline
    \end{tabular}
    \caption{Values of the physical parameters for different cases of the channel flow benchmark.}
    \label{tab:channel_table}
\end{table}
\begin{table}[h!]
    \centering
    \begin{tabular}{|c|c|c|c|}
        \hline
        Bathymetry approximation & $\zeta$ error & $\disc$ error \\
        \hline
        at quadrature nodes, sub-grid & $\num{0.175}$ & $\num{0.108}$ \\
        \hline
        polynomial, interpolated & $\num{1.176} $ & $\num{12.12}$ \\
        \hline
        polynomial, projected & $\num{0.795}$ & $\num{14.03}$ \\
        \hline
        polynomial, limited & $\num{1.207}$ & $\num{13.01}$ \\
        \hline
    \end{tabular}
    \caption{Channel flow, case with irregular bathymetry. $L^{2}$ error for different bathymetry representation for the case of $N = 10$ elements and $r=3$.}
    \label{tab:channel_irregular_errors}
\end{table}

\subsection{Thacker parabolic 2d oscillations}
\label{ssec:Thacker}

We now evaluate the capability of the numerical method to handle moving wet-dry fronts correctly. We also assess the mass conservation property of the wetting and drying algorithm, as well as the preservation of an initial constant concentration, as discussed in Section \ref{sec:numerical}. The shallow water equations admit a periodic solution with wet-dry transition \cite{thacker:1981}. Here we consider the radially symmetric oscillating paraboloid, proposed in the SWASH test suite \cite{delestre:2013} with parabolic bathymetry
$$z_{b}(R)= h_{0}\rpth{1-\frac{R^{2}}{a^{2}}},$$
with $R$ denoting the distance from the center of the domain $\rpth{L/2,L/2}$. The exact solution, in terms of free-surface and velocity, is
\begin{align*}
\zeta(R,t) &=
- h_{0}\spth{\frac{\sqrt{1 - A^{2}}}{1-A\cos\rpth{\omega t}} - 1 - \frac{R^{2}}{a^{2}}\rpth{\rpth{\frac{\sqrt{1 - A^{2}}}{1-A\cos\rpth{\omega t}}}^{2} -1}},
\\
u(x,t) &=
\frac{1}{1 - A\cos\rpth{\omega t}}\rpth{\frac{A\omega}{2}\rpth{x - \frac{L}{2}}\sin\rpth{\omega t}},
\\
v(y,t) &=
\frac{1}{1 - A\cos\rpth{\omega t}}\rpth{\frac{A\omega}{2}\rpth{y - \frac{L}{2}}\sin\rpth{\omega t}},
\end{align*}
where $\omega = \frac{\sqrt{8gh_{0}}}{a}$ and $A = \frac{a^{2} - r_{0}^{2}}{a^{2} + r_{0}^{2}}$.
The initial solution is computed after the exact solution at $t=0$. We consider the following parameters
$$h_{0} = \SI{0.1}{\meter}, \quad L = \SI{3}{\meter}, \quad a = \SI{1}{\meter}, \quad \text{and} \quad r_{0} = \SI{0.8}{\meter},$$
and the final time corresponds to $3T$, the period of one oscillation being $T = \frac{2\pi}{\omega}$. In order to verify numerically that the tracer discretization is consistent with the continuity equation (CWC), we add to this classical test the transport of a tracer with initial concentration $c = 1$. The two depth thresholds for handling wetting-drying are chosen as follows. The threshold $\epsilon$ for the desingularization of the velocity in \eqref{eq:velocity} depends on the mesh size; it is set to $\SI{2.5e-3}{\meter}$ for $N=400$ and is doubled at each coarsening level. The threshold for coarsening the degree in \eqref{eq:criteria_p} is set as $h_{\text{lim}} = \SI{1e-2}{\meter}$.

\begin{figure}[h!]
    \centering
    \includegraphics[scale=0.6]{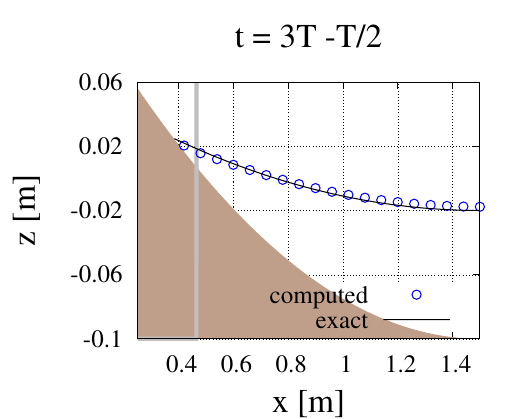}    \includegraphics[scale=0.6]{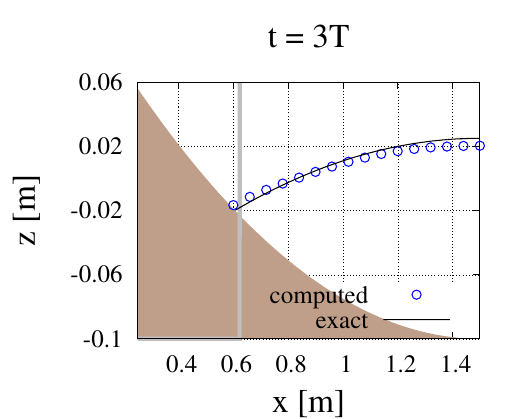}\includegraphics[scale=0.6]{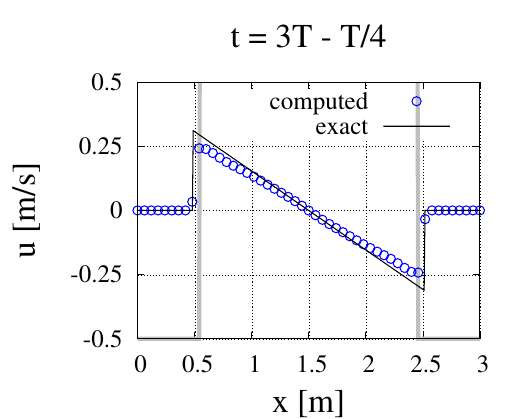}
    \caption{Thacker parabolic oscillations, solution for RK32-DG schemes obtained with $r=1$ and $N=400$ points. Cut of the solution along the line $y=\SI{1.5}{\meter}$ during the third oscillation. Left: free-surface during wetting phase. Middle: free-surface during drying phase. Right: velocity during flood/ebb.
    Black line is the exact solution. The vertical grey line represents the interface between elements flagged for degree adaptivity.}
    \label{fig:thacker2d_lines}
\end{figure}

\begin{figure}[h!]
    \centering
    \includegraphics[scale=0.8]{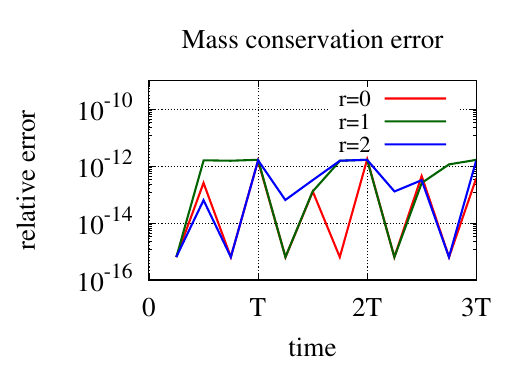}\includegraphics[scale=0.8]{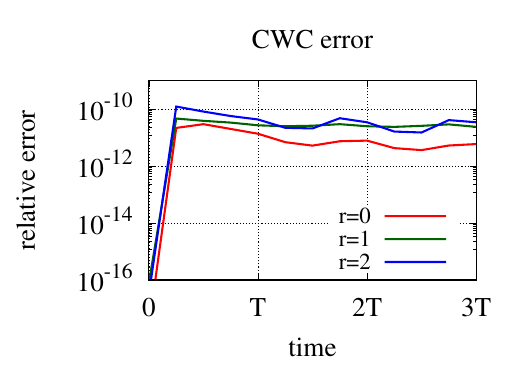}
    \caption{Thacker parabolic oscillations. Time history of the relative mass conservation error (left) and CWC error (right) with RK32-DG and different polynomial orders.}
    \label{fig:thacker2d_error}
\end{figure}

We perform a mesh convergence study employing five Cartesian meshes with $N$ elements in each direction, ranging from $N=25$ to $N=400$.
The errors obtained for different polynomial orders are reported in Table~\ref{tab:thacker2d_table} and demonstrate the correct convergence with mesh refinement. Figure~\ref{fig:thacker2d_lines} compares the numerical and exact solutions at different time instants during the last oscillation corresponding, respectively, to wetting, drying and ebb/flood phases. The solution obtained for $r=1$ and $N=400$ agrees well with the analytical solution. In the same figure, the region close to the wet-dry interface where the scheme is reverted to $r=0$ is also highlighted with a gray line. The use of higher-order polynomials does not provide significant advantages due to the $\mathcal{C}^{1}$ regularity of the velocity field and is not shown. 
Only two iterations of the Newton method are required to ensure mass conservation within numerical round-off errors associated with double precision arithmetic (see Figure~\ref{fig:thacker2d_error}). We can conclude that the overhead associated with the nonlinear iterations is negligible for this test case. In Figure~\ref{fig:thacker2d_error} we show that the CWC property is also verified within numerical round-off errors.
\begin{table}[h!]
    \centering
    \begin{tabular}{|c|c|c|c|}
        \hline
        & \multicolumn{3}{|c|}{$\zeta$ error} \\
        \hline
        $N$ & $r=0$ & $r=1$ & $r=2$ \\
        \hline
        25 & \num{2.253e-02} & \num{2.064e-02} & \num{2.032e-02} \\
        \hline
        50 & \num{2.184e-02} & \num{1.570e-02} & \num{1.715e-02} \\
        \hline
        100 & \num{1.810e-02} & \num{1.032e-02} & \num{1.094e-02} \\
        \hline
        200 & \num{1.281e-02} & \num{6.202e-03} & \num{8.395e-03} \\
        \hline
        400 & \num{7.886e-03} & \num{3.418e-03} & \num{6.234e-03} \\
        \hline
    \end{tabular}
    \caption{Thacker parabolic oscillations. $L^{2}$ errors with mesh refinement for RK32-DG scheme and different polynomial orders. $N$ denotes the number of elements in each direction.}
    \label{tab:thacker2d_table}
\end{table}

\subsection{Travelling vortex}
\label{ssec:travelling_vortex}

The last idealized test consists of a compact support travelling vortex which fulfils the shallow water equations with zero force term on the right-hand side. For this test case, described in detail in \cite{ricchiuto:2021}, an analytical solution is available, which is used to assess the accuracy and correctness of the AMR implementation. The initial conditions read as follows
\begin{align*}
    h(R) &= h_{0} - \frac{4}{g}\rpth{\frac{4\varGamma r_{0}}{\pi}}^{2}\rpth{H\rpth{\pi/2} - H\rpth{\rho/2}},\quad \rho = \frac{\pi R}{r_{0}}, \\
    u &= u_{\infty} - \rpth{y - y_{0}}\omega(\rho), \\
    v &= \rpth{x - x_{0}}\omega(\rho),
\end{align*}
with $R$ denoting the distance from the vortex center $\rpth{x_{0}, y_{0}}$. Moreover, $u_{\infty}$ is the background velocity oriented along the $x-$axis at which the vortex is transported and
$\omega(\rho) = 4\varGamma\cos^{4}\rpth{\rho/2}, $
where $\varGamma$ is a free parameter that controls the vortex strength. It has been selected to have a minimal depth of $h(0) = h_{\text{min}}$ at the center of the vortex. The definition of $H(x)$ is the following
\begin{align*}
    H(x) &= \frac{35\cos(2x)}{384} + \frac{35x\sin(2x)}{192} + \cos^{6}(x)\rpth{\frac{\cos^{2}(x)}{64} + \frac{7}{288}} + \frac{35\cos^{2}(2x)}{3072} + \frac{35x^{2}}{256} \\
    &+ \frac{35x\cos(2x)\sin(2x)}{768} + \frac{x\cos^{5}(x)\sin x\rpth{\cos^{2}(x) + \frac{7}{6}}}{8}.
\end{align*}
Notice that the initial profiles and the solution are $\mathcal{C}^{4}$ functions, which is enough to test a fourth order accurate RK-DG scheme, according to  classical interpolation estimates. The parameter values defining the vortex and the background flow are the following
\begin{equation*}
    h_{0} = \SI{1}{\meter}, \quad h_{\text{min}} = \SI{0.9}{\meter}, \quad u_{\infty} = \SI{6}{\meter\per\second}, \quad r_{0} = \SI{0.25}{\meter}, \quad \text{and} \quad \, \rpth{x_{0}, y_{0}} = \rpth{\SI{0.5}{\meter}, \SI{0.5}{\meter}}.
\end{equation*}
The final time is $T = \SI[parse-numbers=false]{1/6}{\second}$, while the computational domain is given by the rectangle $\SI[parse-numbers = false]{\rpth{0,2} \times \rpth{0,1}}{\meter\squared}$. Given the supercritical flow regime, the exact solution is imposed at the left inflow boundary, while a supercritical outflow is considered at the right boundary, where the internal value is imposed. A wall boundary condition is imposed at the top and bottom boundaries. Only for this test, we employ different fully explicit time discretizations beyond the three-stage second-order Runge-Kutta (RK32), which is the explicit companion method of the IMEX-RK method (see Appendix \ref{app:IMEX_coeffs}). More specifically, we employ the third order Strong Stability Preserving schemes \cite{gottlieb:2001}, denoted as SSP33 (see Table \ref{tab:rk3} in Appendix \ref{app:IMEX_coeffs}), and RK44 scheme already recalled in Section \ref{ssec:channel_friction}. The time discretization is chosen to match the formal order of accuracy of the spatial discretization.

\begin{figure}[h!]
    \centering
    \includegraphics[scale=0.1]{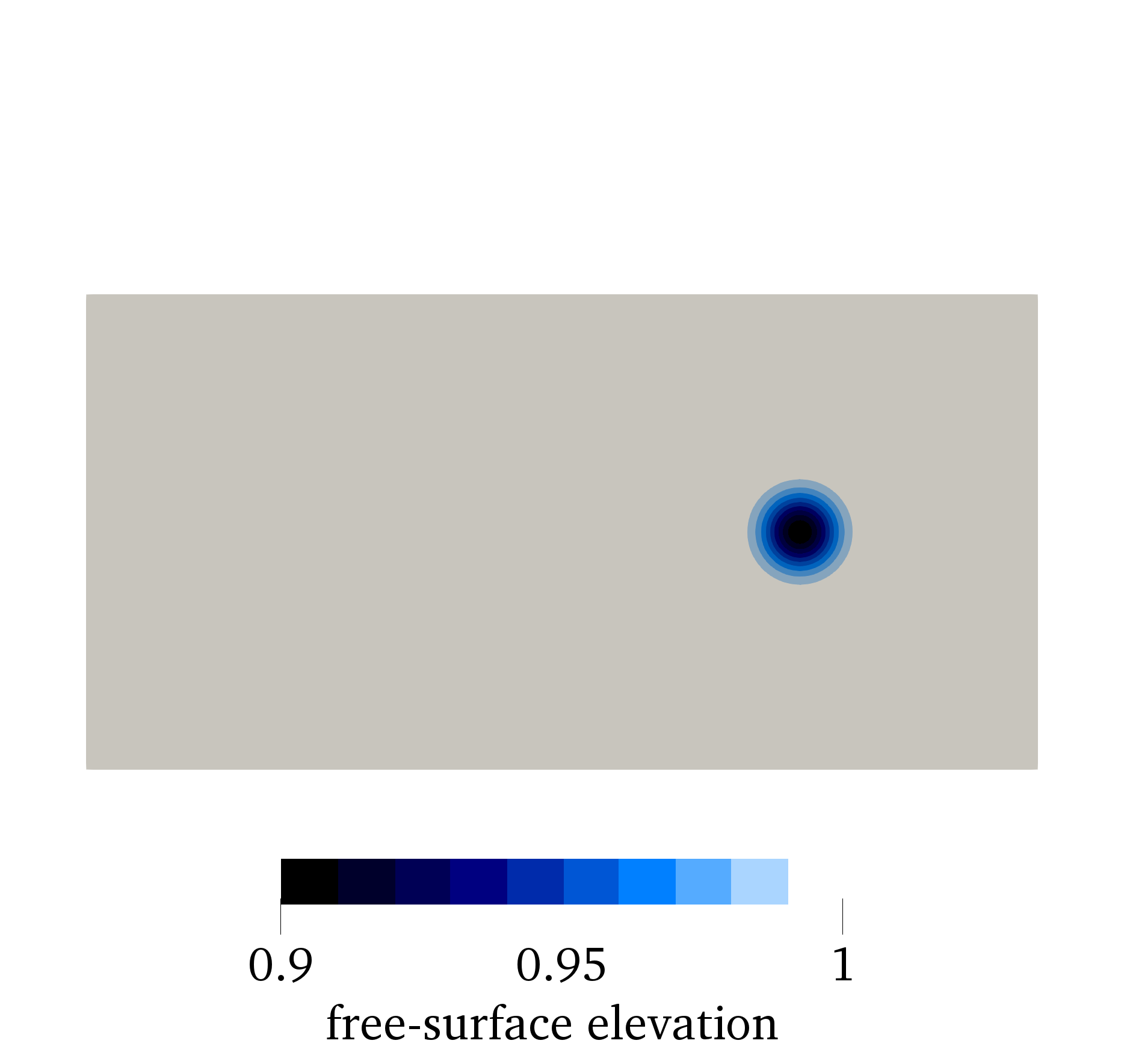}
    \includegraphics[scale=0.1]{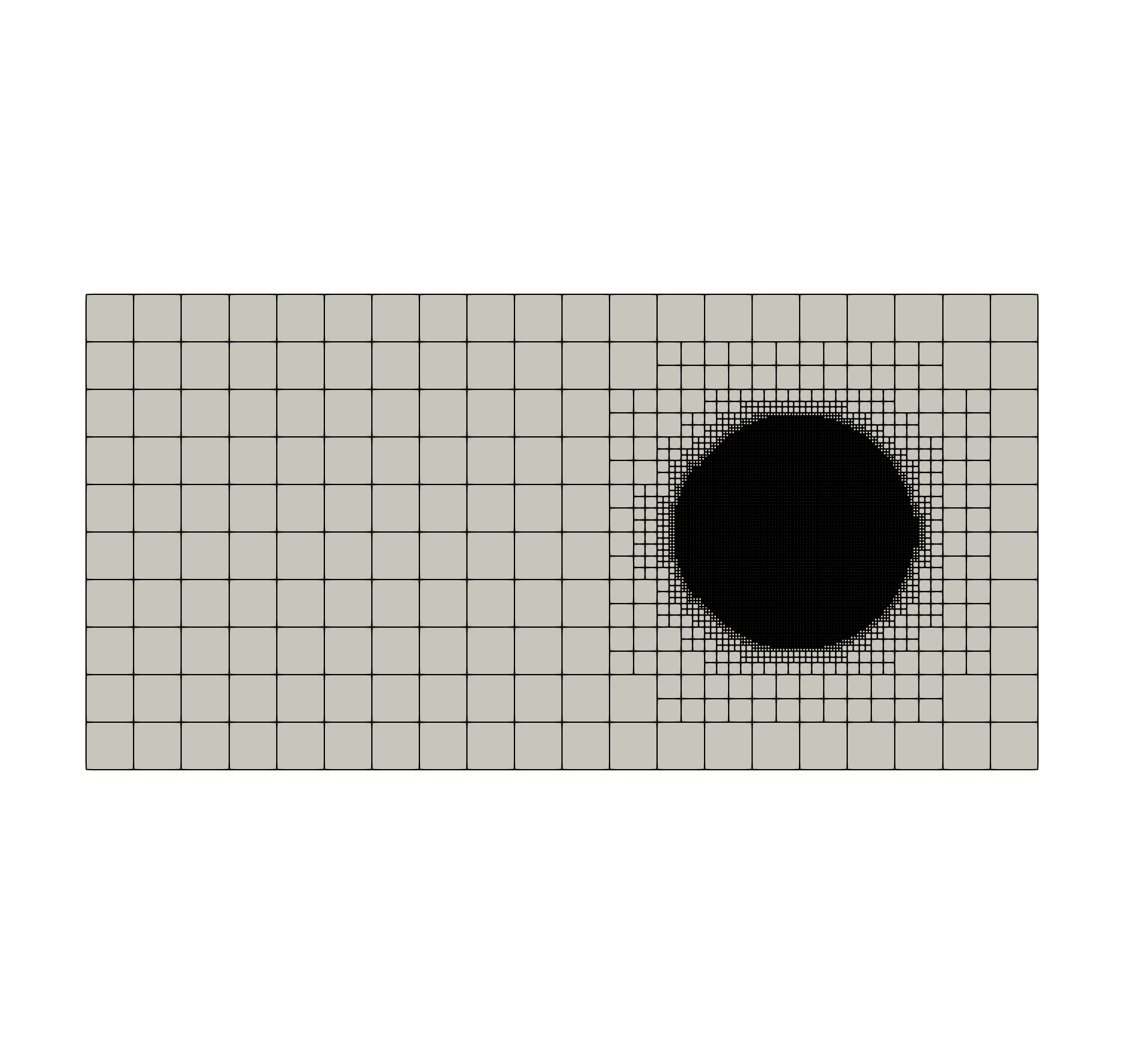}
    \caption{Travelling vortex, contour plots of the free-surface elevation (left) and adapted mesh (right) at final time $T = \SI[parse-numbers=false]{1/6}{\second}$ with $n_{\text{max}} = 5$.}
    \label{fig:vortex_mesh}
\end{figure}

\begin{figure}[h!]
    \centering
    \includegraphics[scale=0.6]{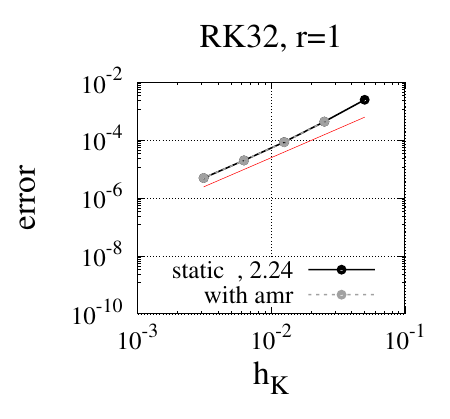}
    \includegraphics[scale=0.6]{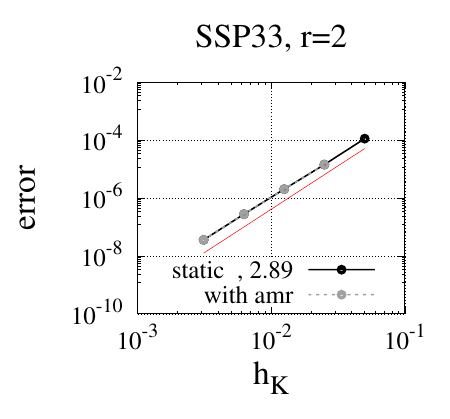}
    \includegraphics[scale=0.6]{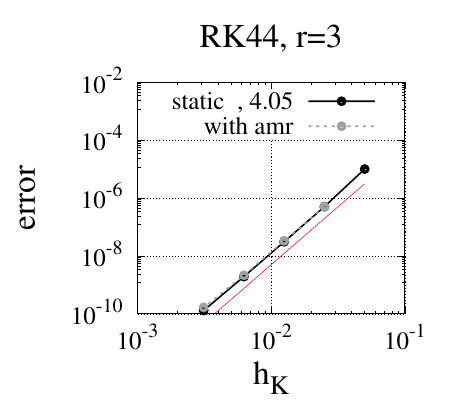}
    \includegraphics[scale=0.6]{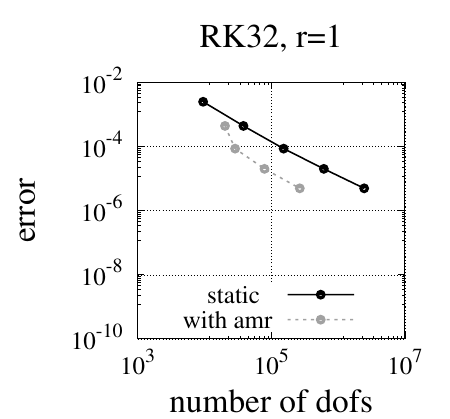}
    \includegraphics[scale=0.6]{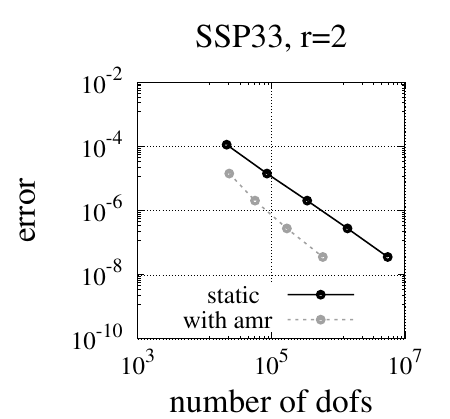}
    \includegraphics[scale=0.6]{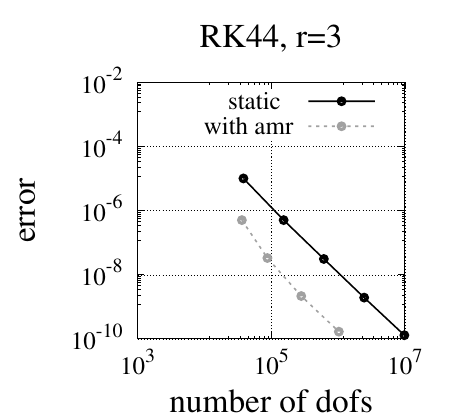}
    \includegraphics[scale=0.6]{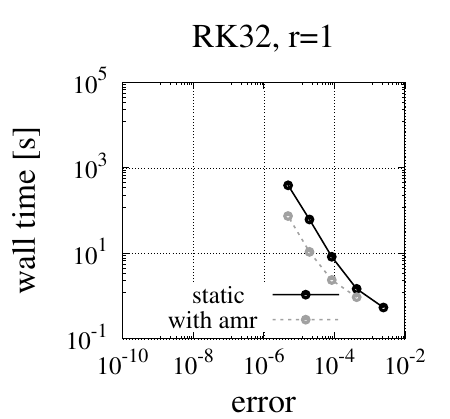}
    \includegraphics[scale=0.6]{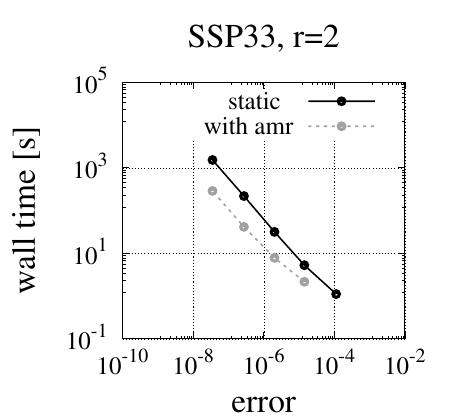}
    \includegraphics[scale=0.6]{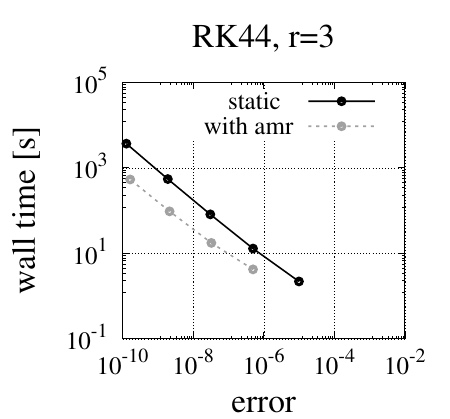}
    \caption{Travelling vortex, AMR performance metrics for different $r-$order accurate RK-DG schemes and time discretization methods. Top: $L^{2}$ error on the free-surface as a function of the mesh size (red lines trace the theoretical convergence rate). Middle: $L^{2}$ error as a function of the number of degrees of freedom. Bottom: wall time as a function of the $L^{2}$ error. The reported wall time corresponds to a simulation executed on eight processors.}
    \label{fig:vortex_metric}
\end{figure}

For the static simulations, the spatial domain is discretized with a Cartesian mesh with $2N \times N$ elements, where $N$ ranges from $N = 20$ at the coarsest mesh level to $N = 360$ for the finest one. For the adaptive simulations, the initial mesh corresponds to the coarsest mesh level with $N = 20$ and adaptation is switched on with the following parameters. The maximum refinement level ranges from $n_{\text{max}} = 2$, one level beyond the coarsest static mesh, to $n_{\text{max}} = 5$, that reaches the resolution of the finest static mesh. The refinement indicator is taken as the absolute value of the vorticity, which allows to detect vortex-like structures
\begin{equation}
    \eta_{K} = \max_{q \in K}\left|\nabla \times \vel(\bm{x}_{q})\right|,
\end{equation}
where we recall that $\bm{x}_{q}$ are the quadrature nodes of the Gaussian formula on the element. The thresholds for coarsening and refining are $\theta_{c} = 5 \times 10^{-5}$ and $\theta_{r} = 10^{-4}$, respectively, and the remeshing procedure is performed every 5 time steps. These choices allow to properly refine the vortex core within a radius $r < r_{0}$ (Figure \ref{fig:vortex_mesh}). In Figure \ref{fig:vortex_metric} we compare the adaptive and static results with three different metrics presented in \cite{beisiegel:2021}: the ratio error over mesh size, the ratio error over number of degrees of freedom, and the ratio of time over error. At all polynomial orders, the adaptive results show the same accuracy level as the static ones, while using a much smaller number of degrees of freedom, which results in a significant reduction of the computational time. In particular, the adaptive simulations reaching the finest refinement level use nine times less degrees of freedom and run from five to seven times faster (Table \ref{tab:vortex_metric}). The total overhead of the AMR operations, including the computation of the refinement indicator, the refinement and coarsening operations, and the transfer of the solutions onto the new mesh, is between 21\% of the total wall time for the case $r=1$ and 8\% for the case $r=3$. 

\begin{table}[h!]
    \centering
    \begin{tabular}{|c|c|c|c|c|c|c|c|c|}
        \hline
        \multicolumn{9}{|c|}{RK32, $r=1$} \\
        \hline
        & \multicolumn{4}{|c|}{static} & \multicolumn{4}{|c|}{with AMR} \\
        \hline
        $n_{\text{max}}$ & n. dofs & $\zeta$ error & $\disc$ error & time & n. dofs & $\zeta$ error  & $\disc$ error & time \\
        \hline
        $1$ & $9,600$ & $\num{2.459e-03}$ & $\num{3.655e-02}$ & $0.54$ & - & - & - & - \\
        \hline
        $2$ & $38,400$ & $\num{4.310e-04}$ & $\num{7.000e-03}$ & $1.47$ & $20,364$ & $\num{4.310e-04}$ & $\num{6.971e-03}$ & $0.94$ \\
        \hline
        $3$ & $153,600$ & $\num{8.494e-05}$ & $\num{1.425e-03}$ & $8.26$ & $28,896$ & $\num{8.489e-05}$ & $\num{1.424e-03}$ & $2.37$ \\
        \hline
        $4$ & $614,400$ & $\num{1.984e-05}$ & $\num{3.339e-04}$ & $61.4$ & $79,296$ & $\num{1.983e-05}$ & $\num{3.338e-04}$ & $10.8$ \\
        \hline
        $5$ & $2,457,600$ & $\num{4.904e-06}$ & $\num{8.239e-05}$ & $386.9$ & $267,828$ & $\num{4.902e-06}$ & $\num{8.238e-05}$ & $74.1$ \\
        \hline
        \multicolumn{9}{|c|}{SSP33, $r = 2$} \\
        \hline
        & \multicolumn{4}{|c|}{static} & \multicolumn{4}{|c|}{with AMR} \\
        \hline
        $n_{\text{max}}$ & n. dofs & $\zeta$ error & $\disc$ error & time & n. dofs & $\zeta$ error & $\disc$ error & time \\
        \hline
        $1$ & $21,600$ & $\num{1.125e-04}$ & $\num{2.207e-03}$ & $1.1$ & - & - & - & - \\
        \hline
        $2$ & $86,400$ & $\num{1.419e-05}$ & $\num{2.977e-04}$ & $5.3$ & $23,625$ & $\num{1.419e-05}$ & $\num{2.976e-04}$ & $2.14$ \\
        \hline
        $3$ & $345,600$ & $\num{2.021e-06}$ & $\num{5.537e-05}$ & $31.6$ & $57,240$ & $\num{2.021e-06}$ & $\num{5.535e-05}$ & $7.79$ \\
        \hline
        $4$ & $1,382,400$ & $\num{2.758e-07}$ & $\num{1.138e-05}$ & $217.0$ & $171,531$ & $\num{2.759e-07}$ & $\num{1.137e-05}$ & $41.3$ \\
        \hline
        $5$ & $5,529,600$ & $\num{3.551e-08}$ & $\num{1.964e-06}$ & $1,516$ & $590,949$ & $\num{3.553e-08}$ & $\num{1.962e-06}$ & $286.7$ \\
        \hline
        \multicolumn{9}{|c|}{RK44, $r = 3$} \\
        \hline
        & \multicolumn{4}{|c|}{static} & \multicolumn{4}{|c|}{with AMR} \\
        \hline
        $n_{\text{max}}$ & n. dofs & $\zeta$ error & $\disc$ error & time & n. dofs & $\zeta$ error & $\disc$ error & time \\
        \hline
        $1$ & $38,400$ & $\num{9.970e-06}$ & $\num{1.677e-04}$ & $2.18$ & - & - & - & - \\
        \hline
        $2$ & $153,600$ & $\num{4.995e-07}$ & $\num{8.514e-06}$ & $12.8$ & $36,384$ & $\num{4.996e-07}$ & $\num{8.457e-06}$ & $4.2$ \\
        \hline
        $3$ & $614,400$ & $\num{3.077e-08}$ & $\num{5.334e-07}$ & $81.0$ & $87,648$ & $\num{3.294e-08}$ & $\num{4.915e-07}$ & $17.4$ \\
        \hline
        $4$ & $2,457,600$ & $\num{1.923e-09}$ & $\num{3.541e-08}$ & $543.9$ & $281,184$ & $\num{2.150e-09}$ & $\num{3.595e-08}$ & $95.2$ \\
        \hline
        $5$ & $9,830,400$ & $\num{1.289e-10}$ & $\num{7.703e-09}$ & $3,658$ & $1,028,256$ & $\num{1.658e-10}$ & $\num{9.212e-09}$ & $531.1$ \\
        \hline
    \end{tabular}
    \caption{Travelling vortex, numerical convergence rates for different $r-$order accurate RK-DG schemes and time discretization methods. The error is measured as the $L^{2}$ error. The computational wall time is expressed in $\SI{}{seconds}$. $n_{\text{max}}$ represents the maximum mesh level (see Section \ref{sec:numerical}), while n. dofs represents the total number of degrees of freedom (free-surface elevation + discharge). The reported wall time corresponds to a simulation executed on eight processors.}
    \label{tab:vortex_metric}
\end{table}

\section{A benchmark with realistic bathymetry}
\label{sec:realistic}

\begin{figure}[h!]
    \centering
    \includegraphics[scale=0.23]{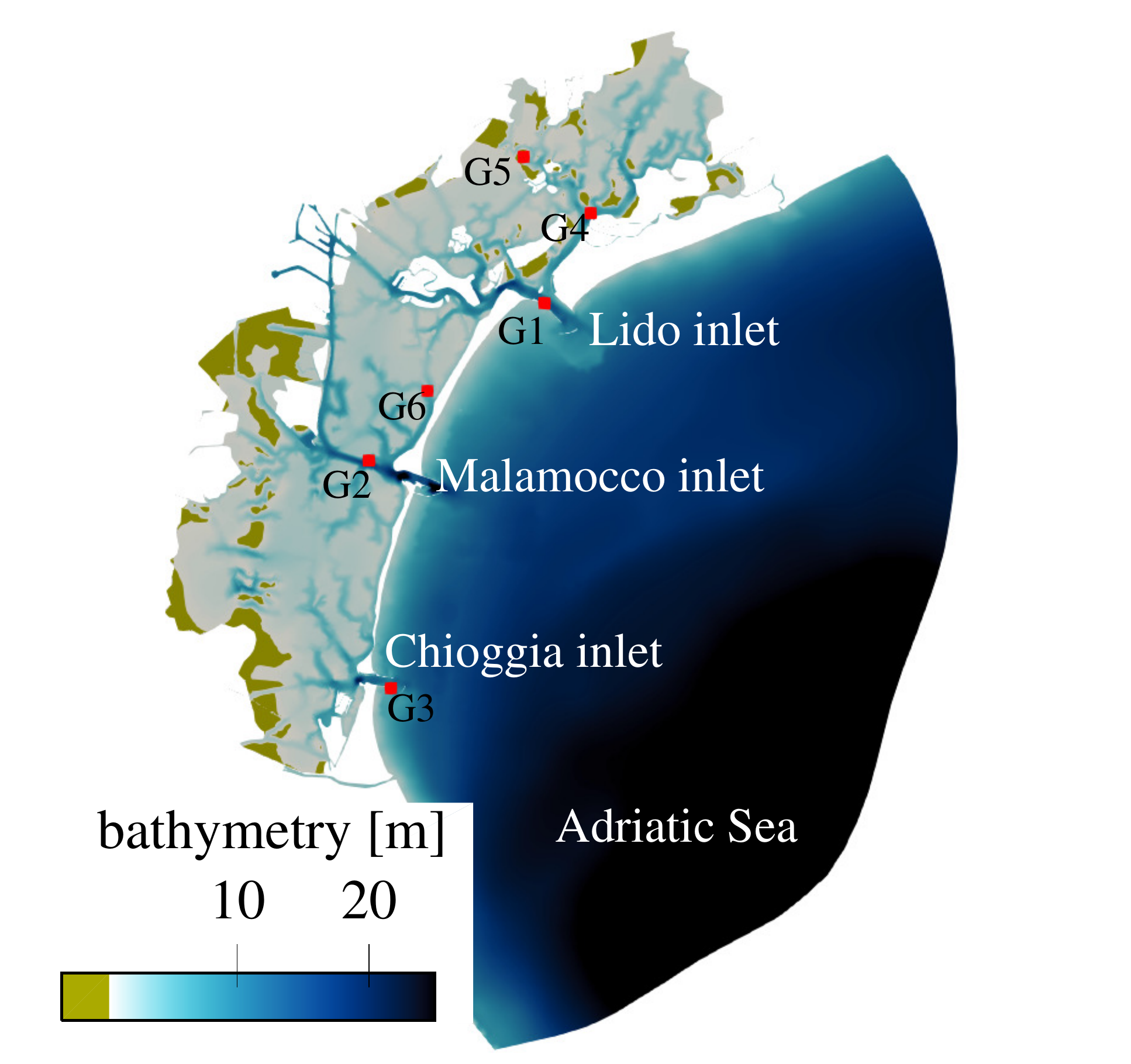}\includegraphics[scale=0.23]{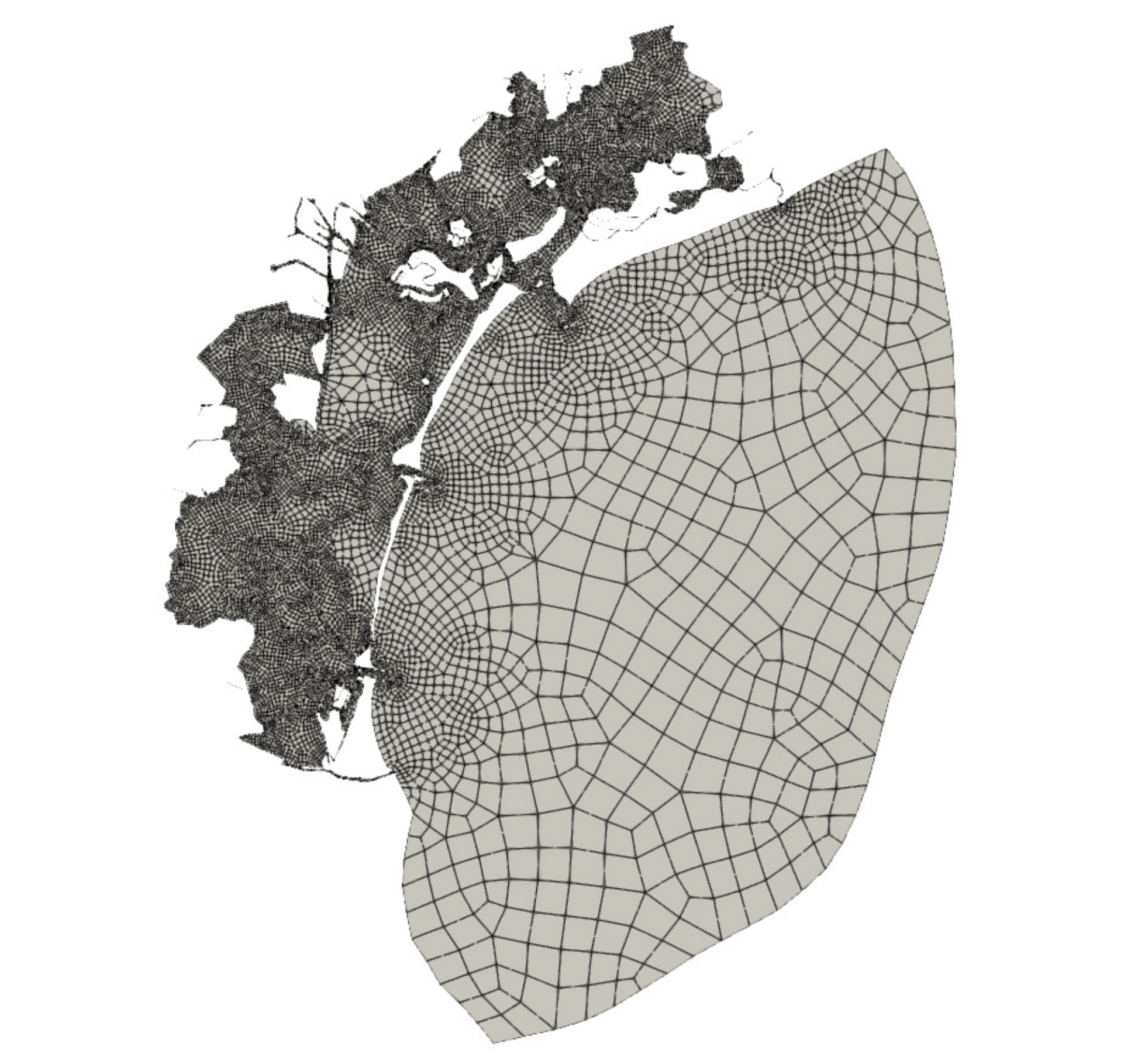}
    \caption{Benchmark with realistic bathymetry. Left: computational domain with the bathymetry and salt-marshes (green areas) for the Venice Lagoon. Right: initial coarse mesh of quadrilaterals.}
    \label{fig:veniceLagoon_mesh}
\end{figure}

We now consider a benchmark with a more realistic bathymetry and complex geometry. More specifically, we consider the tidal circulation in the Venice Lagoon and the induced tracer transport. The Venice Lagoon is a very complex sea water basin, whose area is about $\SI{50}{\kilo\meter\squared}$ and which includes several interconnected narrow channels with a maximum width of $\SI{1}{\kilo\meter}$ and depth of $\SI{50}{\meter}$ surrounding large and flat shallow areas. The Lagoon is connected to the Adriatic Sea through three narrow inlets, namely Lido, Malamocco, and Chioggia (see Figure \ref{fig:veniceLagoon_mesh}). The city of Venice is situated upon the largest island, near Lido inlet. Tides propagate from the Adriatic Sea into the lagoon through the three inlets, and then from the inlets into the lagoon’s interior through major channels such as the Treporti and Petroli channels. We consider a coarse mesh with $17,084$ elements with the edges aligned to the main channels. The resolution varies from $\SI{2}{\km}$ at open sea to $\SI{200}{\meter}$ at the inlets and in the main channels. Close to salt marshes we use a resolution of $\SI{400}{\meter}$. Smaller salt-marshes and channels are therefore unresolved at such a grid scale. The coarse mesh of quadrilaterals was generated using the software \texttt{gmsh} \cite{geuzaine:2009}. It is worth mentioning that, at present day, generating suitable coarse meshes for complex boundary geometries composed exclusively of quadrilaterals is not fully automated and can be time-consuming.

The only dissipation mechanism is the bottom friction, for which we set a uniform Manning coefficient $n = \SI{0.055}{\meter^{-1/3}\second}$. Given this setting, we consider a semi-realistic test in which an M2 tide with amplitude of $\SI{0.3}{\meter}$ and period of $\SI{12.41}{\hour}$ has been specified at the Adriatic Sea boundary. Zero wind stress and zero mass flux at the internal boundaries of the Lagoon are assumed. The two threshold coefficients necessary for wetting and drying, namely the coefficient to avoid division by zero $\epsilon$ and the threshold value for degree coarsening $h_{\mathrm{lim}}$ (see formula \eqref{eq:velocity} and formula \eqref{eq:p-criteria}) are set to $\epsilon = h_{\mathrm{lim}} = \SI{0.1}{\meter}$.

\begin{figure}[h!]
    \centering
    \includegraphics[trim={0.2cm 0 0.55cm 0},clip,scale=1.2]{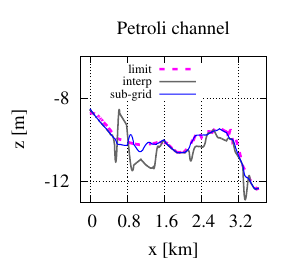}\includegraphics[ trim={0.55cm 0 0.55cm 0},clip,scale=1.2]{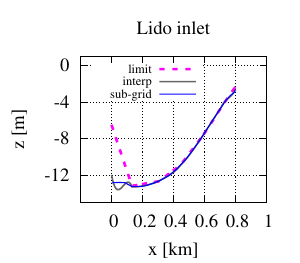}\includegraphics[trim={0.55cm 0 0.55cm 0}, clip, scale=1.2]{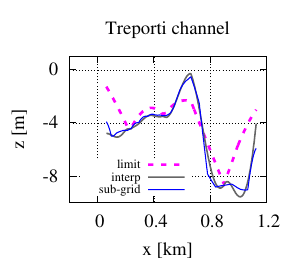}
    \caption{Benchmark with realistic bathymetry. Comparison of the bathymetric profiles obtained with different approaches: sub-grid bathymetry with $r=3$ (blue line), interpolated bathymetry with $r=3$ (grey line), limited interpolated bathymetry (dashed magenta line). Left: Petroli along-channel profile. Middle: Lido cross-channel profile. Right: Treporti cross-channel profile.}
    \label{fig:venice_lagoon_bathy}
\end{figure}

We use the same bathymetry dataset employed in \cite{bajo:2021,ferrarin:2010}, performing linear interpolation within each pixel of dimension $\SI{30}{\meter}\times\SI{30}{\meter}$, to recover the bathymetry data at the quadrature points, without any need for further modification of these values. For comparison, we perform a simulation using a polynomial bathymetry, in which the bathymetry nodal values are computed with an interpolation approach. None of the simulations with the high-order interpolated bathymetry and with $r>1$ end correctly, due to instabilities that develop immediately after the first time steps of the simulation in the Petroli channel. The left picture in Figure \ref{fig:venice_lagoon_bathy} compares the along-channel bathymetric profile in the Petroli channel obtained with our approach (labelled sub-grid) with that obtained using the interpolation-based approach (labelled interpolation). The figure shows that the interpolated bathymetry exhibits localized spurious oscillations that make the along-channel profile very noisy, suggesting that smoothing of the bathymetry would be necessary to improve the method's robustness in this case. In order to remove the oscillations, we tested a limited bathymetry where the nodal values are computed to match a piecewise linear profile over the element. In this case, the simulation ends correctly but the sub-grid variability is no longer captured and the channel cross-section is poorly represented, as show for the Lido inlet and Treporti Channel in Figure \ref{fig:venice_lagoon_bathy}.

\begin{figure}[h!]
    \centering
    \includegraphics[scale=0.2]{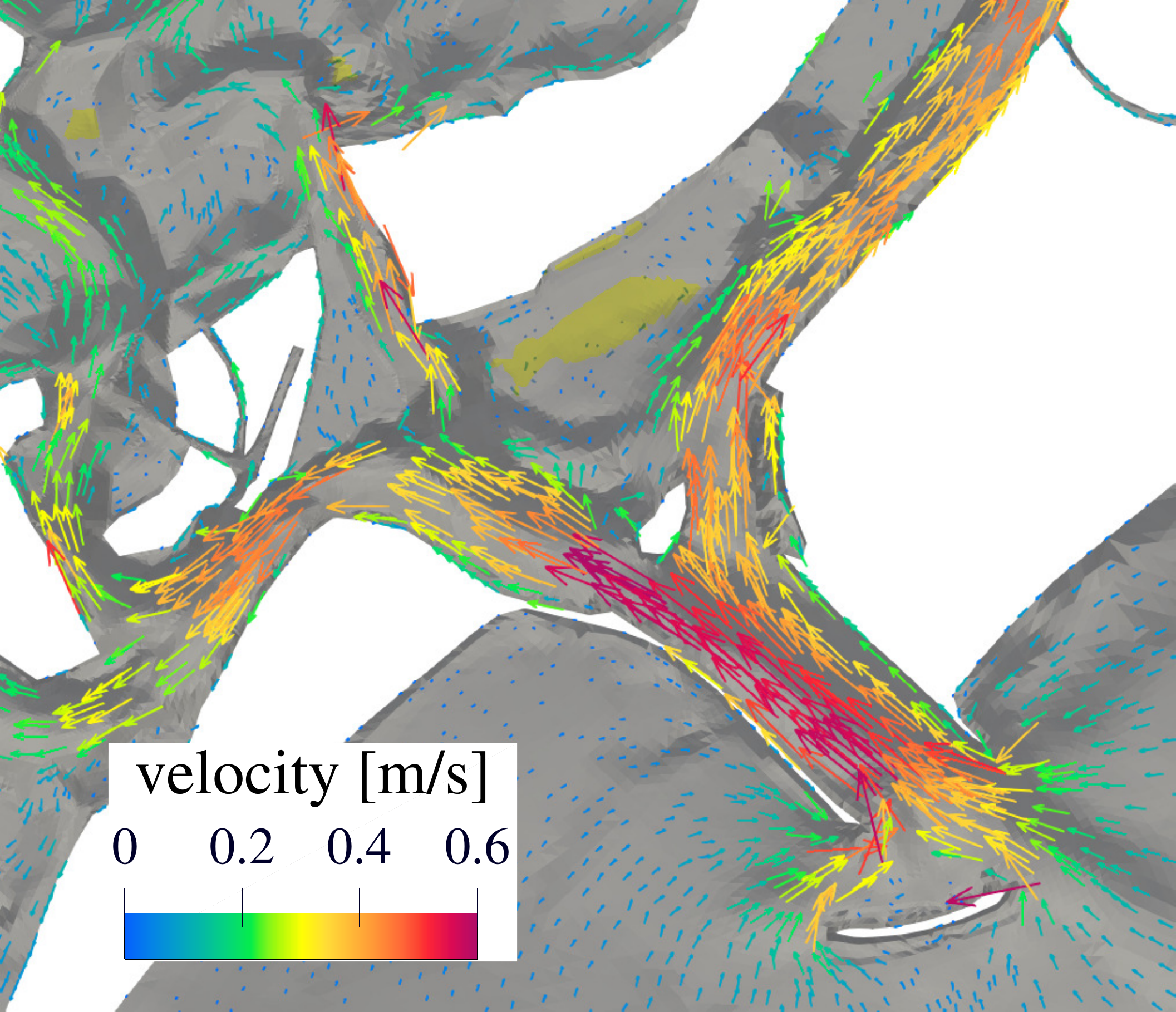}\hspace{0.5cm}\includegraphics[scale=0.2]{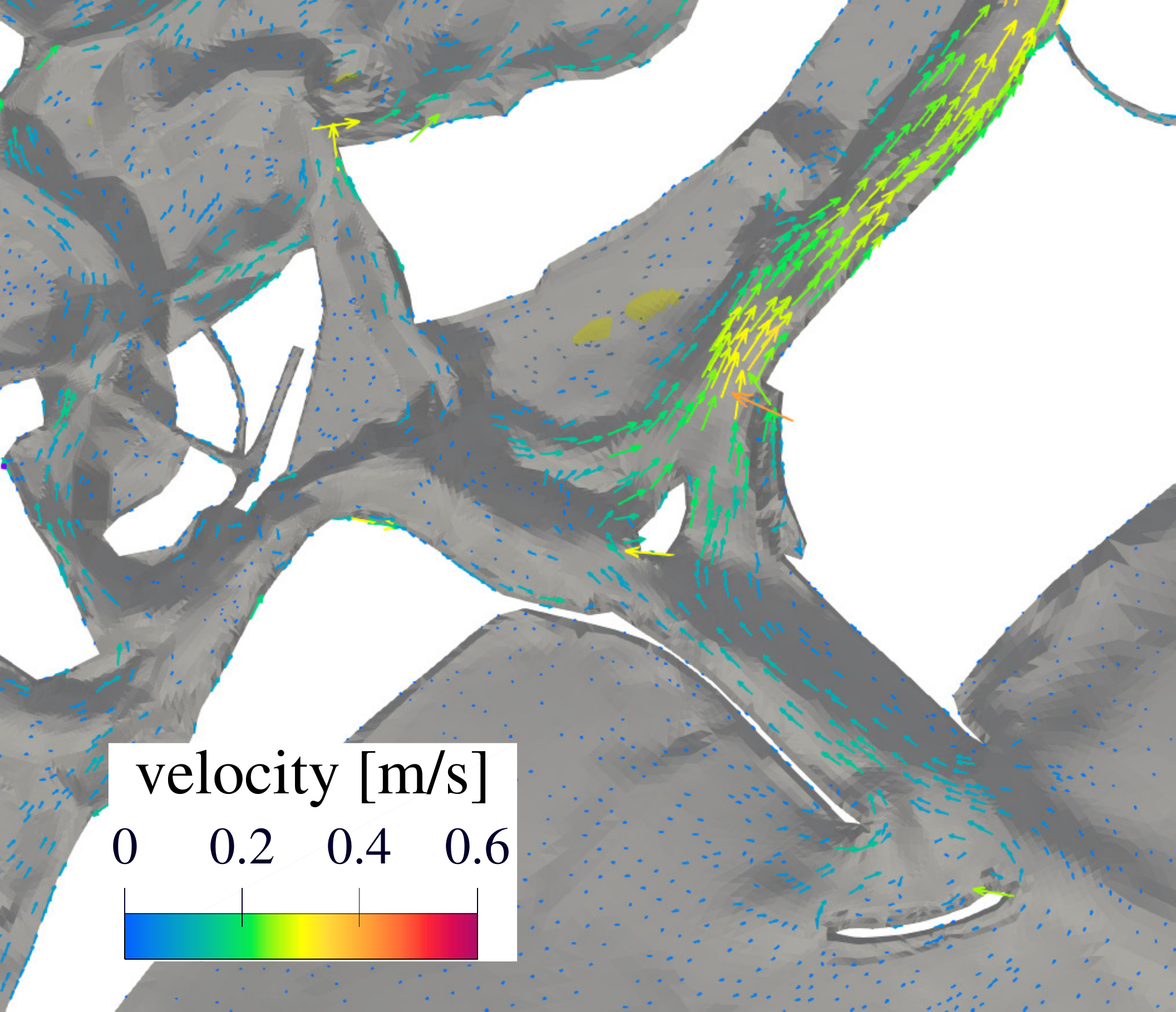}
    \vspace{0.5cm}

    \includegraphics[scale=0.2]{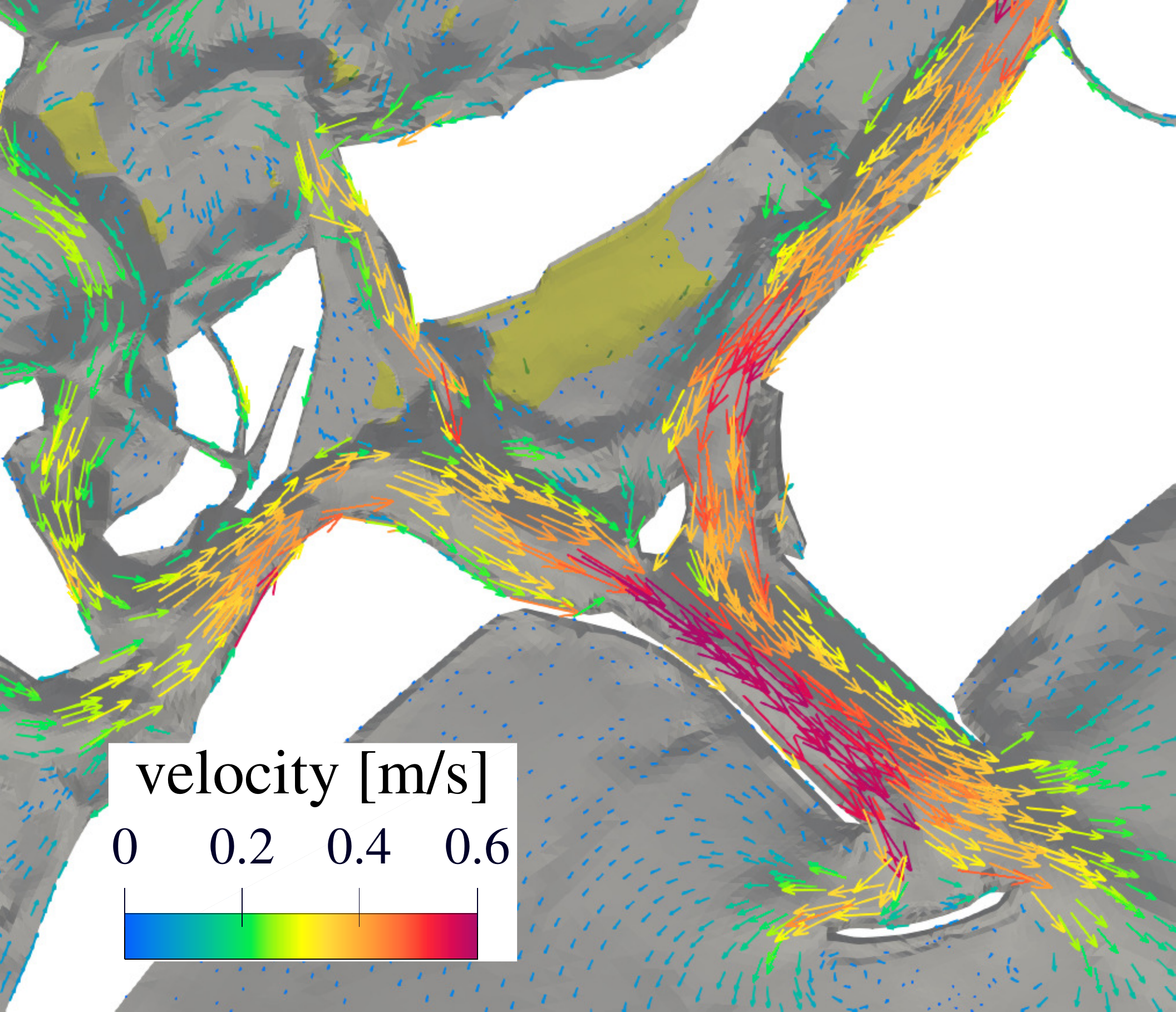}\hspace{0.5cm}\includegraphics[scale=0.2]{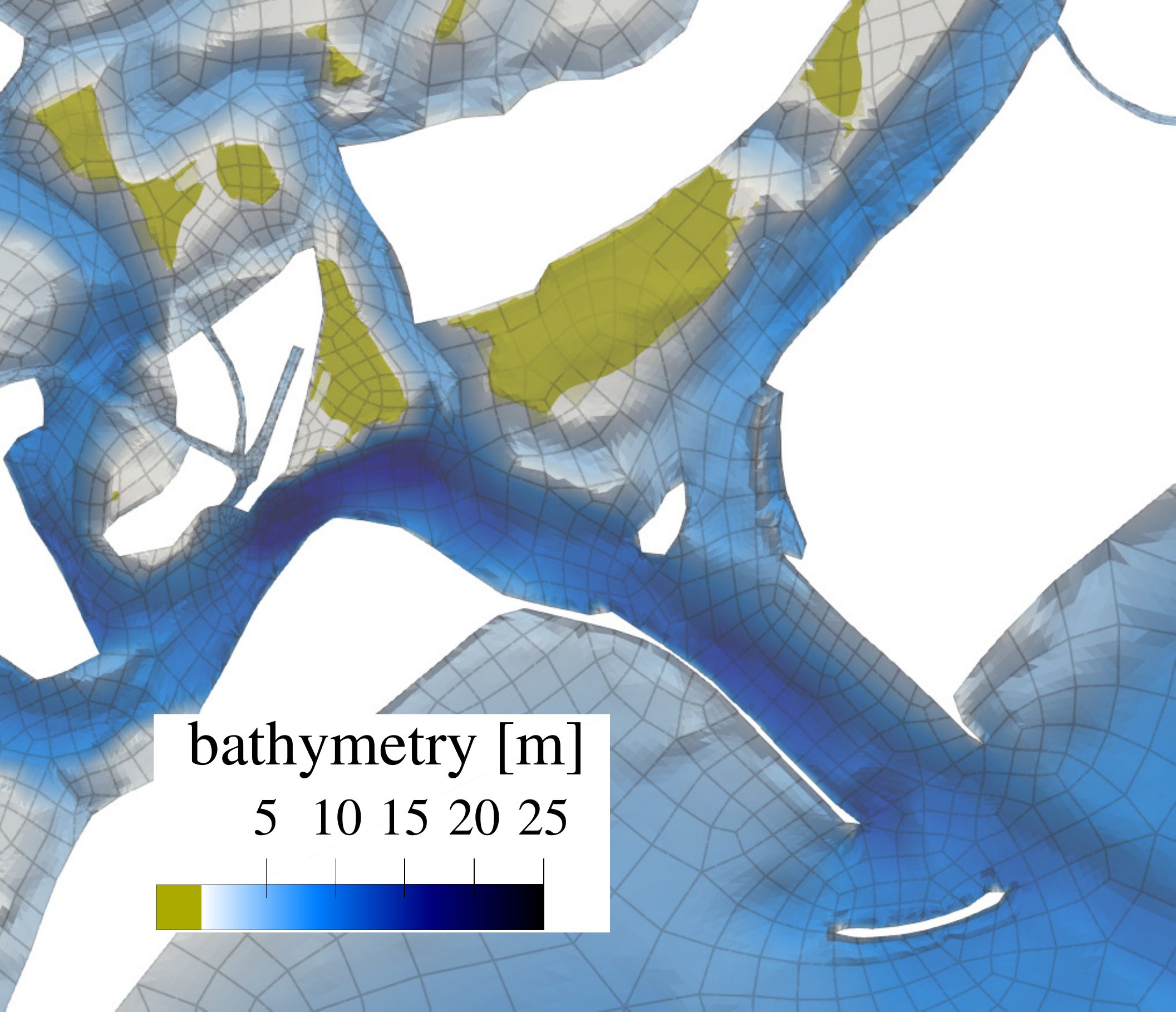}
    \caption{Benchmark with realistic bathymetry. Velocity vector map at the Lido inlet for the simulation with $r=3$. Top left: flood ($t=\SI{39}{\hour}$). Top right: high tide ($t=\SI{42}{\hour}$). Bottom left: ebb ($t=\SI{45}{\hour}$). Bottom right: mesh and associated bathymetry. Green areas represent salt-marshes.}
    \label{fig:venice_lagoon_velocity}
\end{figure}

\begin{figure}[h!]
    \centering
    \includegraphics[scale=0.2]{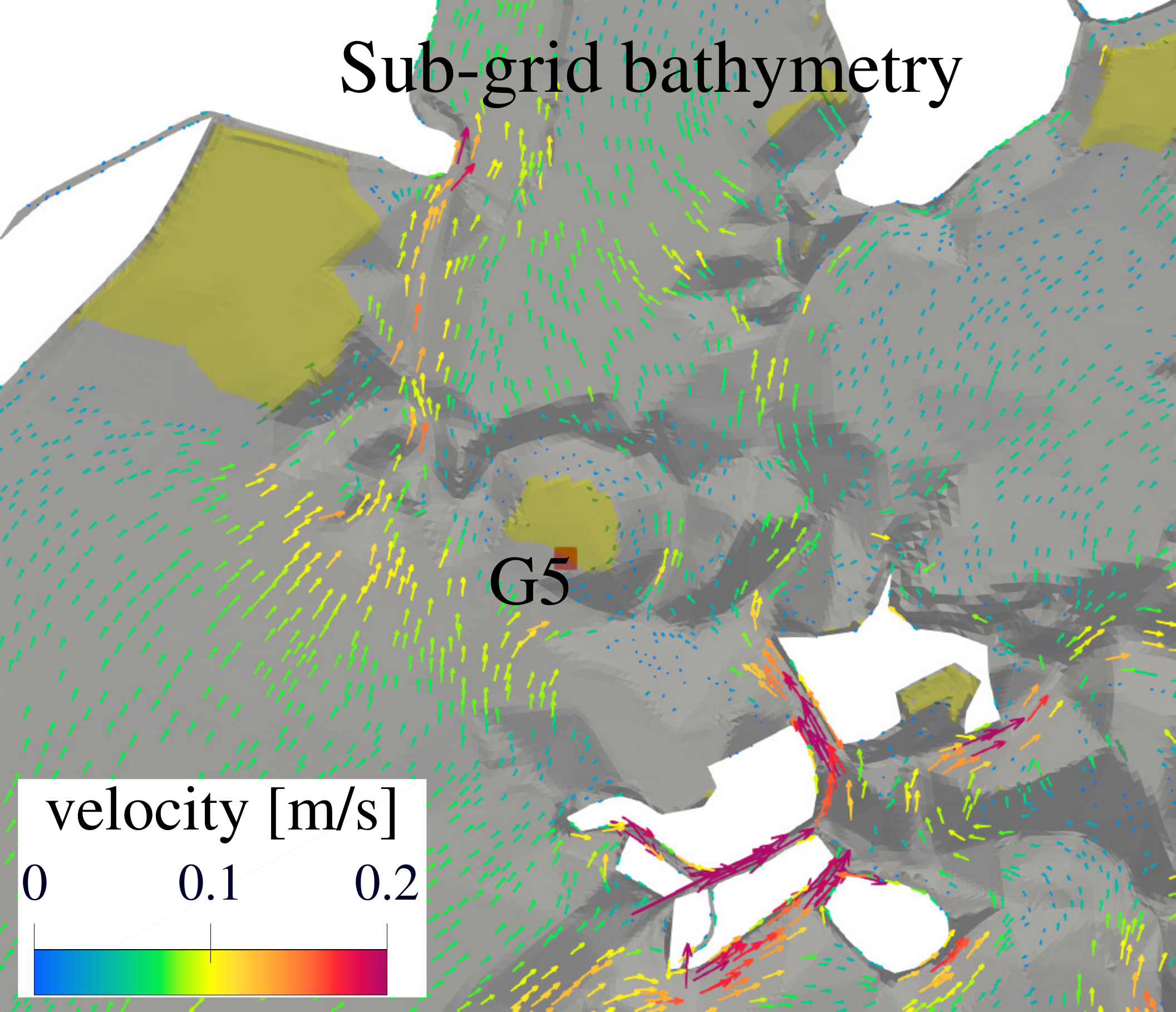}\hspace{0.5cm}\includegraphics[scale=0.2]{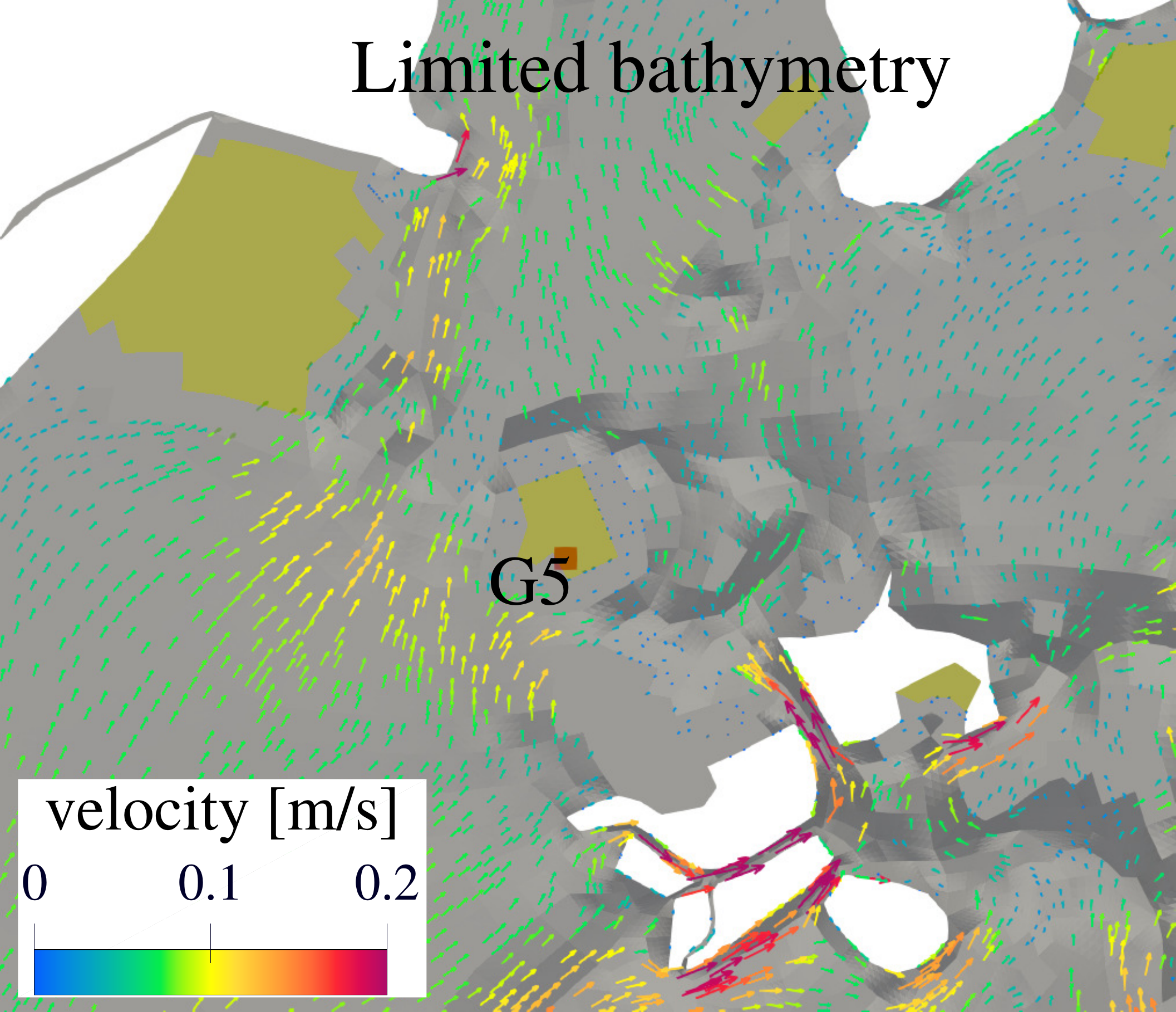}
    \caption{Benchmark with realistic bathymetry. Velocity vector map in the north of the Lagoon for the simulation with $r=3$ at high tide ($t=\SI{42}{\hour}$). Top left: sub-grid bathymetry. Top right: limited bathymetry. Green areas represent salt-marshes.}
    \label{fig:venice_lagoon_comparison}
\end{figure}

\begin{figure}[h!]
    \centering
    \includegraphics[scale=0.6]{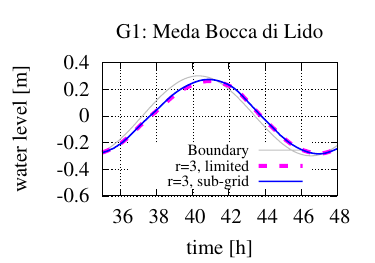}\includegraphics[scale=0.6]{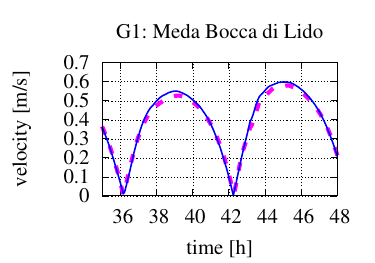}
    \includegraphics[scale=0.6]{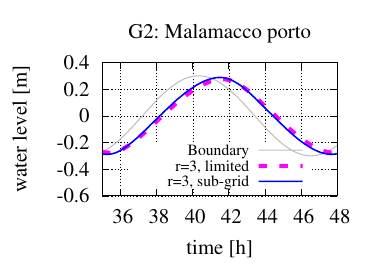}\includegraphics[scale=0.6]{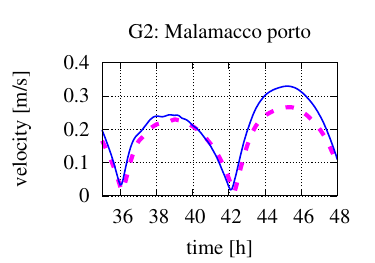}
    \includegraphics[scale=0.6]{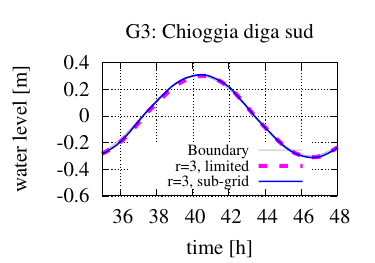}\includegraphics[scale=0.6]{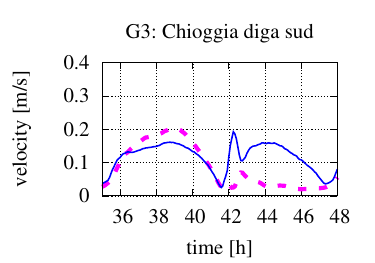}
    \includegraphics[scale=0.6]{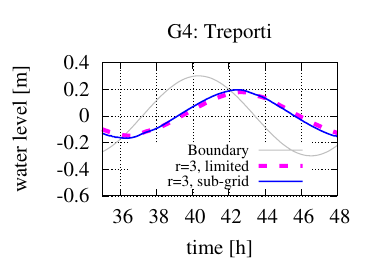}\includegraphics[scale=0.6]{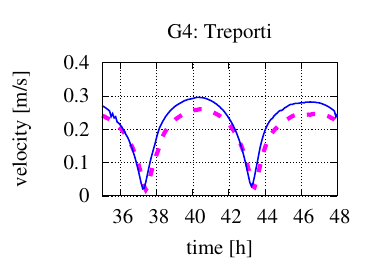}
    \includegraphics[scale=0.6]{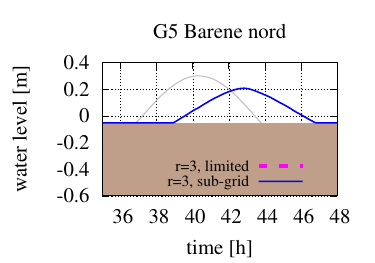}\includegraphics[scale=0.6]{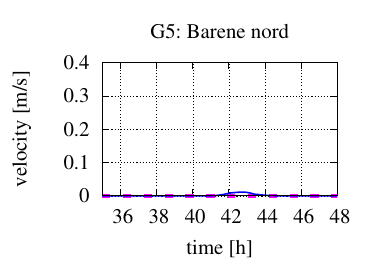}
    \includegraphics[scale=0.6]{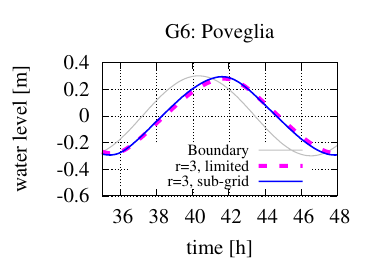}\includegraphics[scale=0.6]{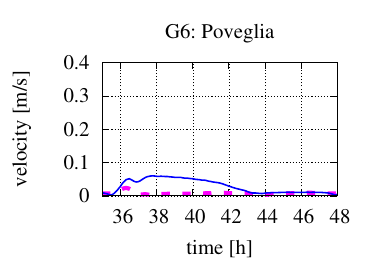}
    \caption{Benchmark with realistic bathymetry. Time series of the free-surface (left) and velocity (right) recorded at the gauge stations for different bathymetry approximations. The gray line represents the free-surface elevation imposed at the boundary.}
    \label{fig:venice_lagoon_gauges}
\end{figure}

We perform three simulations, with $r=1 $, $r=2 $ and $r=3, $ respectively. Table \ref{tab:veniceLagoon} summarizes all the parameter values that define the simulations. In Figure \ref{fig:venice_lagoon_velocity} we show, for $r=3$, the circulation at the Lido inlet, the corresponding sub-grid bathymetry and the mesh. These figures highlights the potential of the sub-grid approach in reproducing the channel geometry as well as the geometry of the salt-marshes, approximated by coarse elements  but with a variable underlying bathymetry. In Figure \ref{fig:venice_lagoon_comparison}, we compare the circulations in the northern lagoon, with $r=3$, obtained using our implementation of the sub-grid method with those obtained with the limited piecewise linear bathymetry. Although the velocities are similar, along the channels the sub-grid approach produces higher velocities due to its improved representation of the bottom geometry. In Figure \ref{fig:venice_lagoon_gauges} we show the time series of the solution, recorded at specific points in correspondence of the three inlets and inside the lagoon. We compare the results obtained with a sub-grid representation of the bathymetry against a limited piecewise linear approximation. Although the results are similar to each other, it is confirmed that the velocities are stronger with a sub-grid representation of the bathymetry. At the points inside the lagoon (G4 and G6), the tidal wave travels slightly faster and with a higher amplitude. There is also a difference close to wet-dry fronts, such as at location G5. Indeed, the sub-grid approach shows flooding while with a the limited bathymetry the same point remains dry.

\begin{table}[h!]
    \centering
    \begin{tabular}{|c|c|c|c|c|c|}
        \hline
        Name & $n_{max}$ & N. of elements & $r$ & N. of dofs & Time (s)\\
        \hline
        static & $1$ & $17,084$ & $1$ & $228,812$ & $6,209$ \\
        \hline
        static & $1$ & $17,084$ & $2$ & $495,728$ & $10,410$ \\
        \hline
        static & $1$ & $17,084$ & $3$ & $868,196$ & $44,160$ \\
        \hline
        with AMR & $3$ & $19,469-47,567$ & $1$ & $311,504-761,072$ & $27,200$ \\
        \hline
    \end{tabular}
    \caption{Benchmark with realistic bathymetry, data of the computational meshes and wall time obtained for a parallel run on 36 processors.}
    \label{tab:veniceLagoon}
\end{table}

\begin{figure}[h!]
    \centering
    \includegraphics[trim={0cm 0cm 0cm 0cm},clip,scale=0.10]{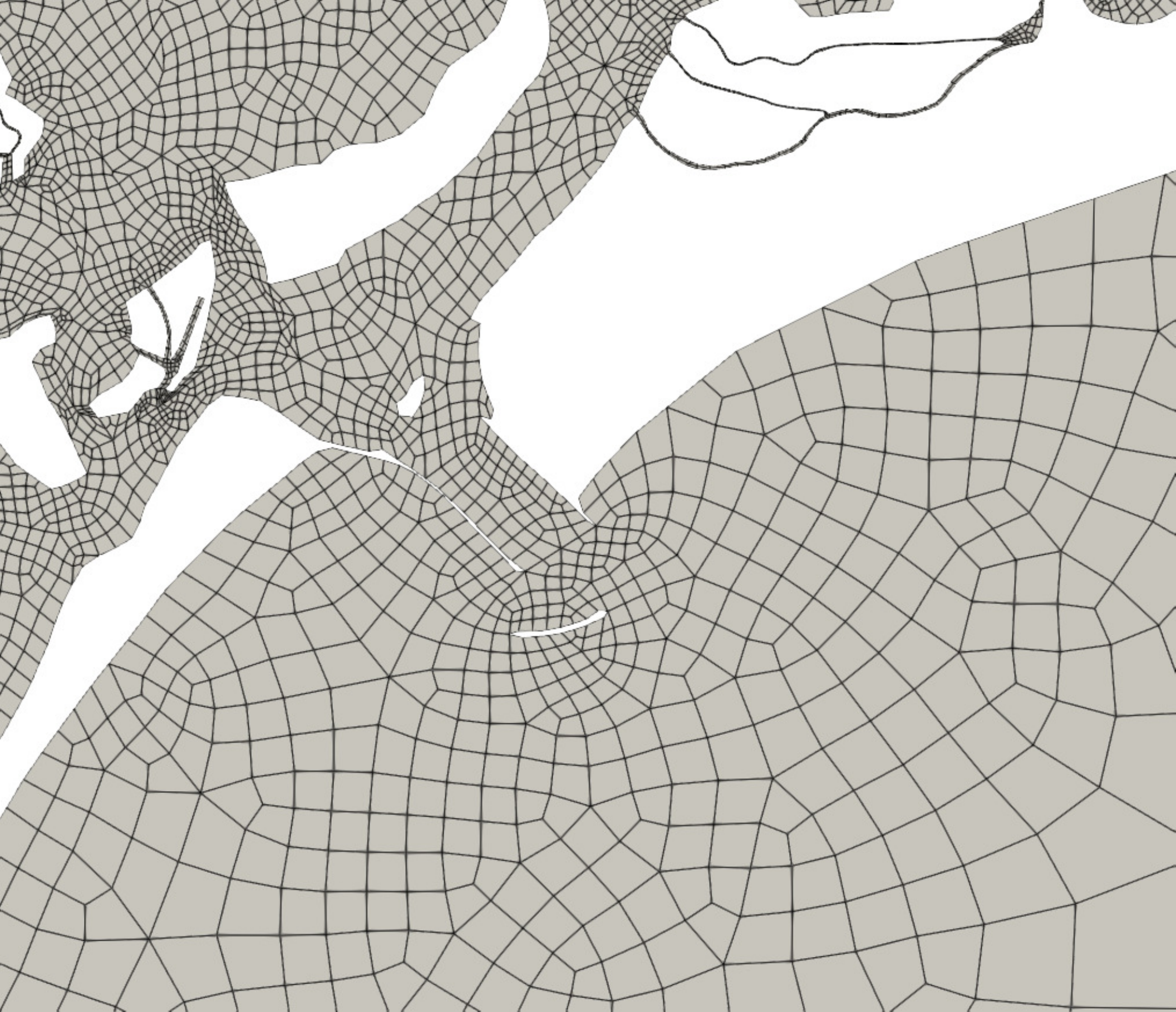}
    \hspace{0.15cm}
    \includegraphics[trim={0cm 0cm 0cm 0cm},clip,scale=0.10]{images/veniceLagoonWithTracer/67maxlevref0_fedegree1/solution_mesh_lido_042-eps-converted-to.pdf}
    \hspace{0.15cm}
    \includegraphics[trim={0cm 0cm 0cm 0cm},clip,scale=0.10]{images/veniceLagoonWithTracer/67maxlevref0_fedegree1/solution_mesh_lido_042-eps-converted-to.pdf}
    \hspace{0.15cm}
    \includegraphics[trim={0cm 0cm 0cm 0cm},clip,scale=0.10]{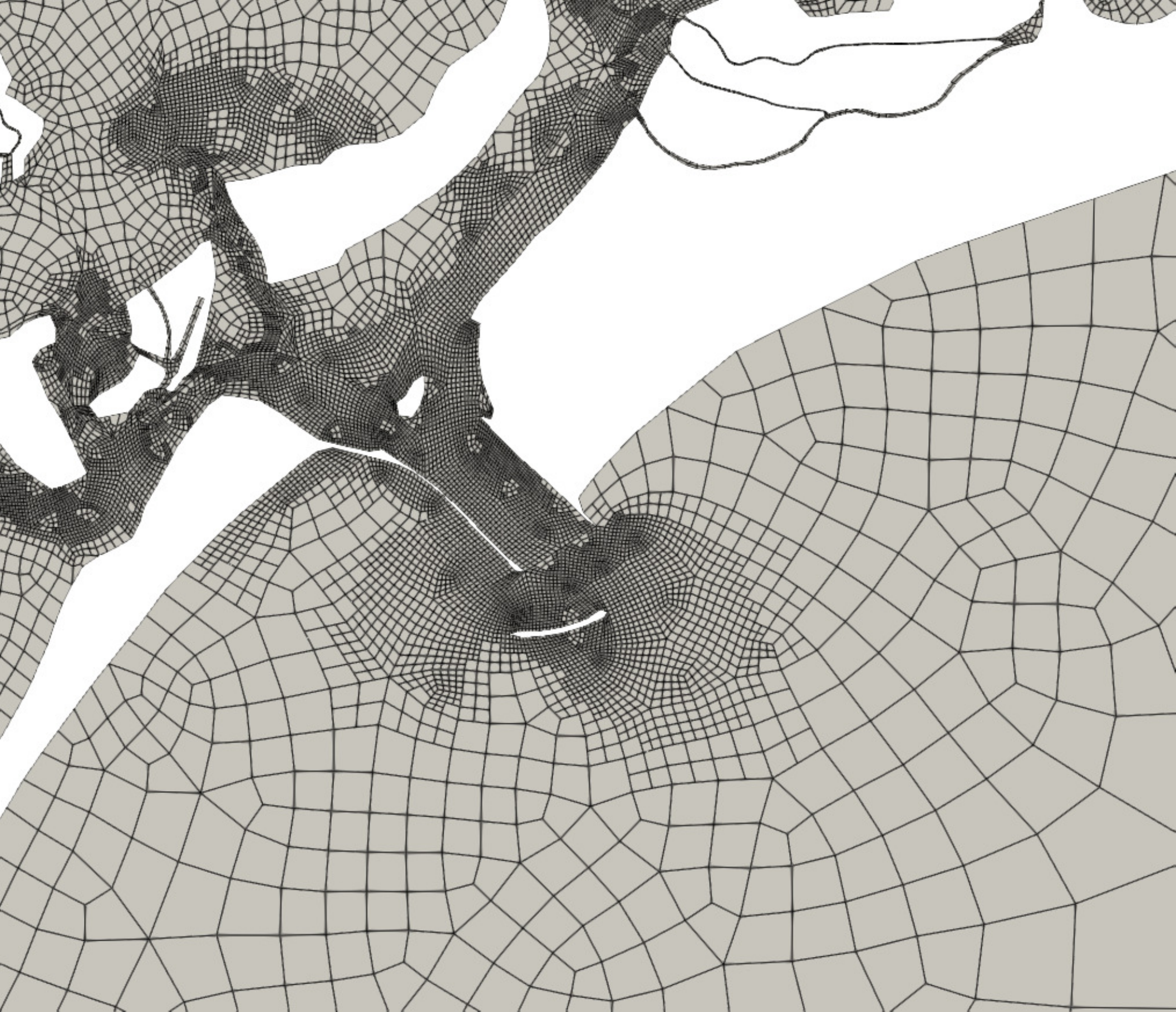}
    \vspace{0.15cm}

    \includegraphics[trim={0cm 0cm 0cm 0cm},clip,scale=0.10]{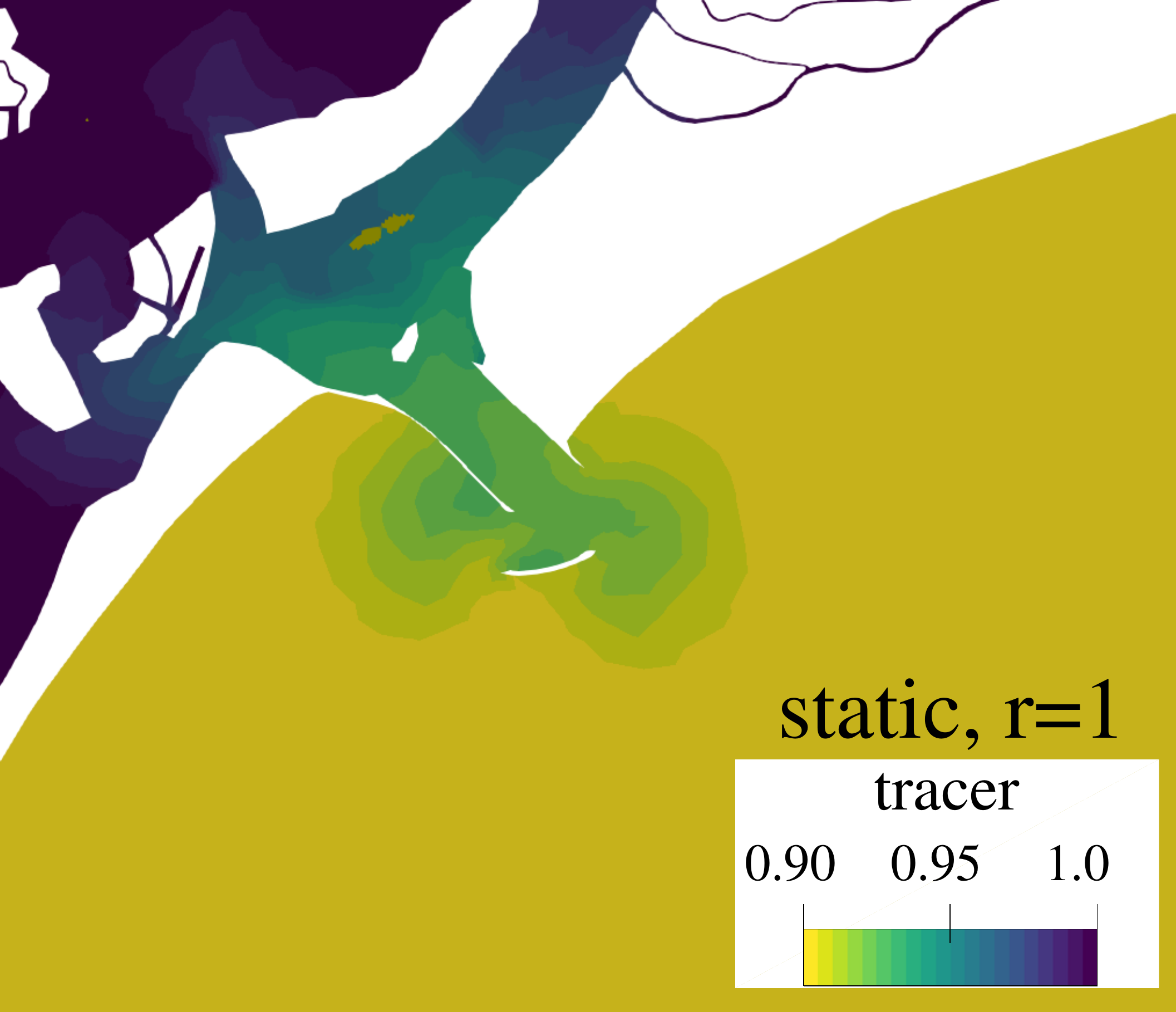}
    \hspace{0.15cm}
    \includegraphics[trim={0cm 0cm 0cm 0cm},clip,scale=0.10]{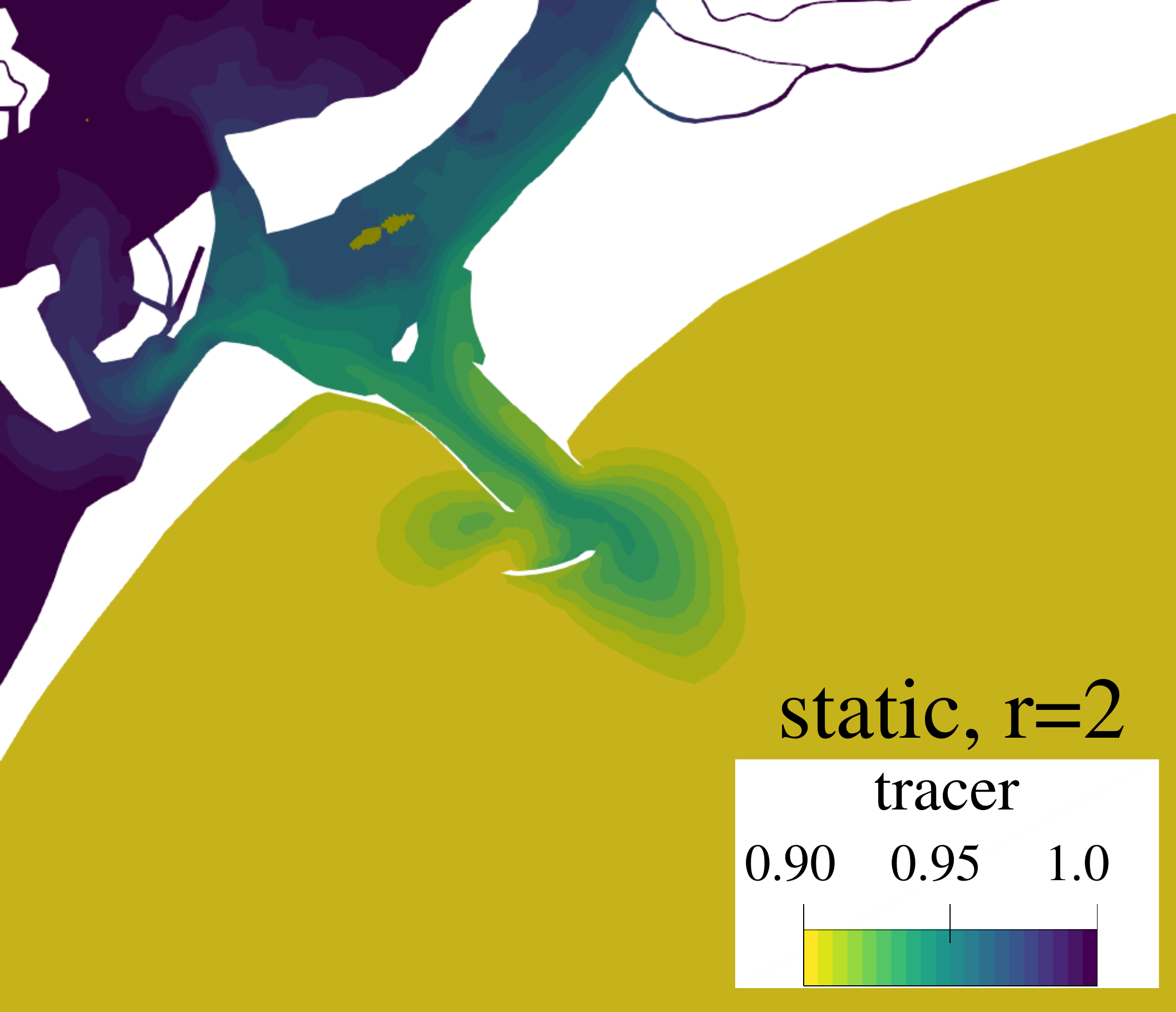}
    \hspace{0.15cm}
    \includegraphics[trim={0cm 0cm 0cm 0cm},clip,scale=0.10]{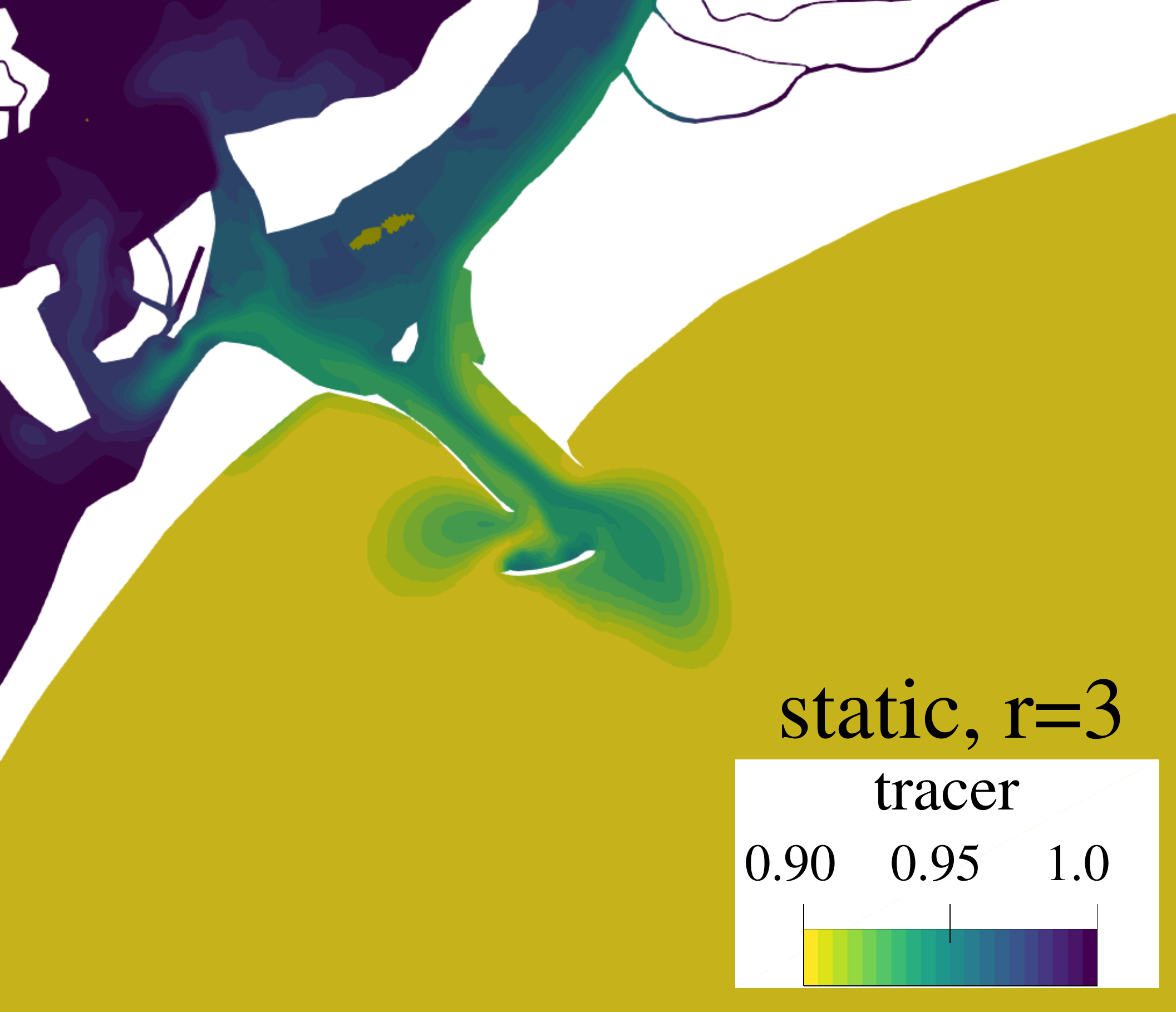}
    \hspace{0.15cm}
    \includegraphics[trim={0cm 0cm 0cm 0cm},clip,scale=0.10]{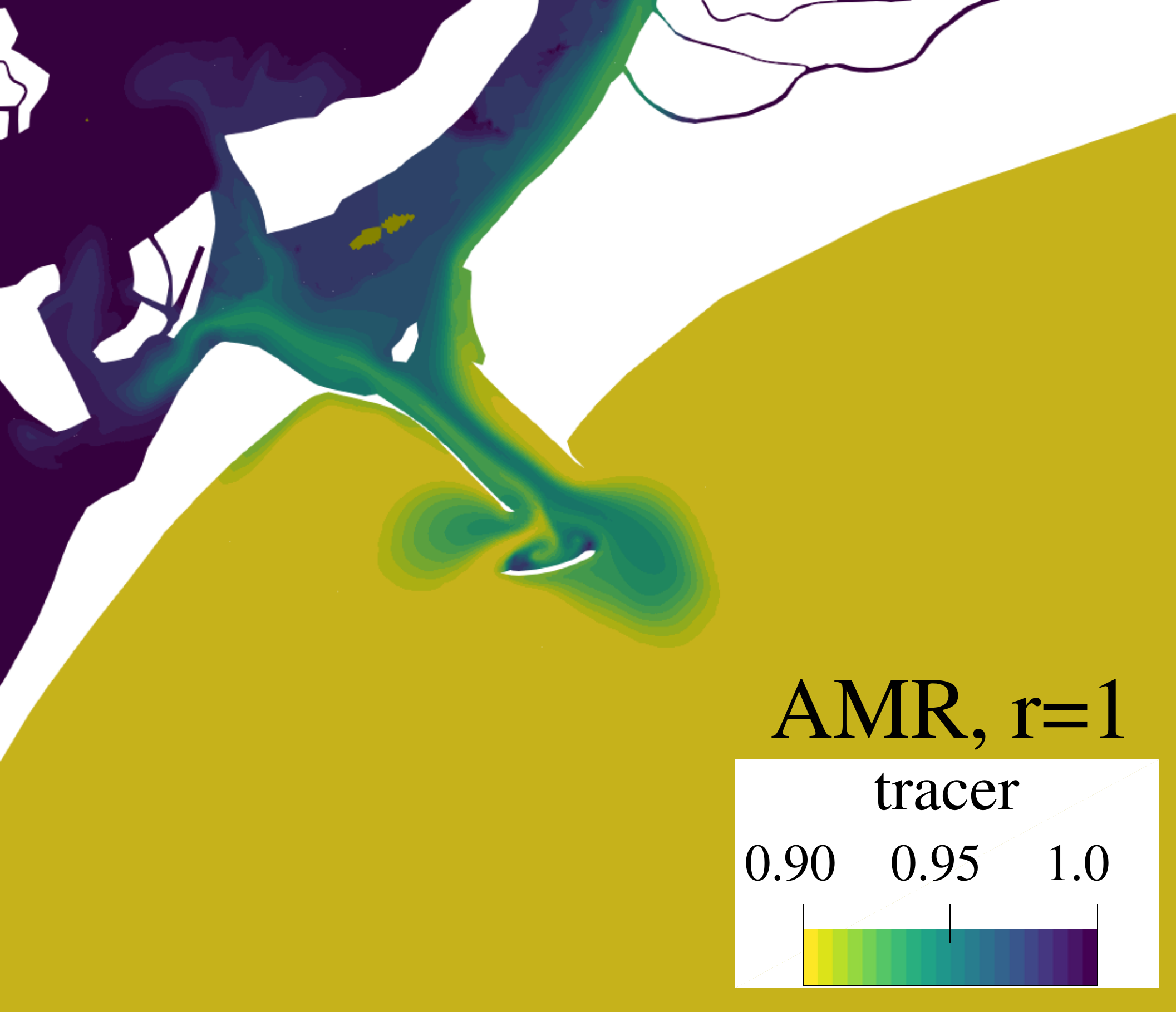}
    \vspace{0.15cm}

    \includegraphics[trim={0cm 0cm 0cm 0cm},clip,scale=0.10]{images/veniceLagoonWithTracer/67maxlevref0_fedegree1/solution_mesh_lido_042-eps-converted-to.pdf}
    \hspace{0.15cm}
    \includegraphics[trim={0cm 0cm 0cm 0cm},clip,scale=0.10]{images/veniceLagoonWithTracer/67maxlevref0_fedegree1/solution_mesh_lido_042-eps-converted-to.pdf}
    \hspace{0.15cm}
    \includegraphics[trim={0cm 0cm 0cm 0cm},clip,scale=0.10]{images/veniceLagoonWithTracer/67maxlevref0_fedegree1/solution_mesh_lido_042-eps-converted-to.pdf}
    \hspace{0.15cm}
    \includegraphics[trim={0cm 0cm 0cm 0cm},clip,scale=0.10]{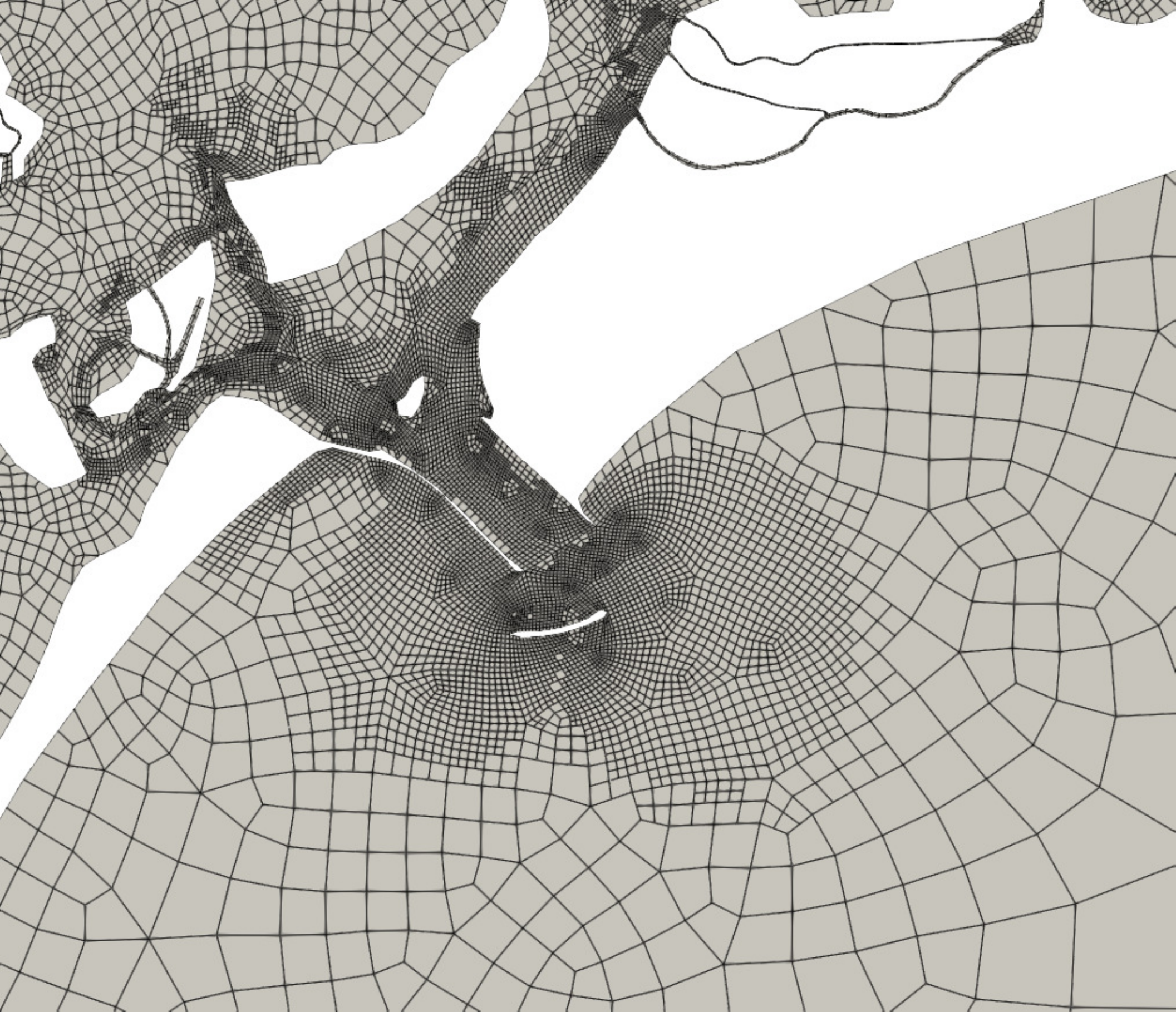}
    \vspace{0.15cm}

    \includegraphics[trim={0cm 0cm 0cm 0cm},clip,scale=0.10]{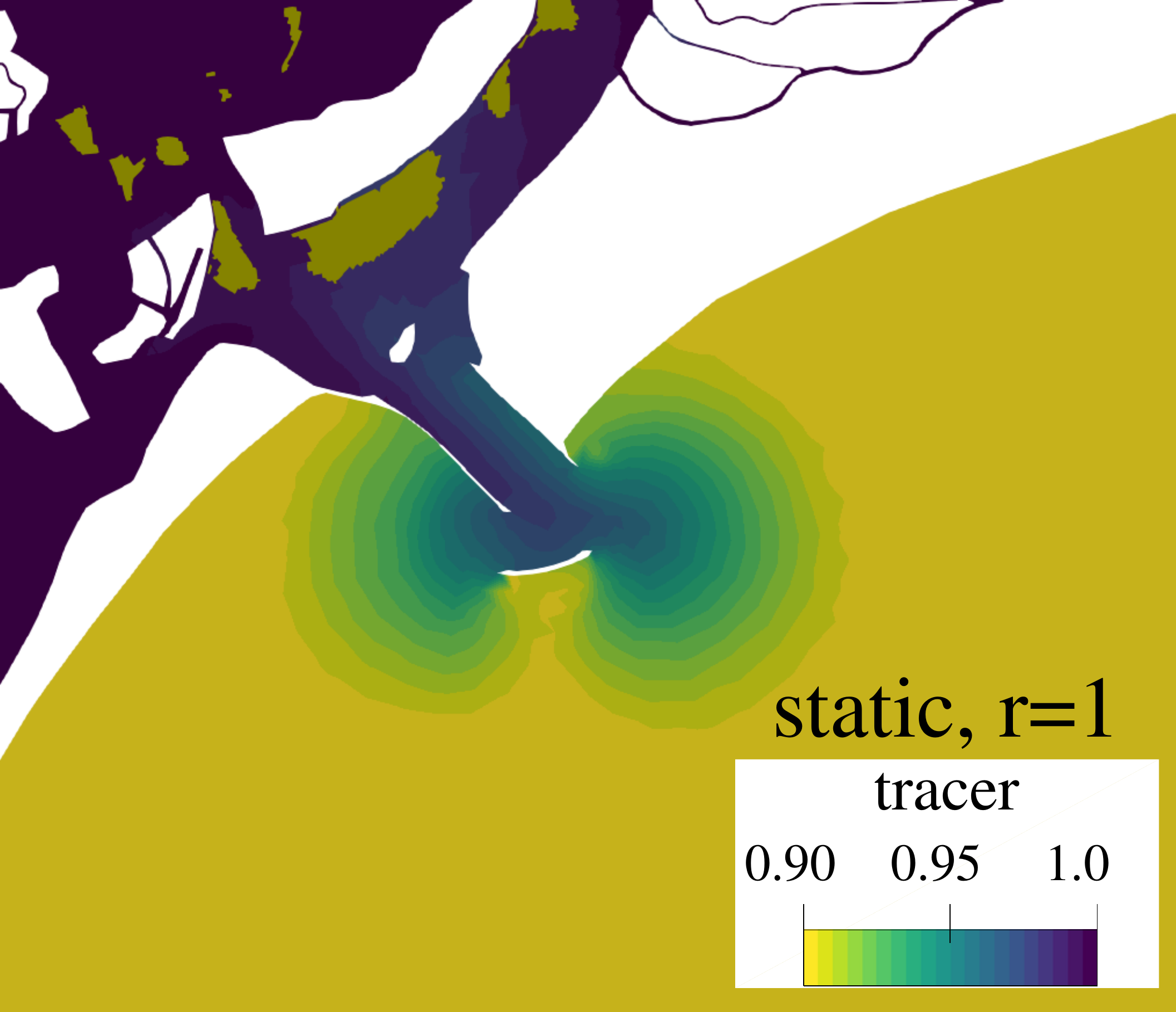}
    \hspace{0.15cm}
    \includegraphics[trim={0cm 0cm 0cm 0cm},clip,scale=0.10]{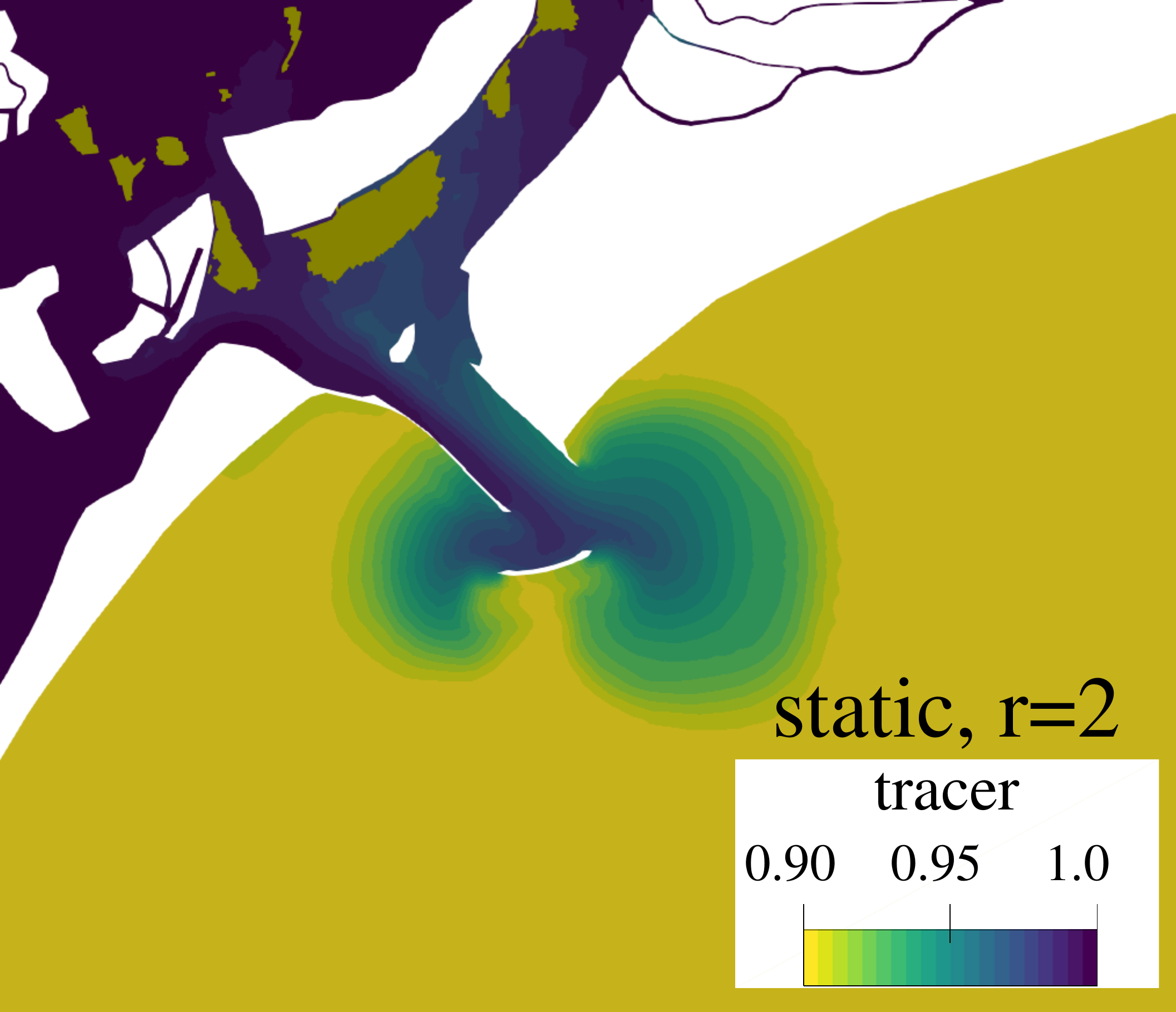}
    \hspace{0.15cm}
    \includegraphics[trim={0cm 0cm 0cm 0cm},clip,scale=0.10]{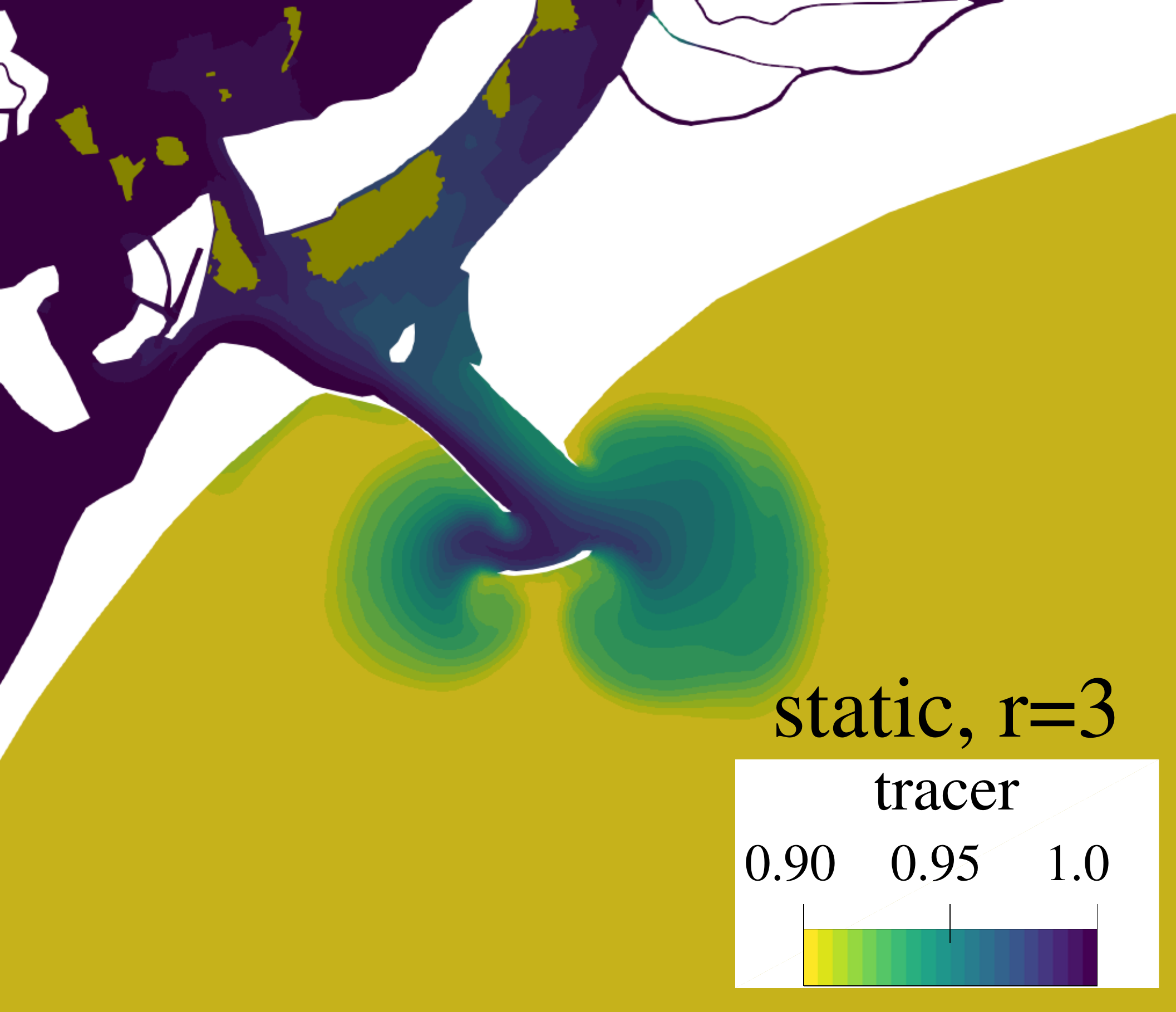}
    \hspace{0.15cm}
    \includegraphics[trim={0cm 0cm 0cm 0cm},clip,scale=0.10]{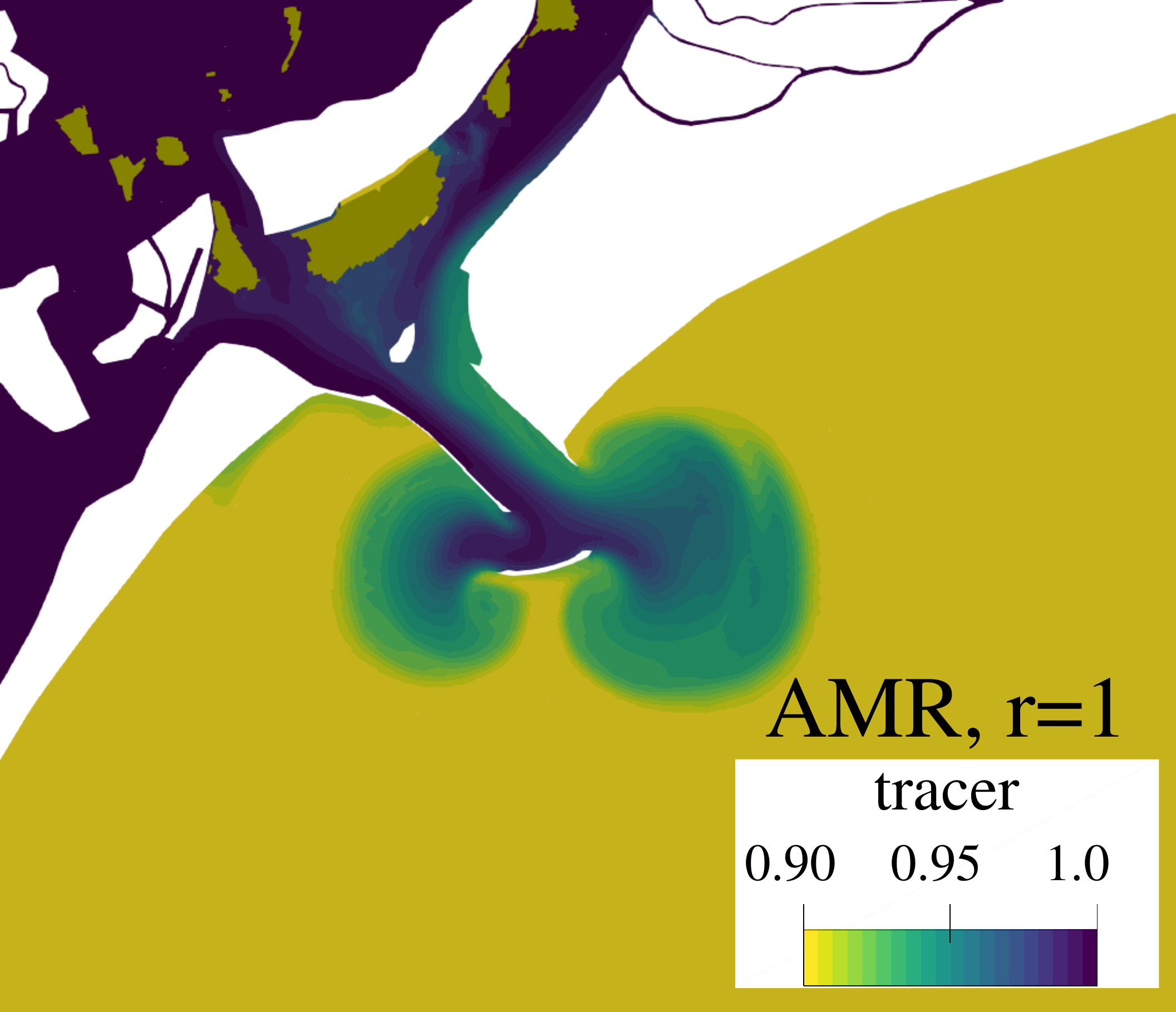}
    \caption{Benchmark with realistic bathymetry, contour plots of the tracer and adaptive mesh at the Lido inlet. Comparison between static mesh with $r=1,2$ and $3$ and adaptive mesh with $r = 1$. First two rows corresponds to high tide ($t = \SI{42}{\hour}$), last two rows to low tide ($t = \SI{48}{\hour}$).}
    \label{fig:veniceLagoon_tracer}
\end{figure}

Then, we present results obtained coupling a tracer concentration equation to the shallow water flow.
For the tracer only, the numerical diffusion described in Equation \eqref{eq:numdiff} has been added, setting $C = 0.33$ based on a sensitivity study. We have simulated a semi-realistic configuration so as to test the capability of our dynamic AMR procedure to improve the accuracy of tracer fronts in coastal areas. To model two different water masses, the tracer is initialized as $c = 1$ inside the lagoon or $c = 0.9$ outside, specifying at the sea boundary a tracer value equal to the initial one $c = 0.9$. 
At each tidal cycle, lower tracer concentrations intrude into the lagoon through the main channels, while higher tracers values form a plume outside the inlets. We provide some qualitative arguments highlighting the advantages of high-order polynomials or the AMR technique in the simulation of tracer fronts. The comparison between the static simulations with $r=1$, $r=2$ and $r=3$, is shown in the first three columns of Figure \ref{fig:veniceLagoon_tracer}, for the Lido inlet and for two different flow conditions corresponding to high and low tide. Using higher order methods allows to compute sharper plumes, fronts and filaments and to resolve small-scale structures that are absent in the second order run. At the Lido inlet, one can observe a sharper gradient formed by lower concentration water intruding into the lagoon from the northern side. This is associated with a stronger and deeper intrusion along the Treporti channel.   
We then run an adaptive simulation, always with $r=1$, with a dynamically adaptive non-conforming mesh. The dynamic mesh has two levels of refinement beyond the coarse mesh, so that $n_{\text{max}} = 3$, assuming a minimum target mesh size $\mathcal{H}_{\text{min}} = \SI{50}{\meter}$. The refinement indicator is based on the gradient of the concentration
\begin{equation}
    \eta_{K} = \max_{q \in K} \nabla c(\bm{x}_{q}),
\end{equation}
where $q$ denotes the index of the set of the quadrature nodes of the Gaussian formula on each element. The thresholds for coarsening and refining are $\theta_{c} = 10^{-2}$ and $\theta_{r} = 5 \times 10^{-3}$. With these choices, we have been able to track the tracer fronts with a variable mesh size between $\SI{50}{\meter}$ and $\SI{200}{\meter}$.
Always in Figure \ref{fig:veniceLagoon_tracer} we compare the results of the static simulations with those of the adaptive simulation. It is apparent that AMR simulation is close to, or even more accurate than, the static simulation with $r=3$ with a reduction of almost $40\%$ of the computation time, as shown in Table \ref{tab:veniceLagoon}.

\begin{figure}[h!]
    \centering
    \includegraphics[trim={0cm 0cm 0cm 0cm},clip,scale=1.1]{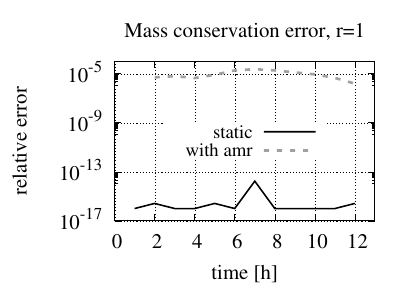}\includegraphics[trim={0cm 0cm 0cm 0cm},clip,scale=1.1]{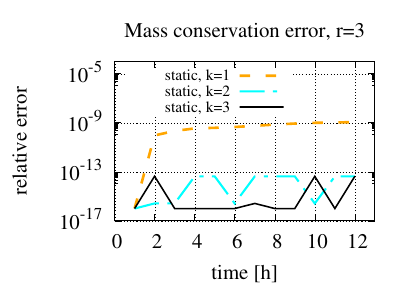}
    \caption{Benchmark with realistic bathymetry, time history of the relative mass conservation errors. Left: Comparison of static (black line) and dynamic adaptive simulations with $r=1$(gray dashed line). Right: Comparison of different numbers of Newton iterations $k$ for the static simulation with $r=3$.}
    \label{fig:veniceLagoon_cons}
\end{figure}

Since, as discussed in Section \ref{ssec:spacedisc_prop}, the simulation with dynamic AMR violates global mass conservation, we have inspected the relative mass conservation error, computed as $\sum\limits_{m=1}^{n+1}\left|Q^{m}\right|/H_{0}$, where the spurious source/sink term $Q^{m}$ is defined as in \eqref{eq:mass-loss} and $H_{0}$ is the total mass at the initial time. The time history of the relative mass conservation error for the static and adaptive simulations is shown in the left picture of Figure \ref{fig:veniceLagoon_cons}. The simulation with AMR exhibits a violation in the conservation of the total mass, of the order of $10^{-5}$. These values are found to be acceptable since they are several orders of magnitude below the uncertainties on the boundary data that would be used, for example, for a forecast system of the circulations in the Venice Lagoon. As found in the original work \cite{casulli:2009}, when using a static mesh, only a few iterations are needed to conserve the total mass up to machine precision, as shown in the right panel of Figure \ref{fig:veniceLagoon_cons}. More specifically for all polynomial degrees ($r=3$ shown in Figure), only two iterations, corresponding at maximum to an overhead of only $2.4\%, $ are required to achieve mass conservation up to machine precision.


\section{Conclusions}
\label{sec:conclu}

We have proposed and validated a high-order Discontinuous Galerkin (DG) discretization for the shallow water equations which exploits high-resolution realistic bathymetry data without any regularity assumption. The proposed discretization is based on an extension of the sub-grid method \cite{casulli:2009} to high-order finite elements. Some modifications of the DG method, tailored for general bathymetries that are not necessarily approximated by polynomials (as is instead the free-surface elevation) have been discussed. Also, these modifications allow to handle wetting and drying with a general bathymetry in a mass-conserving fashion. We prove several mathematical properties: well-balancing, positivity preservation under a mild CFL condition and 
consistency of a conservative discretization for passive tracers. The solver has been implemented in the \texttt{deal.II} library, whose $hp-$capabilities have been exploited to implement dynamic AMR approaches and to reduce the polynomial order in dry regions. We have demonstrated that this strategy maintains a high level of accuracy at a reduced computational cost.

We have performed a wide range of numerical experiments on idealized benchmarks, showing the verification of the mathematical properties and also the robustness in presence of irregular bathymetries with under-resolved features at the grid scale. Results obtained on realistic bathymetries and complex domains show the potential of the method for accurate and efficient adaptive simulations of coastal flows.

In this work we have only tested Gaussian quadrature formulae that do not introduce aliasing errors in the discretization of terms involving polynomials. One could consider further enhancing the sub-grid capabilities as in \cite{casulli:2019} by employing more accurate quadrature formulae. However, we have observed that, when using Gaussian quadrature, the time-step restriction for positivity becomes too severe. As an alternative, simple equispaced quadrature with a large number of nodes could be tested. Only discretization methods have been employed that treat the free-surface gradients in an explicit way, as a first step towards the application of advanced and high-order IMEX techniques. We plan to implement efficient Implicit-Explicit Runge-Kutta (IMEX-RK) time discretization schemes with an implicit treatment of the pressure gradient terms, as already done in \cite{orlando:2022, orlando:2023a} for atmospheric models, thus guaranteeing time step restrictions based on velocity rather than celerity values.

\section*{Acknowledgements}

We are grateful to Georg Umgiesser for several useful discussions that helped us to configure the numerical experiments presented in this paper. L.A has been supported by the project PNRR-HPC ``SPOKE 4 - Earth Climate" of the National Centre for HPC, Big Data and Quantum Computing. G.O. is part of the INdAM-GNCS National Research Group. The simulations have been run at CINECA thanks to the computational resources made available through the ISCRA-C project FemCoast - HP10CKVNZ6. We acknowledge the CINECA award for the availability of high-performance computing resources and support.

\appendix

\section{Coefficients of time discretization schemes}
\label{app:IMEX_coeffs}

We report here for the reader's convenience the detailed description of the employed time discretization methods. The IMEX-RK method is based on the combination of the \texttt{L}-stable TR-BDF2 scheme for the implicit part ($\chi = 1 - \frac{\sqrt{2}}{2}$ in Table \ref{tab:imex_rk2_butch}, left), as discussed in \cite{hosea:1996}, and an explicit three stages second order Runge-Kutta scheme designed to match the coupling and order conditions, which is denoted in the text as RK32 (Table \ref{tab:imex_rk2_butch}, right). The order and coupling conditions for IMEX-RK methods of second order are \cite{kennedy:2016, pareschi:2005}
\begin{align*}
    &\sum_{l=1}^{s}b_{l} = 1, \qquad \sum_{l=1}^{s}\tilde{b}_{l} = 1, \\
    &\sum_{l=1}^{s}b_{l}c_{l} = \frac{1}{2}, \quad \sum_{l=1}^{s}\tilde{b}_{l}c_{l} = \frac{1}{2}, \quad \sum_{l=1}^{s}b_{l}\tilde{c}_{l} = \frac{1}{2}, \quad \sum_{l=1}^{s}\tilde{b}_{l}\tilde{c}_{l} = \frac{1}{2}.
\end{align*}
Under the usual assumption for Runge-Kutta schemes
$$\sum_{m=1}^{s}a_{lm} = c_{l}, \qquad l=1,\dots,s,$$
and assuming $\mathbf{b} = \tilde{\mathbf{b}}$ and $\mathbf{c} = \tilde{\mathbf{c}}$, the explicit companion method is characterized by one free parameter, which we identify with $a_{32}$. The condition $\mathbf{b} = \tilde{\mathbf{b}}$ is necessary for the preservation of linear invariants \cite{giraldo:2013}. Following the analysis on the absolute monotonicity reported in \cite{orlando:2022}, we take $a_{32} = \frac{1}{2}$.

\begin{table}[pos=H]
    \begin{minipage}{0.45\textwidth}
	\begin{center}
		\begin{tabular}{c|ccc}
			$0$ & $0$ & $0$ & $0$  \\
			$2\chi$ & $2\chi$ & $0$ & $0$ \\
            $1$ & $1 - a_{32}$ & $a_{32}$ & $0$ \\
			\hline \\[-0.3cm]
			& $1 - \frac{1 - 2\chi}{4\chi} - \chi$ & 
            $\frac{1 - 2\chi}{4\chi}$ & $\chi$
		\end{tabular}
	\end{center}
    \end{minipage}
     \begin{minipage}{0.45\textwidth}
        \begin{center}
		\begin{tabular}{c|ccc}
			$0$ & $0$ & $0$ & $0$ \\
			$2\chi$ & $\chi$ & $\chi$ & $0$ \\[1.5mm]
			1 & $1 - \frac{1 - 2\chi}{4\chi} - \chi$ &
            $\frac{1 - 2\chi}{4\chi}$ & $\chi$ \\
			\hline \\[-0.3cm]
			& $1 - \frac{1 - 2\chi}{4\chi} - \chi$ & 
            $\frac{1 - 2\chi}{4\chi}$ & $\chi$
		\end{tabular}
	\end{center}
    \end{minipage}
    \caption{Butcher tableaux of the the IMEX-RK method. Left: explicit scheme (RK32). Right: implicit scheme.}
    \label{tab:imex_rk2_butch}
\end{table}

The third order Strong Stability Preserving scheme \cite{gottlieb:2001}, denoted in the text as SSP3, is identified by the Butcher tableau in Table \ref{tab:rk3}, while the classical fourth order Runge-Kutta method \cite{lambert:1991}, denoted in the text as RK44, is identified by the  Butcher tableau in Table \ref{tab:rk4}.

\begin{table}[pos=H]
    \begin{minipage}{0.45\textwidth}
        \centering  
        \begin{tabular}{c|ccc} 
            0 & & & \\[1.5mm]
            $1$ & $1$ & & \\[1.5mm]
            $\frac{1}{2}$ & $\frac{1}{4}$ & $\frac{1}{4}$ & \\[1.5mm]
            \hline \\[-0.3cm]
            & $\frac{1}{6}$ & $\frac{1}{6}$ & $\frac{2}{3}$
        \end{tabular}
        \caption{Butcher tableau of the SSP3 method.}
        \label{tab:rk3}
    \end{minipage}
    \begin{minipage}{0.45\textwidth}
        \centering
        \begin{tabular}{c|cccc} 
            0 & & & & \\[1.5mm]
            $\frac{1}{2}$ & $\frac{1}{2}$ & & & \\[1.5mm]
            $\frac{1}{2}$ & $0$ & $\frac{1}{2}$ & & \\[1.5mm]
            $1$ & $0$ & $0$ & $1$ & \\
            \hline \\[-0.3cm]
            & $\frac{1}{6}$ & $\frac{2}{6}$ & $\frac{2}{6}$ & $\frac{1}{6}$ \\
        \end{tabular}
        \caption{Butcher tableau of the RK44 method.}
        \label{tab:rk4}
    \end{minipage}
\end{table}

\bibliographystyle{cas-model2-names}
\bibliography{shwdeal2.bib}
	
\end{document}